\documentclass{aa} 

\usepackage{graphicx}
\usepackage[varg]{txfonts}
\usepackage{amsmath}
\usepackage{mathrsfs} 
\usepackage{multirow}
\usepackage{lscape}
\usepackage{nccmath}
\usepackage{placeins}
\usepackage{color}
\usepackage{xcolor}
\usepackage{hyperref}
\usepackage{siunitx}
\usepackage[normalem]{ulem}
\hypersetup{
  unicode=true,      
  pdftoolbar=true,    
  pdfmenubar=true,    
  pdffitwindow=true,   
  pdftitle={Mass-ratio distribution of contact binary stars},
  pdfauthor={Pe\v{s}ta \& Pejcha},
  pdfkeywords={},     
  pdfnewwindow=true,   
  colorlinks=true,    
  linkcolor=blue,     
  citecolor=blue,     
  filecolor=gray,     
  urlcolor=blue      
}

\makeatletter
\renewcommand*\aa@pageof{, page \thepage{} of \pageref*{LastPage}}
\makeatother

\newcommand{\qmin}{\ensuremath{q_\text{min}}}
\newcommand{\lambdak}{\ensuremath{\lambda_\text{K}}}

\graphicspath{{./}{figures/}}

\begin{document}

\title{Mass-ratio distribution of contact binary stars\thanks{The full Table \ref{tab:online_data} is available
at the CDS via anonymous ftp to \url{cdsarc.cds.unistra.fr} (130.79.128.5)
or via \url{https://cdsarc.cds.unistra.fr/cgi-bin/qcat?J/A+A/}.}}
  \titlerunning{Mass-ratio distribution of contact binary stars}
  \authorrunning{Pe\v{s}ta \& Pejcha}
\author{Milan Pe\v{s}ta
\and Ond\v{r}ej Pejcha}

\institute{Institute of Theoretical Physics, Faculty of Mathematics and Physics, Charles University, V Hole\v{s}ovi\v{c}k\'{a}ch 2, Praha 8, 180 00, Czech Republic}

\date{Received 2022}

\abstract{
The mass ratio $q$ of a contact binary star evolves through mass transfer, magnetic braking, and thermal relaxation oscillations to low values until it crosses a critical threshold $q_\text{min}$. When this occurs, the binary undergoes the tidal Darwin instability, leading to a rapid coalescence of the components and to an observable brightening of the system. The distribution of $q$ has not been measured on a sufficiently large population of contact binary stars so far because determining $q$ for a single contact binary usually requires spectroscopy. As was shown previously, however, it is possible to infer the mass-ratio distribution of the entire population of contact binaries from the observed distribution of their light-curve amplitudes. Employing Bayesian inference, we obtained a sample of contact binary candidates from the Kepler Eclipsing Binary Catalog combined with data from \textit{Gaia} and estimates of effective temperatures. We assigned a probability of being a contact binary of either late or early type to each candidate. Overall, our sample includes about 300 late-type and 200 early-type contact binary candidates. We modeled the amplitude distribution assuming that mass ratios are described by a power law with an exponent $b$ and a cutoff at $q_\text{min}$. We find $q_\text{min}=0.087^{+0.024}_{-0.015}$ for late-type contact binaries with periods longer than 0.3 days. For late-type binaries with shorter periods, we find $q_\text{min}=0.246^{+0.029}_{-0.046}$, but the sample is small. For early-type contact binary stars with periods shorter than one day, we obtain $q_\text{min}=0.030^{+0.018}_{-0.022}$. These results indicate a dependence of $q_\text{min}$ on the structure of the components, and they are broadly compatible with previous theoretical predictions. We do not find any clear trends in $b$. Our method can easily be extended to large samples of contact binaries from TESS and other space-based surveys.}


\maketitle

\section{Introduction} \label{sec:intro}
A contact binary system consists of two stars that have filled their Roche lobes and started sharing a single envelope. The luminosity generated by the individual stars is efficiently distributed through the envelope, leading to a nearly constant temperature across the whole shared surface, regardless of the masses of the components \citep{lucy68a,lucy68b,shu76,shu81}. If the orbital inclination is sufficiently high, contact binaries can be observed as eclipsing W Ursae Majoris (W UMa) or EW variables. In the rest of the paper, we use the terms contact binary, W UMa variable, and EW variable interchangeably. The class of W UMa variables is characterized by equal-depth primary and secondary eclipses and by periods from approximately 0.22 days up to about one day \citep[e.g.,][]{rucinski07,jiang12}. There are two subclasses of W UMa variables: A type (or early type) and W type (or late type). In A-type systems, both components are A or F stars, in contrast to G-K stars, which make up W-type systems. \citet{Jayasinghe_2020J} showed that this dichotomy of W UMa variables is most likely related to their location relative to the Kraft break, which is a sudden drop in the average rotation rate of stars in the temperature range 6200--6700 K. The drop is caused by the different efficiency of magnetic braking for stars possessing or lacking subsurface convection zones \citep{kraft67}. The two subclasses also differ in the slopes of their period--luminosity--color (PLC) relations, which result from the blackbody relation applied to Roche-lobe filling stars \citep{Rucinski_1994,Rucinski_2004,Pawlak_2016}. Additionally, \citet{Stepien_2012} argued that W-type systems with periods shorter than $0.3$\,days form a distinct population that is different from both A types and longer-period W types. 

Observations suggest that the majority of contact binaries originate in triple systems \citep[e.g.,][]{pribulla06,dangelo06,Hwang_2023}. Some of these triples might condense directly out of the star-forming region, but they are more likely the result of dynamical interaction between independent binaries or higher multiples followed by the ejection of the excess stars \citep[e.g.,][]{bate02,tokovinin14,Antognini_2016}. Under the right conditions, the inner binary in the triple system is subject to the von Zeipel--Lidov--Kozai mechanism, forcing the orbital eccentricity and inclination to undergo long oscillation cycles. The cycles lead to the extraction of orbital energy through tidal friction and the inner orbit gradually shrinks \citep{Lidov_1962,Kozai_1962,eggleton01,fabrycky07,naoz16}. This process is no longer efficient when the orbital period of the inner binary reaches about 1--3 days. At that point, either magnetic braking or nuclear evolution of the more massive component takes over, and coupled with tidal friction, it reduces the period even further, leading to the formation of a contact binary system \citep{Eggleton_2006,Hwang_2020}. The relative importance of the two mechanisms depends on the mass of the stars in the binary. Magnetic braking is thought to be the driving force in the formation of W-type systems, while nuclear evolution is most likely the dominant mechanism in the precontact phase of A-type systems \citep{Yildiz_2014}. 

After it is formed, a contact binary evolves toward low mass ratios $q$ on the timescale of the dominant evolutionary process. The evolution to small $q$ is not linear, and the binary goes through a series of thermal relaxation oscillations (TROs), during which the flow of mass is temporarily reversed and the contact is broken \citep{Lucy_1976,Flannery_1976,Robertson_1977,Yakut_2005,Paczynski_2006}. The cycle length of TROs is set by the thermal timescale of the secondary, which grows as $q$ decreases. As a result, contact binaries pile up at small $q$ \citep{Rucinski_2001}. \citet{Stepien_2006,Stepien_2011} and \citet{Stepien_2012} proposed an alternative to the TRO model, which assumes rapid initial mass transfer followed by a mass-ratio inversion and linear evolution toward small $q$. Regardless of the actual mechanism, the trend toward unequal masses continues until the binary becomes unstable due to the tidal Darwin instability, which occurs when the spin angular momentum of the more massive component exceeds one-third of the orbital angular momentum of the system \citep[e.g.,][]{Darwin_1879,Hut_1980}. The angular momentum criterion translates into a minimum mass ratio $\qmin$ \citep{Webbink_1976,Rasio_1995}. The exact value of $\qmin$ depends on the stellar structure and masses of the components, but theoretical models generally predict values below 0.1 \citep{Rasio_1995,Li_2006,Arbutina_2007,Arbutina_2009,wadhwa21}. Alternatively, the binary can expand and overflow its outer critical surface before it reaches $\qmin$, leading to a rapid mass and angular momentum loss through the vicinity of the L2 point \citep{webbink77,shu79,Stepien_2012,pejcha16a,pejcha16b,hubova19}.
When the contact binary becomes unstable due to either of the two mechanisms, it enters the dynamical common envelope phase, which is accompanied by a luminous red nova transient \citep[e.g.,][]{Tylenda_2011,ivanova13_rev,ivanova13_sci,pejcha14,pejcha17,macleod17,blagorodnova21} and leads to a single, rapidly rotating remnant \citep{paczynski07}. 

The effects of all the evolutionary processes are imprinted on the mass-ratio distribution of contact binaries. Since many of these processes, such as magnetic braking and thermal and tidal instabilities, are not completely understood, we might be able to illuminate them by studying the observed mass-ratio distribution \citep{vilhu81}. Surprisingly, little work has been done to observationally constrain the distribution of $q$ on sufficiently large and homogeneous samples of contact binaries with a well-understood selection function. The reason is that to accurately estimate $q$ of a contact binary system, spectroscopy of both components is typically required. Another option is to infer $q$ directly from photometry, but this method does not yield reliable results due to the degeneracy of contact binary light curves with respect to $q$ and the orbital inclination. An exception to this is the special case of totally eclipsing contact systems, for which the degeneracy is lifted and $q$ can be reliably estimated, but precise photometry is required to resolve the shape of the minimum for the low amplitudes expected from systems with small $q$ \citep{Rucinski_2001,Terrell_2005,Hambalek_2013}. \citet{Yakut_2005} investigated parameters of about $100$ binaries close to contact, but their systems were collected from the literature and could be a very biased representation of the actual population. More efforts have focused on the determination of $\qmin$. For several contact binaries, $q$ is close to the theoretically predicted minimum value \citep[e.g.,][]{paczynski07,Li_2021,wadhwa21,Popov_2022,Christopoulou_2022}, but it is unclear how these detections relate to the entire population. Recently, \citet{Kobulnicky_2022} performed a computationally expensive Monte Carlo exploration of the light-curve parameter space for about 200 contact binaries and found that $\qmin$ increases with orbital period from $0.044$ at 0.74\,days to $0.15$ at 2\,days. However, none of these approaches scale well to the large amounts of astronomical data that have recently become available or will become available in the future.

To overcome these issues, \citet{Rucinski_2001} developed an independent method for the inference of the mass-ratio distribution using only photometric amplitudes extracted from contact binary light curves. The method does not require modeling of each contact binary system in the sample individually, but rather it exploits the strong correlation between the shape of the mass-ratio distribution and the photometric amplitude distribution of contact binary stars. \citet{Rucinski_2001} constructed a sample of contact binaries from ground-based data and modeled the mass-ratio distribution as a power law with a cutoff at $\qmin$, but could not obtain any reasonable constraint on $\qmin$ due to the insufficient sensitivity of ground-based photometry, blending, and the limited size of their sample. Another complication is that at low photometric amplitudes, contact binary samples can be contaminated by unresolved companions, ellipsoidal variables, or various types of pulsating stars \citep{skarka22}. The advantage of the method of \citet{Rucinski_2001} is that it requires neither spectroscopy nor tedious modeling of individual objects, and it is therefore very well suited for current and future massive high-precision photometric surveys.

Our goal is to characterize the mass-ratio distribution and $\qmin$ of contact binaries with the help of high-precision photometric amplitudes that are available from space-borne telescopes. In Section~\ref{sec:method} we formulate the ideas of \citet{Rucinski_2001} in the framework of Bayesian inference. In Section~\ref{sec:data} we describe our initial sample of contact binary candidates from the Kepler Eclipsing Binary Catalog. In Section~\ref{sec:contaminants} we present a Bayesian model for selecting a clean sample of contact binaries using the PLC relation. In Section~\ref{sec:results} we infer the mass-ratio distribution of contact binary stars, and we investigate its dependence on various parameters of our model. Finally, in Section \ref{sec:conclusions} we summarize our findings and discuss possible future extensions and applications of the method.

\section{Method} \label{sec:method}
The light curves of contact binary stars are special because their shapes depend more strongly on the geometrical features of the system than on the intrinsic properties of the stellar components. Due to the transfer of mass and energy in the system, the two components have nearly identical effective temperatures, rendering the light curve amplitude $a$ almost exclusively dependent on the orbital inclination $i$, the fill-out factor $f$, and the mass ratio $q$, that is, $a = a(i, f, q)$. The fill-out factor is usually defined by linearly relating the photospheric Roche potential to the potentials of the L1 and L2 points, giving $f=0$ for stars barely touching at L1 and $f=1$ for stars starting to overflow L2. The mass ratio is defined as the ratio of the less massive star to the more massive component, that is, $q\le 1$. Higher-order physical effects such as gravitational and limb darkening do not significantly affect $a$, but rather influence the shape of the light curve. These effects can be used to alleviate the degeneracy between $i$, $f$, and $q$, but their usefulness is reduced by the necessary time-consuming modeling and the effects of stellar spots. Depending on their distribution on the surface, spots can deform the light curve and decrease or increase the observed $a$. The position and size of spots changes over time, which suggests that estimates of $a$ can be improved by averaging data taken over longer periods of time. Furthermore, not the amplitude of any single system, but the overall distribution constructed from many systems is important.

In this section, we give a general overview of the procedure to obtain the distribution of $q$ from the observed distribution of $a$ based on \citet{Rucinski_2001} (Sect.~\ref{sec:method_overview}). We describe the functional form of the mass-ratio distribution (Sect.~\ref{sec:q_prescription}) and the construction of light curves and amplitude distributions (Sect.~\ref{sec:amplitude_distribution}), and we present the Bayesian procedure for finding posteriors (Sect.~\ref{sec:likelihood_construction}).

\subsection{Overview of the method} \label{sec:method_overview}

\begin{figure*}
\centering
\includegraphics[width=0.705\textwidth]{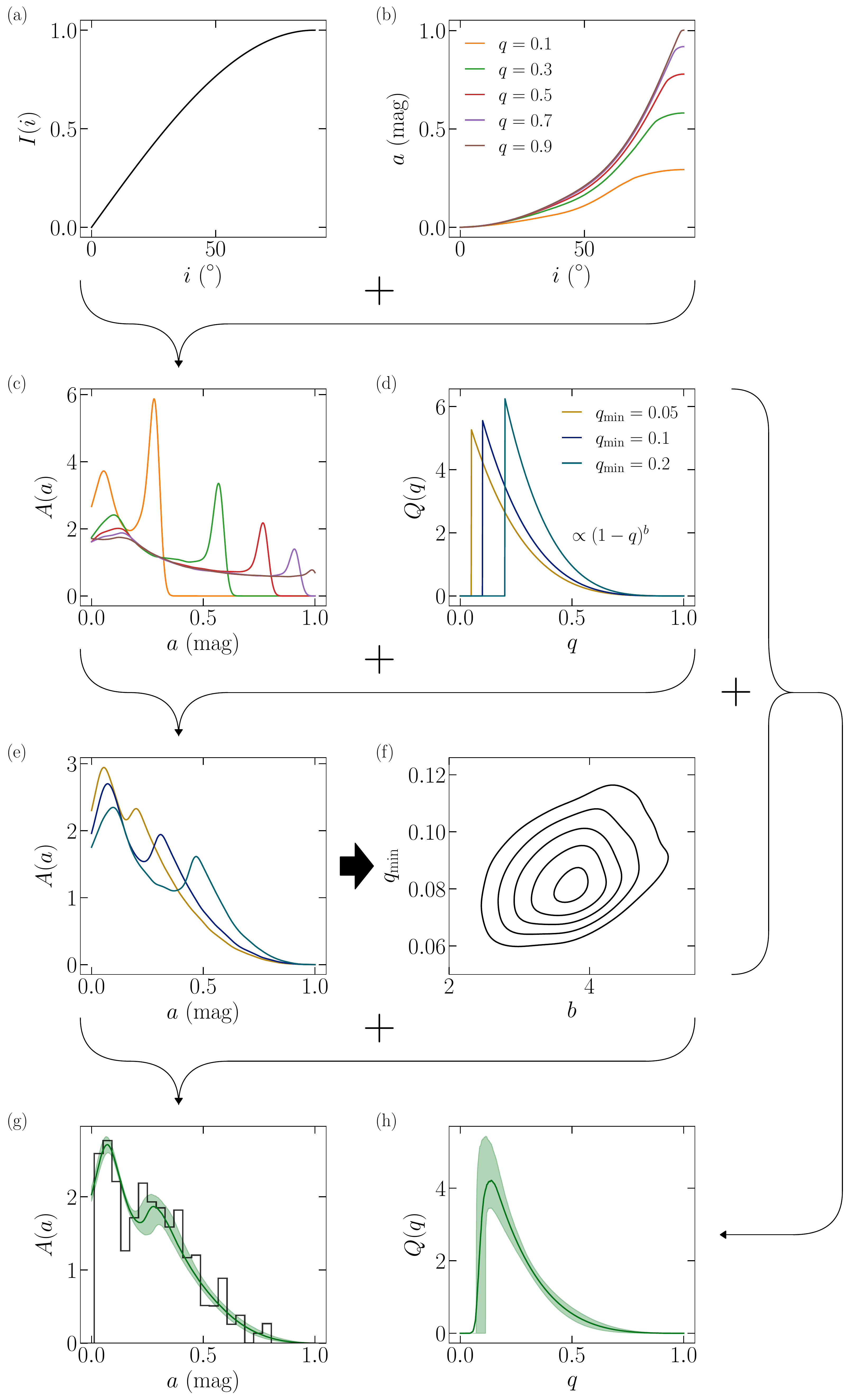}
\caption{Summary of our method for the inference of the mass-ratio distribution of contact binary stars. \emph{(a)} Assuming the orbits of contact binary systems are randomly distributed in space, the probability of observing a system with an inclination $i$ is proportional to $\sin{i}$. \emph{(b)} By using light-curve synthesis models, we derive the contact binary photometric amplitude $a$ as a function of $i$ for different values of the mass ratio $q$. \emph{(c)} By marginalizing out the inclination, we obtain $a$ as a function of $q$. \emph{(d)} We approximate the mass-ratio distribution by a power law with index $b$ and a sharp cutoff at the minimum mass ratio $\qmin$. \emph{(e)} By using the power law to marginalize out $q$, we construct the full photometric amplitude distribution, with its shape strongly depending on the value of $\qmin$. \emph{(f)} We apply Bayesian inference to fit the amplitude distribution to a sample of contact binary stars, yielding the posterior distribution of the parameters of the model. \emph{(g)} We marginalize over the posterior and get the mean amplitude distribution (solid green line) and its $1\sigma$ credible interval (green band). \emph{(h)} Repeating the same procedure, we obtain the mean mass-ratio distribution and its $1\sigma$ credible interval.}
\label{fig:model_summary}
\end{figure*}

The observed distribution of light-curve amplitudes $A(a)$ for a constant $f$ is given by
\begin{equation}
\label{eq:master}
\begin{split}
  A(a) & = \int \delta\left(a'(i,q) - a\right) I(i) Q(q) \text{d}i \text{d} q = \\
     & = \int I(i) Q(q(i,a)) \frac{\partial q}{\partial a}(a,i) \text{d} q,  
\end{split}
\end{equation}
where we assume that the joint probability distribution of $(i,q,f)$ can be separated into individual components, specifically, $I(i)$ is the distribution of $i,$ and $Q(q)$ is the distribution of $q$, which we aim to obtain. The second line of Eq.~\eqref{eq:master} works out an explicit form for $A(a)$ by assuming that $a(i,q)$ can be inverted to give a function $q(i,a)$. This form of $A(a)$ is similar to what was derived by \citet{Rucinski_2001}. For most of this work, we suppress the dependence of $a(i,q)$ on $f$ by assuming that all contact binaries have the same value of $f$. Detailed light-curve models as well as the TRO theory suggest that $f$ is typically small, $0\lesssim f \lesssim 0.5$, and that its distribution has a poorly defined maximum around $f\approx 0.25$ \citep{Lucy_1973,Rucinski_1973,Rucinski_1997}. However, there are some indications that early-type binaries have higher $f$ than late-type ones \citep{mochnacki81}. Still, we chose $f=0.25$ as the default value, and we investigate the sensitivity of our results to $f$ in Sect.~\ref{sec:dependence_fill_out}. 

In Fig.~\ref{fig:model_summary} we outline our procedure for obtaining $Q(q)$ from $A(a)$. The key assumptions are that the inclinations are distributed isotropically, that is, $I(i) \propto \sin i$ (Fig.~\ref{fig:model_summary}a), and that the function $a(i,f,q)$ can be calculated with a binary light-curve synthesis code (Fig.~\ref{fig:model_summary}b). For a constant $q$, the distribution of $a$ is peaked near the maximum $a$ achievable for the given $q$ (Fig.~\ref{fig:model_summary}c). Following \citet{Rucinski_2001}, we modeled $Q(q)$ as a power law with a cutoff at $\qmin$ (Fig.~\ref{fig:model_summary}d, Sect.~\ref{sec:mass_ratio_distribution}). The existence of $\qmin$ gives rise to a local maximum in $A(a)$ and prevents $A(a)$ from diverging as $a\xrightarrow{}0$. The value of $\qmin$ is directly related to the location of the local maximum, while the exponent in the power law controls the overall shape and slope of $A(a)$ (Fig.~\ref{fig:model_summary}e).

By employing a specific goodness-of-fit metric, it is possible to find the optimal value of the two parameters that best match the observed data. \citet{Rucinski_2001} used $\chi^2$-minimization applied to the binned histogram of $A(a)$. This is problematic, because binning leads to loss of information and the result might depend on the specific choice of the bins. In this work, we use Bayesian inference, which is applicable to both binned and continuous data and works even with small samples. By applying the Bayes theorem, we obtain the posterior distributions of the mass-ratio distribution parameters (Fig.~\ref{fig:model_summary}f). By marginalizing over the posteriors, we smooth out the amplitude distribution and the mass-ratio distribution, obtaining their $1\sigma$ credible intervals in the process (Figs.~\ref{fig:model_summary}g and \ref{fig:model_summary}h). 

\subsection{Mass-ratio distribution} \label{sec:q_prescription}
Motivated by \citet{Rucinski_2001}, we considered two different power-law prescriptions for $Q(q)$, which should capture the essential manifestations of contact binary evolution, specifically, the pile-up of objects at small $q$ and a cutoff at $\qmin$. The two prescriptions are 
\begin{subequations}
\label{eq:Q}
\begin{eqnarray} \label{eq:Q_1}
Q_1(q;\Theta)&=&
  \begin{cases}
  \frac{1}{K} q^{-b} & \text{if } \qmin< q \le 1,\\
  0 & \text{else},
  \end{cases}\\
Q_2(q;\Theta)&=&
  \begin{cases}
  \frac{1}{K} (1-q)^b & \text{if } \qmin< q \le 1,\\
  0 & \text{else},
  \end{cases}
  \label{eq:Q_2}
\end{eqnarray}
\end{subequations}
where $\qmin$ represents the theoretical minimum $q$ cutoff due to the Darwin instability, $b$ controls the slope of the power law, and $\Theta=(\qmin,b)$. The parameter $b$ encodes the effect of nuclear evolution, magnetic braking, and TROs. The normalization constant $K$ ensures that $Q_1$ and $Q_2$ integrate to unity. 

\subsection{Amplitude distribution and light-curve synthesis} \label{sec:amplitude_distribution}

\begin{figure}
\begin{center}
\includegraphics[width=0.48\textwidth]{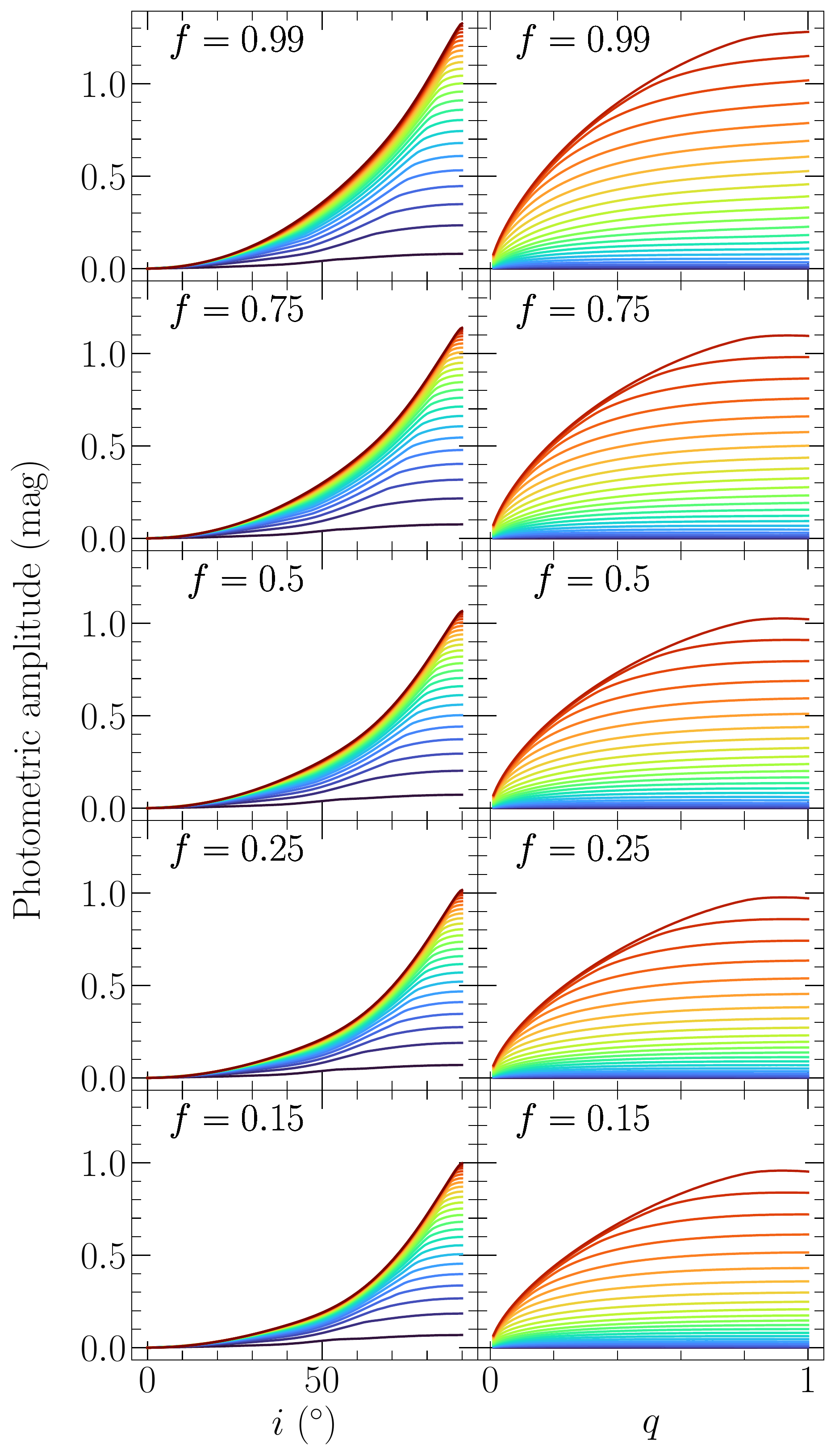}
\end{center}
\caption{Dependence of the contact binary photometric amplitude on the orbital inclination $i$ for different fixed values of the mass ratio $q$ (left) and vice versa (right), conditional on distinct values of the fill-out factor $f$. For each panel, the transition from blue to red indicates the gradual increase in the fixed parameter from its minimum to its maximum value.
 \label{fig:amplitude_dependence}}
\end{figure}

After defining $I(i)$ and $Q(q;\Theta)$, we could use Eq.~\eqref{eq:master} to obtain $A(a;\Theta)$, but this requires inverting $a(i,q)$ to give $q(i,a)$ and calculating its derivative with respect to $a$. This is difficult, because $a(i,q)$ is not an analytic function of its arguments and a forward light-curve synthesis model is needed to get from $(i,q,f)$ to $a$. Moreover, the derivative must be calculated numerically, which amplifies any numerical noise introduced in the calculation of $q(i,a)$. As \citet{Rucinski_2001} showed, a much simpler option is to sample the joint distribution $I(i)\times Q(q;\Theta)$ and obtain $A(a;\Theta)$ by repeated evaluation of $a(i,q)$.

We used PHOEBE version 2.3.58 \citep{prsa05,prsa16,conroy20} to derive the functional form of $a$. We started by initializing the default contact binary supplied by PHOEBE. To make $f$ unconstrained, we flipped the constraints on the potential of the envelope and the equivalent radius. Next, we changed the effective temperature of both components to $5700$\,K and we set the passband to ``Kepler:mean''. We did not change the default limb darkening and gravitational brightening coefficients. We evaluated this model on a three-dimensional grid of $i$, $q$, and $f$, and for each value of $f,$ we performed linear interpolation of $a$ as a function of $i$ and $q$. The grid covers $0 \le i \le 90^\circ$ with a step of $0.5^\circ$ and $0.01 \le q \le 1$ with a step of $0.01$. We do not expect $f$ to have a significant impact on the shape of the light curves, therefore, we considered only six discrete values $f = (0.15, 0.25, 0.5, 0.75,\text{and }0.99)$. We show the linearly interpolated $a$ as a function of $i$ and $q$ in Fig.~\ref{fig:amplitude_dependence}.

Next, we obtained the synthetic distribution of $a$ from $Q(q;\Theta)$ by randomly sampling the joint distribution of $i$ and $q$, $I(i)Q(q)$, for a given $f,$ and we evaluated $a$ for each sample using linear interpolation. To get from the synthetic amplitude distribution to $A(a;\Theta)$, we employed kernel density estimation (KDE), which is a nonparametric approach for estimating the probability density function from a finite sample by replacing each localized observation with a delocalized kernel function. By performing KDE with a normal kernel of bandwidth $h$, we obtained an analytical approximation of $A(a;\Theta)$.

For our KDE, we chose to use a linear combination of Gaussians centered on the drawn amplitudes with standard deviations given by $h$. For this to converge to $A(a;\Theta)$ as the number of drawn samples goes to infinity, we ought to correct for the boundary effect, which is a downward bias near the boundaries of the KDE support. The effect is present only when the modeled distribution is nonzero near the boundaries, which is the case for $A(a\xrightarrow{}0),$ as demonstrated by \citet{Rucinski_2001}. To mitigate the boundary effect, we followed the procedure for a first-degree boundary correction outlined by \citet{Jones_1993}. We also adopted a sufficiently high value for the right boundary, so that $A(a;\Theta)$ effectively goes to zero as $a$ approaches this value ($a\gtrsim3$ mag). This left us with a correctly calibrated analytical approximation of $A(a;\Theta)$, which we denote by $\hat{A}(a;\Theta)$.

The stochastic process involved in the generation of the synthetic distribution entering the KDE means that $\hat{A}(a;\Theta)$ is not a deterministic function of $\Theta$. In other words, repeated evaluation of $\hat{A}(a;\Theta)$ for the same $\Theta$ will give slightly different results depending on the number of samples entering the KDE. We explore the nondeterministic property of $\hat{A}(a;\Theta)$ in more detail in Appendix~\ref{app:likelihood}.

\subsection{Likelihood construction} \label{sec:likelihood_construction}
Given a sample of contact binaries, we can infer the mass-ratio distribution by assuming a specific $Q(q;\Theta)$ and fitting the resulting $A(a;\Theta)$ to the observed distribution of photometric amplitudes. In the Bayesian framework, this is achieved by evaluating the Bayes theorem and updating the prior distribution of the model parameters based on the observed data, resulting in the posterior distribution of the parameters. 

The most essential ingredient entering the Bayes theorem is the likelihood function $\mathscr{L}(\text{model\ parameters}|\text{data})$, which is the probability of observing the data given the parameters of the model, viewed as a function of these parameters. Assuming the observed amplitudes are drawn independently from $A(a;\Theta)$, the likelihood $\mathscr{L}(\Theta|\{a_k\}_{k=1}^N)$ of $\Theta$ given a sample of $N$ amplitudes $\{a_k\}_{k=1}^N$ is the product of the individual generative distributions weighted by the probability of each object being a contact binary star,
\begin{equation} \label{eq:Theta_likelihood}
  \mathscr{L}(\Theta|\{a_k\}_{k=1}^N) = \prod_{k=1}^N p_{\text{CB},k} \int \hat{A}(a;\Theta) \mathscr{N}(a; a_k, \sigma_{a_k})\text{d}a,
\end{equation}
where $p_{\text{CB},k}$ is the weight of the $k$-th object and we use the KDE approximation $\hat{A}(a;\Theta)$ instead of the true distribution $A(a;\Theta)$. In addition to the sampling noise of $\hat{A}(a;\Theta)$, each observed $a_k$ also comes with its own uncertainty $\sigma_{a_k}$, as we discuss in more detail in Sect.~\ref{sec:amplitude_determination}. To factor the uncertainties into the model, we convolved $\hat{A}(a;\Theta)$ with normalized Gaussians $\mathscr{N}(a;a_k,\sigma_{a_k})$, which we used to model the uncertainty of the observed amplitudes. This procedure smears the likelihood in Eq.~\eqref{eq:Theta_likelihood} even further. Fortunately, the specific form of $\hat{A}(a;\Theta)$ given by a sum of Gaussians makes it relatively fast and straightforward to perform the convolutions.

We used \emph{emcee} \citep{Foreman_Mackey_2013} to sample the posterior distribution of the parameters. As a compromise between accuracy and computational cost, most runs were carried out with a KDE smoothing bandwidth $h=0.02$ and KDE number of Gaussians $n=10000$. Running on 16 logical cores in parallel, a typical run with 16 walkers and 2500 steps takes 30 to 50 minutes to complete. We discarded the first 500 steps of each chain as burn in. We also thinned the chains by a factor of $20$, which is higher than the autocorrelation time of most of our runs (Table \ref{tab:runs}).

\section{Data} \label{sec:data}
Our method for the inference of the mass-ratio distribution requires highly precise photometric measurements of contact binary light curves. At the time of writing, no catalog of contact binaries is available that would satisfy this requirement. For this reason, we constructed our own sample of contact binaries based on the photometry from \textit{Kepler} \citep{borucki10}. First, we took the Kepler Eclipsing Binary Catalog \citep[KEBC;][]{Prsa11,Kirk_2016,abdulmasih16} (Sect.~\ref{sec:KEBC}) and combined it with data from \textit{Gaia} and other catalogs (Sect.~\ref{sec:cross_match}). Finally, we determined the photometric amplitude of each object in the sample using detrended \textit{Kepler} fluxes (Sect.~\ref{sec:amplitude_determination}). 

\subsection{Kepler Eclipsing Binary Catalog} \label{sec:KEBC}
The third revision of the KEBC contains 2920 eclipsing and ellipsoidal systems in the primary mission field of view of \textit{Kepler}. The online version of the catalog\footnote{\url{http://keplerEBs.villanova.edu}} also includes the data from the K2 mission \citep[K2 Engineering and C1--C5;][]{howell14}, increasing the total number of observed systems to 3584. The selection function of stars observed by \textit{Kepler} is well understood \citep{batalha10}, and we mitigated its effects by considering late- and early-type contact binaries separately (Sect.~\ref{sec:contaminants}). The construction of the KEBC involved manual filtering of objects that might affect the selection efficiency as a function of amplitude. However, the photometric precision of \textit{Kepler} allows comfortable detection of signals with $a \ll 0.01$\,mag, which is much smaller than $a \approx 0.2$\,mag, where we expect the local maximum of the amplitude distribution (Fig.~\ref{fig:model_summary}). We therefore did not perform any correction of the sample, effectively assuming that the selection efficiency is $100\%$ in the range of amplitudes of our interest.

In addition to basic astrometric and photometric data, the catalog also contains output from the Kepler Eclipsing Binary Pipeline, which is a collection of several methods that are used to extract additional information from the data. For instance, locally linear embedding \citep[LLE;][]{Matijevic_2012} is used to obtain the \emph{morph} parameter that quantifies the detachedness of binary systems. Values of \emph{morph} between 0 and 0.5 indicate a detached binary system, while over-contact systems usually have values from 0.7 to 0.8. The interval from 0.5 to 0.7 is occupied by semidetached systems, and values above 0.8 but below 1 correspond to ellipsoidal variables. The pipeline also includes polyfit, which is a polynomial-chain approximation used for light-curve fitting \citep{Prsa_2008}. Polyfit yields normalized fluxes, which can be used to calculate light-curve amplitudes.

\subsection{Cross-match with other catalogs} \label{sec:cross_match}
Using the \emph{CDS XMatch Service} with a matching radius of $5\arcsec$, we combined the KEBC with \textit{Gaia} DR2 \citep{Gaia_DR2} and \textit{Gaia} EDR3 \citep{Gaia_EDR3}. \textit{Gaia} DR3 was not available at the time when we finalized the data we used for this study. After matching, we had multiple \textit{Gaia} objects for some of the KEBC objects. To ensure uniqueness of the match, we calculated the relative fluxes of the individual \textit{Gaia} sources and retained only the objects with relative \textit{Gaia} EDR3 fluxes higher than 99\%. We obtained luminosities from \textit{Gaia} DR2 (field \textit{lum\_val}). These luminosities are based on the Apsis--FLAME pipeline, which assumes single stars. This is not entirely appropriate for contact binaries, and we discuss possible improvements in Sect.~\ref{sec:conclusions}. Next, we cross-matched the sample with a catalog of stellar effective temperatures for objects in \textit{Gaia} DR2 constructed by \citet{Bai_2019}. They obtained the temperatures by performing regression on stars from four spectroscopic surveys: the Large Sky Area Multi-Object Fiber Spectroscopic Telescope, the Sloan Extension for Galactic Understanding and Exploration, the Apache Point Observatory Galactic Evolution Experiment, and the Radial Velocity Extension. \citet{Bai_2019} found that the temperatures estimated in this way are precise to about $200$\,K. Finally, we excluded any system for which information about its period, luminosity, or effective temperature was lacking, which reduced our sample to 2353 objects.

\subsection{Determination of amplitudes} \label{sec:amplitude_determination}
The KEBC does not specify the photometric amplitudes of the systems, but it is possible to calculate them from polyfit, which is available for most objects in the catalog. However, polyfit does not return amplitude uncertainties resulting from time variation of light curves, and in some cases, it even yields incorrect amplitudes. For this reason, we chose to estimate the amplitudes directly from the detrended fluxes included in the catalog. For each object observed during the primary mission of \textit{Kepler},  we took the long-cadence data and divided them into blocks corresponding to 18 \textit{Kepler} quarters (Q0–Q17). Most quarters represent $\sim$90 days of observation. For systems observed during the K2 mission, we divided the data into $\text{ten}$ equal-size blocks. Before we tried to estimate the amplitudes, we ran some checks to ensure the completeness of the phase-folded light curves observed during the individual blocks, and we excluded the blocks with very few data around phase $0$. 

Within each block, we estimated the mean minimum flux. First, we sorted the data by ascending flux. Then, we iterated over the individual data points and applied local sigma clipping, taking only a close neighborhood of each data point in the phase and flux space into account. When the flux of the data point was not within three standard deviations from the mean of its neighbors, we excluded it as an outlier. In the opposite case, we stopped the iteration and proceeded to the next step. 

\begin{figure}
\centering
\includegraphics[width=0.47\textwidth]{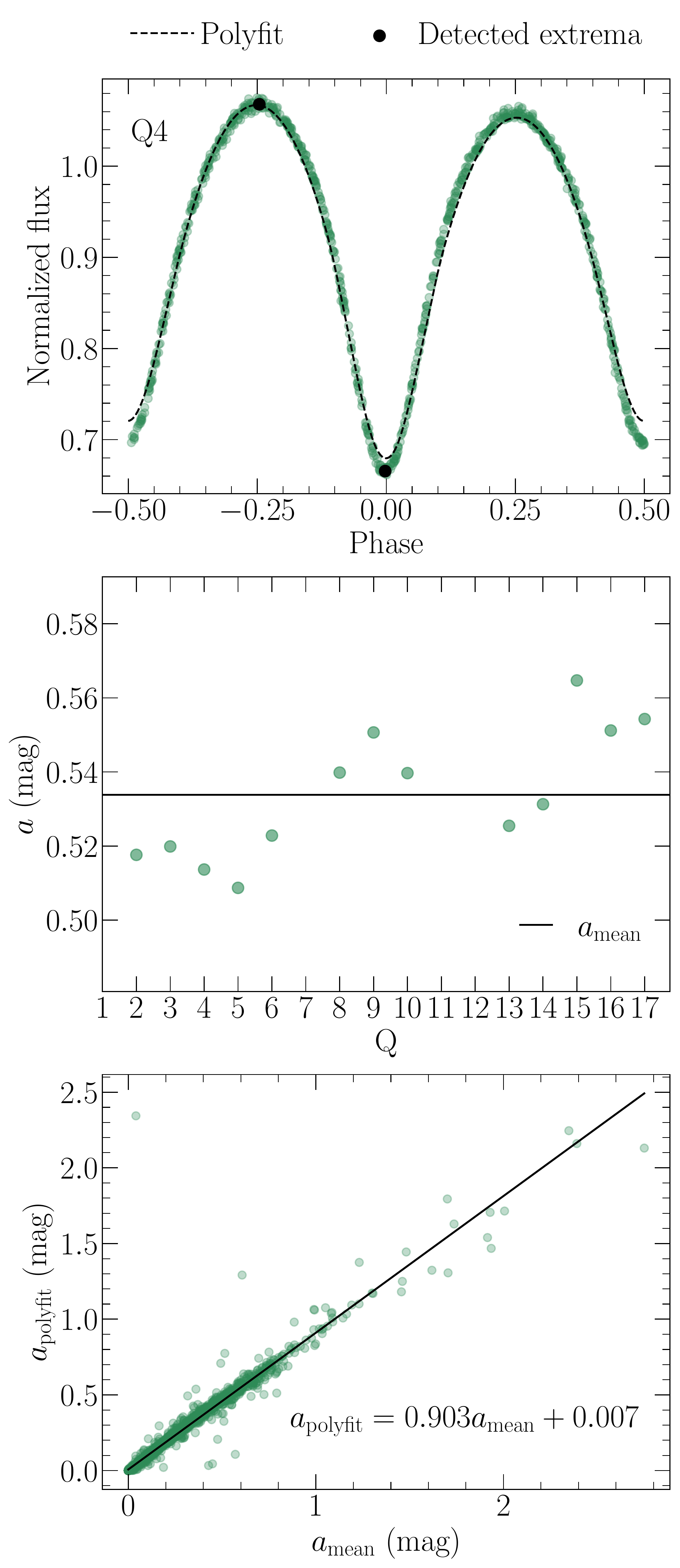}
\caption{Procedure for estimating light-curve amplitudes. Top panel: Detrended \textit{Kepler} light curve of the contact binary KIC 7871200 observed during quarter Q4. The dashed line is the result of polyfit. The data points labeled ``Detected extrema'' correspond to the minimum and maximum normalized fluxes resulting from the procedure described in Sect.~\ref{sec:amplitude_determination}. Middle panel: Photometric amplitudes extracted from the light curves observed during the individual \textit{Kepler} quarters. The solid black line represents the arithmetic average of the values. Bottom panel: Comparison of photometric amplitudes resulting from our procedure ($x$-axis) and polyfit ($y$-axis). The slope of the line indicates a downward bias in the estimates from polyfit, most likely resulting from its tendency to underestimate the depth of light-curve minima.
\label{fig:amplitude_calculation}}
\end{figure}

Second, based on the distribution of the fluxes in the neighborhood of the selected data point, we distinguished between light curves with wide, narrow, and sharp minima. For each type of light curve, we adopted a different method for the calculation of the minimum flux. In the case of wide minima, which are characteristic of continuously varying light curves relevant to our analysis, we approximated the vicinity of the data point selected in the previous step with a second-degree polynomial and we localized its minimum.

We estimated the mean maximum flux in a similar manner, with the exception of some systems, for which we took the median of the observed fluxes as the maximum. The top panel in Fig.~\ref{fig:amplitude_calculation} illustrates the procedure on the light curve of contact binary KIC 7871200. Using the Pogson equation, we then calculated the amplitude within each block and obtained the mean amplitude and its standard deviation over all blocks (Fig.~\ref{fig:amplitude_calculation}, middle panel). For some systems, only one or two blocks passed the completeness checks, making it difficult to reliably estimate the amplitude uncertainty. For these systems, we calculated the amplitude on the full data set and assumed an uncertainty of 4\%, which is comparable to the median uncertainty of $\sim$3.6\% obtained from the systems with more than two complete blocks.

After we determined the mean amplitude for each object in the sample, we compared our method with polyfit (Fig.~\ref{fig:amplitude_calculation}, bottom panel), and we fit the relation between the two with a straight line. The slope of the line indicates a downward bias in the estimates from polyfit. The origin of the bias is not obvious, but visual inspection of randomly chosen light curves reveals the tendency of polyfit to underestimate the depth of light-curve minima.

\section{Identification of contaminants} \label{sec:contaminants}
When a classification based on light-curve morphology is attempted, a sample of contact binaries can become contaminated by various types of pulsating variable stars, rotating spotted stars, or ellipsoidal variables. By ellipsoidal variables we mean binaries with at least one tidally deformed star, but without a Roche-overflowing shared envelope. This contamination is especially prominent at low amplitudes, where a clean sample is crucial for determining $\qmin$. The KEBC has a disproportionately large number of objects with $a<0.01$\,mag, and it is unlikely that most of them are true contact binaries.

To obtain a clean sample of contact binaries, we employed additional information in the form of a PLC relation, which is a combined constraint based on the Roche geometry, the third Kepler law, and the Stefan--Boltzmann law. For example, ellipsoidal variables at a given color and period will appear as underluminous compared to contact binaries because the area of their stellar surface is smaller. Rotating spotted stars at a given luminosity and temperature can have a range of rotational periods that are often longer than the corresponding Keplerian orbital period. Similarly, genuine contact binaries with bright unresolved companions that contaminate the \textit{Kepler} photometry and reduce the observed amplitude also appear as outliers to the PLC relation. This is important because a large fraction of contact binaries should have a companion \citep{pribulla06,dangelo06}. Because of the steepness of the mass--luminosity relation on the main sequence, only companions with masses similar to or higher than the mass of the contact binary influence the amplitude and cause the object to deviate from the PLC relation. By identifying these outliers and removing them from our sample, we mitigated the effect of third light on the observed distribution of photometric amplitudes. We already removed the stars with bright companions resolved by \textit{Gaia} from our sample in Sect.~\ref{sec:cross_match}. 

By modeling the population of contact binaries as a tube in the PLC space, we filtered out most of the contaminants. We followed the general approach for Bayesian data fitting and mixture modeling outlined in \citet{data_fitting_2010}. That is, we viewed our sample as a mixture of a genuine signal (contact binaries) and background noise (everything except contact binaries). We modeled the components using different generative models, with each model conditional on its own set of parameters. Our analysis was based on luminosities from \textit{Gaia} DR2, which are obtained under the assumption that the sources are single stars. Despite this drawback, we show in the following sections that our filtering method works even with these data.

We give a general overview of the method in Sect.~\ref{sec:intrinsic_scatter}, and we construct the generative model of the problem in Sect.~\ref{sec:generative_model}. In Sect.~\ref{sec:posterior_sampling} we obtain the posteriors of the model parameters, and we calculate the probability of being a contact binary of either late or early type for each object in our sample in Sect.~\ref{sec:probability_calculation}. Finally, we present our clean sample of contact binaries in Sect.~\ref{sec:clean}.

\subsection{Intrinsic scatter of the PLC relation} \label{sec:intrinsic_scatter}
We model the PLC relation as a straight line in the $\pi\lambda\tau$-space, where $\pi = \log (P/\text{d})$, $\lambda = \log (L/L_\odot)$, and $\tau = \log (T_\text{eff}/\text{K})$. The relation is not exact, but rather has an intrinsic scatter, which means that the data points can depart from the relation even if we were able to observe all variables with perfect accuracy. Intrinsic scatter is generically present whenever some additional unmeasured quantities affect the measurements but are not accounted for in the relation.

The standard practice is to model the observed data as a single realization of a sequence of independent and identically distributed random variables, in which case the probability distribution of the whole sample is simply the product of the probability distributions from which the individual data points are drawn. In reality, noise is also present, and each data point is measured with a finite uncertainty. This implies that unlike the idealized noise-free data, the actually observed data are not distributed according to the intrinsic-scatter distribution, but rather each observed data point is drawn from an effective distribution given by the convolution of the intrinsic scatter with the uncertainty distribution of the data point. This is true if the individual data points are drawn independently, which we assumed implicitly.

We modeled the effective distribution in the framework of Bayesian inference. In principle, the effective distribution is different for each data point, but when we assume that all data points share the same uncertainty distribution, then formally, they are all drawn from the same effective distribution. In the absence of uncertainty measurements or a physically motivated prescription for the intrinsic scatter, it is more practical to directly model the effective distribution than the two convolution components individually. This allows us to view the originally noisy data generated from the intrinsic scatter as though they were noise free, but generated from the effective distribution, which implicitly reflects the uncertainty properties of the data.

\subsection{Generative model} \label{sec:generative_model}
We constructed the generative model of the signal as the product of the effective distributions evaluated for each data point, but since the data are effectively noise free, we may, at least formally, fix the coordinates of the data points along a chosen dimension and instead construct a generative model conditional on these values. The difference is that the conditional generative model yields the probability of drawing a random sample with the same observed coordinates along the chosen dimension as the original sample, while the full generative model imposes no such constraint. This reduces the complexity of our problem significantly, as it is much easier to parametrically model a sequence of normalized cuts through the effective distribution than the effective distribution itself. However, this is efficient only when the parameters of the normalized cuts vary continuously with the independent variable, as otherwise the number of the parameters of the conditional generative model would scale with the size of the sample.

Traditionally, the variable that is known with the smallest uncertainty is treated as the independent variable. If we were to follow this approach for the signal, we would model the effective distribution of $\lambda$ and $\tau$ conditional on $\pi$ because $\pi$ can be measured with the highest accuracy. However, the choice of the independent variable is not that essential when the effective distribution rather than the intrinsic scatter is modeled directly. Motivated by this, we constructed the generative model for the signal conditional on $\lambda$ instead of $\pi$. We emphasize that our point here is not an accurate characterization of the PLC relation, but rather efficient distinction between contact binaries and contaminants.

\begin{figure}
\centering
\resizebox{\hsize}{!}{\includegraphics{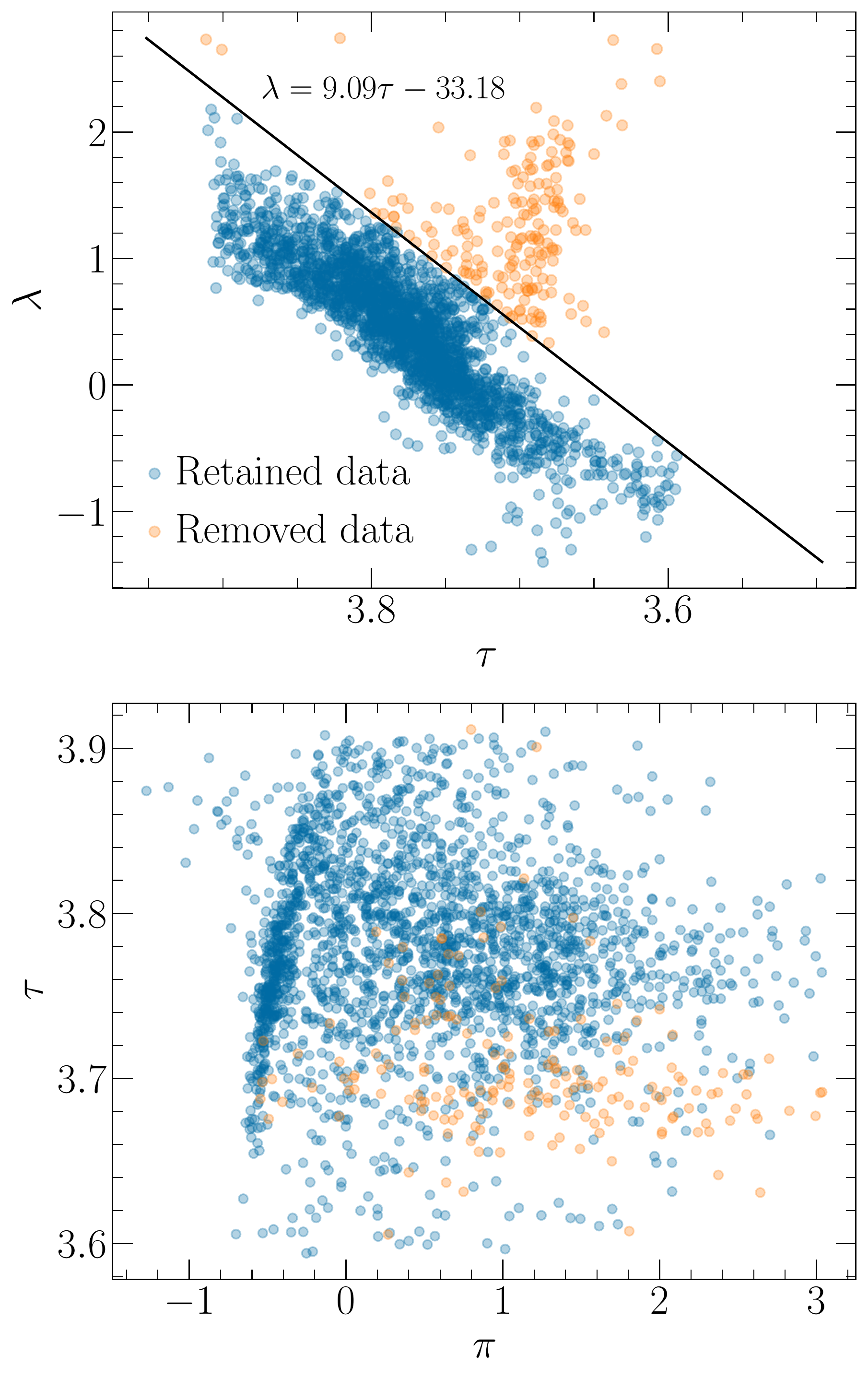}}
\caption{Projection of our sample to the $\tau\lambda$-plane (Hertzsprung--Russell diagram, top panel) and $\pi\tau$-plane (bottom panel). Most objects in the KEBC are constrained to the main sequence perpendicular to the $\tau\lambda$-plane. We remove objects lying above the solid black line drawn in the top panel. The bottom panel shows that the periods of most excised objects are inconsistent with the object being a typical contact binary star.\label{fig:outlier_clipping}}
\end{figure}

In Fig.~\ref{fig:outlier_clipping} we show the Hertzsprung--Russell diagram of our sample, which reveals that most data points lie on the main sequence. We excised the data points that are located above the line $\lambda=9.09\tau-33.18$ (solid black line in Fig.~\ref{fig:outlier_clipping}). The periods of most of the removed objects are longer than a few days (bottom panel in Fig.~\ref{fig:outlier_clipping}), which indicates subgiant or giant components. In addition, visual inspection of the light curves reveals that many of the removed objects are similar to heartbeat stars or have other light curve peculiarities. This reduces the number of the objects in the sample to $2172$ and allows us to adopt a particularly simple conditional generative model for the background, where the background data points are generated from a plane with a nonzero Gaussian thickness in $\tau$. 

We modeled the conditional effective distribution of the signal as a two-dimensional uncorrelated Gaussian distribution in $\pi$ and $\tau$ centered on the PLC relation. The total conditional effective distribution $p(\pi,\tau|\lambda,\theta)$ of a single data point is a weighted sum of the conditional effective distributions for the signal $p_\mathrm{S}(\pi,\tau|\lambda,\theta)$ and the background noise $p_\mathrm{B}(\pi,\tau|\lambda,\theta)$,
\begin{equation} \label{eq:total_model}
  p(\pi,\tau|\lambda,\theta) = X p_\mathrm{S}(\pi,\tau|\lambda,\theta)+(1-X)p_\mathrm{B}(\pi,\tau|\lambda,\theta),
\end{equation}
where $X$ is the weight parameter, and $\theta$ is a vector of all the model parameters. We write the conditional effective distributions as
\begin{eqnarray}
p_\mathrm{S}(\pi,\tau|\lambda,\theta) &=& \mathscr{N}(\pi;\mu_{\mathrm{S}\pi},\sigma_{\mathrm{S}\pi})\mathscr{N}(\tau;\mu_{\mathrm{S}\tau},\sigma_{\mathrm{S}\tau}) \label{eq:eff_distr_contact_binaries},\\
p_\mathrm{B}(\pi,\tau|\lambda,\theta) &=&
  \begin{cases}
  \frac{\mathscr{N}(\tau;\mu_{\mathrm{B}\tau},\sigma_{\mathrm{B}\tau})}{\pi_\text{max}-\pi_\text{min}} & \text{if } \pi_\text{min} \le \tau \le \pi_\text{max},\\
  0 & \text{else},
  \end{cases} \label{eq:eff_distr_noise}
\end{eqnarray}
where we suppressed the dependence of the parameters of the Gaussians on $\lambda$. The specific form of Eq.~\eqref{eq:eff_distr_noise} comes from the assumption that the background noise is distributed uniformly between the minimum and maximum observed values of $\pi$, denoted by $\pi_\text{min}$ and $\pi_\text{max}$.

To account for a possible change in the slope of the PLC relation, we considered a separate effective distribution for late-type and early-type contact binaries. The two distributions are formally the same, but each has its own set of parameters and is applicable only to a subset of the signal, disjunct from the other subset. We suspect that the dividing line between the samples will be along the Kraft break, but instead of using previously determined dividing lines such as the one from \citet{Jayasinghe_2020J}, we included the location of the break in our model as one of the parameters. The only constraint was that the transition in the PLC relation from late types to early types is smooth. From the modeling point of view, it is more convenient to model the break along the $\lambda$-axis rather than $\tau$-axis, as $\lambda$ acts as the independent variable. This is possible because the PLC relation provides us with a unique mapping between $\tau$ and $\pi$. In general, we need two separate models for the background as well, because the parameters of the noise may also change at the Kraft break. However, in the case of the noise, we did not require that the transition is continuous.

Using subscripts $1$ and $2$ to refer to the parameters of the models below and above the break, we assumed the following functional dependencies for the parameters of the Gaussians:
\begin{eqnarray}
  \mu_{\mathrm{S}\pi}&=&
  \begin{cases}
  \alpha_{\pi1}+\beta_{\pi1}\lambda & \text{if } \lambda\le \lambdak,\\
  \alpha_{\pi1}+(\beta_{\pi1}-\beta_{\pi2})\lambdak + \beta_{\pi2}\lambda & \text{if } \lambda>\lambdak,
  \end{cases} \label{eq:mu_pi}\\
  \sigma_{\mathrm{S}\pi}&=&
  \begin{cases}
  \alpha_{\sigma_{\pi1}}+\beta_{\sigma_{\pi1}}\lambda & \text{if } \lambda\le \lambdak,\\
  \alpha_{\sigma_{\pi2}}+\beta_{\sigma_{\pi2}}\lambda & \text{if } \lambda>\lambdak,
  \end{cases} \label{eq:sigma_pi}\\ 
  \mu_{\mathrm{S}\tau}&=&
  \begin{cases}
  \alpha_{\tau1}+\beta_{\tau1}\lambda & \text{if } \lambda\le \lambdak,\\
  \alpha_{\tau1}+(\beta_{\tau1}-\beta_{\tau2})\lambdak + \beta_{\tau2}\lambda & \text{if } \lambda>\lambdak,
  \end{cases} \label{eq:mu_tau}\\
  \sigma_{\mathrm{S}\tau}&=&
  \begin{cases}
  \alpha_{\sigma_{\tau1}}+\beta_{\sigma_{\tau1}}\lambda & \text{if } \lambda\le \lambdak,\\
  \alpha_{\sigma_{\tau2}}+\beta_{\sigma_{\tau2}}\lambda & \text{if } \lambda>\lambdak,
  \end{cases} \label{eq:sigma_tau}\\
  \mu_{\mathrm{B}\tau}&=&
  \begin{cases}
  m_1+l_1\lambda & \text{if } \lambda\le \lambdak,\\
  m_2+l_2\lambda & \text{if } \lambda>\lambdak,
  \end{cases}\\
  \sigma_{\mathrm{B}\tau}&=&
  \begin{cases}
  w_1 & \text{if } \lambda\le \lambdak,\\
  w_2 & \text{if } \lambda>\lambdak,
  \end{cases}
\end{eqnarray}
where $\lambdak$ denotes the location of the break along $\lambda$. The prescriptions for $\mu_{\mathrm{S}\pi}$ and $\mu_{\mathrm{S}\tau}$ above the break ($\lambda>\lambdak$) derive from the requirement that the PLC relation is continuous at the transition. 

Since we employed conditional generative models instead of full generative models, we only required that the total conditional effective distribution is normalized within a given slice of constant $\lambda$ and not in the whole parameter space. This allowed us to model the weight parameter $X$ as a function of $\lambda$, and the simplest nontrivial choice is to assume linear dependence,
\begin{equation} \label{weight_dependence}
  X =
  \begin{cases}
  \alpha_{X1} + \beta_{X1}\lambda & \text{if } \lambda \le \lambdak,\\
  \alpha_{X2} + \beta_{X2}\lambda & \text{if } \lambda > \lambdak.
  \end{cases}
\end{equation}
We emphasize that this choice does not imply that the probability of being a contact binary for each individual object depends linearly on $\lambda$. Instead, $X$ sets the relative weight of the signal and the noise for the objects within a given slice, and the linear dependence on $\lambda$ is just the simplest nontrivial model that can be employed. In total, our model has 25 parameters, which are listed together with their definitions in Table~\ref{tab:model_parameters}.

Denoting the $j$th object in our sample of $M$ objects ($M\ge N$) with the subscript $j$, we can write the likelihood of the total conditional generative model as 
\begin{equation}
  \mathscr{L}(\theta|\{\pi_j,\tau_j\}_{j=1}^M) = \prod_{j=1}^M p(\pi_j,\tau_j|\lambda_j,\theta), 
\end{equation}
where the curly brackets are shorthand for iterating over all objects in the sample. 

\subsection{Posterior sampling} \label{sec:posterior_sampling}
Following the Bayesian approach, we modeled the parameters of the model as random variables. We assumed that the prior distribution $p(\theta)$ is separable, and we assigned a uniform prior to each parameter. Using the Bayes theorem,
\begin{equation}
  p(\theta|\{\pi_j,\tau_j\}_{j=1}^M)=\frac{\mathscr{L}(\theta|\{\pi_j,\tau_j\}_{j=1}^M) p(\theta)}{p(\{\pi_j,\tau_j\}_{j=1}^M)}, 
\end{equation}
we then arrive at the joint posterior distribution $p(\theta|\{\pi_j,\tau_j\}_{j=1}^M)$.

We used \emph{emcee} to sample the posterior distribution. With a total of 50 walkers, we ran the sampler for 160 000 steps to ensure that the chains converge. The steps in the chains were generated via differential evolution, and we discarded the first 10000 steps as burn-in. To reduce autocorrelation, we only considered every 300th sample in the chains. We optimized the efficiency by sampling the distribution in two steps. First, we prescribed rather broad priors for the parameters and ran the sampler for a few thousand steps. Then we restricted the priors based on the results from the initial run, and we ran the sampler again for the full 160 000 steps. Fig.~\ref{fig:data_model_chains} and \ref{fig:data_model_corner} show the chain plots and corner plots resulting from the run. We present the median values of the parameters together with their $16th^\mathrm{}$ and $84th^\mathrm{}$ percentiles in Table~\ref{tab:model_parameters}.

\subsection{Probability calculation} \label{sec:probability_calculation}
After we obtained the posterior distribution, we assigned the probability of being a contact binary star to each data point in our sample. The idea is that given a sample in a slice of constant $\lambda$, the probability of being a contact binary system is simply the conditional probability of being drawn from the conditional effective distribution of the signal. In general, the probability of the $j$th data point being a contact binary depends on the value of $\theta$, and a straightforward derivation for a mixture of two distributions with a fixed $\theta$ within a slice of constant $\lambda$ gives us
\begin{equation}
p_{\mathrm{CB},j}(\theta) = \frac{X_j p_\mathrm{S}(\pi_j,\tau_j|\lambda_j,\theta)}{X_j p_\mathrm{S}(\pi_j,\tau_j|\lambda_j,\theta) + (1-X_j) p_\mathrm{B}(\pi_j,\tau_j|\lambda_j,\theta)},
\end{equation}
where $X_j\equiv X(\lambda_j,\theta)$. When we replace the total contact binary effective distribution $p_\mathrm{S}(\pi_j,\tau_j|\lambda_j,\theta)$ with the part purely below or above the break and assume that the distribution is zero for the data points on the opposite side of the break, we obtain the probabilities $p_{\mathrm{CBL},j}(\theta)$ and $p_{\mathrm{CBE},j}(\theta)$ of being a contact binary of either late or early type.

To remove the dependence on $\theta$, we calculated the average probability using the thinned posterior sample that we obtained from \emph{emcee}. This amounts to marginalizing out the parameters of the conditional generative model using the posterior probability distribution, that is,
\begin{equation}
p_{\mathrm{CB},j} = \int p_{\mathrm{CB},j}(\theta)p(\theta|\{\pi_j,\tau_j\}_{j=1}^M) {\rm d}\theta. 
\end{equation}
Again, if we replaced $p_{\mathrm{CB},j}$ with $p_{\mathrm{CBL},j}$ or $p_{\mathrm{CBE},j}$, we would obtain the marginalized probabilities of being a contact binary of either late or early type.

\begin{table}
\caption{Photometric amplitudes and probabilities of being a contact binary of either late or early type for the objects in our sample.
\label{tab:online_data}}
\begin{center}
\resizebox{\columnwidth}{!}{
\begin{tabular}{ccccc}
\hline\hline
KIC/EPIC & $a$ (mag) & $\sigma_a$ (mag) & $p_\text{CBL}$ & $p_\text{CBE}$\\
\hline
  1026032 & 0.085360 & 0.001293 & 0.0000 & 0.0000\\
  1026957 & 0.001115 & 0.000429 & 0.0000 & 0.0000\\
  1432214 & 0.098101 & 0.000707 & 0.0000 & 0.0000\\
  1571511 & 0.022127 & 0.000116 & 0.0000 & 0.0000\\
  1572353 & 0.116464 & 0.004660 & 0.9474 & 0.0000\\
\vdots\hfil & \vdots\hfil & \vdots\hfil & \vdots\hfil & \vdots\hfil\\
212163353 & 0.098400 &    ---   & 0.0000 & 0.0000\\
212175535 & 0.080279 &    ---   & 0.9402 & 0.0000\\
\hline
\end{tabular}
}
\end{center}
\tablefoot{The full table is available online. For some systems, we were not able to reliably estimate their amplitude uncertainties. In our analysis, we adopted an uncertainty of $4\%$ for these systems (Sect.~\ref{sec:amplitude_determination}).}
\end{table}

\subsection{Clean sample of contact binaries}
\label{sec:clean}

\begin{figure}
\centering
\includegraphics[width=0.48\textwidth]{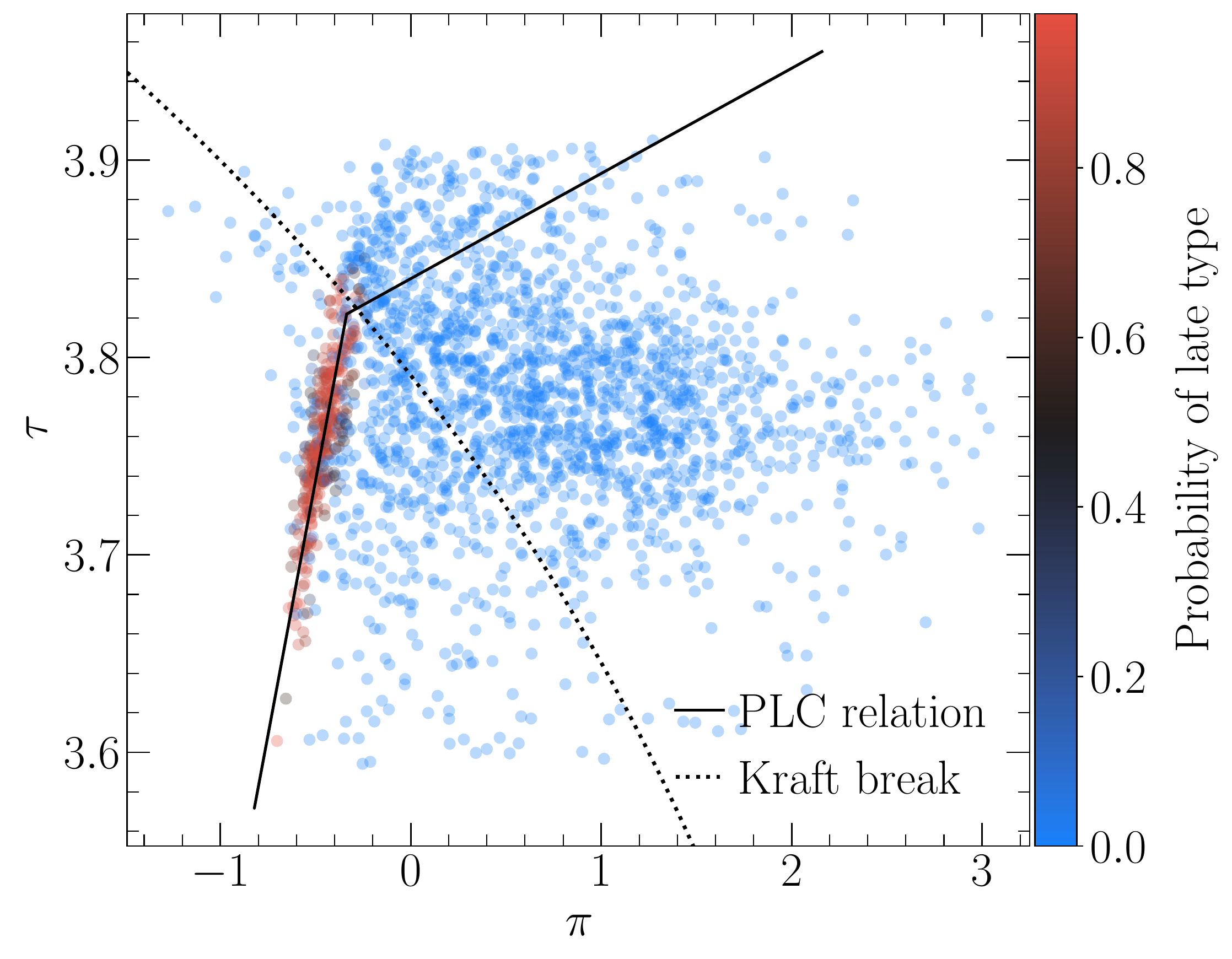}
\includegraphics[width=0.48\textwidth]{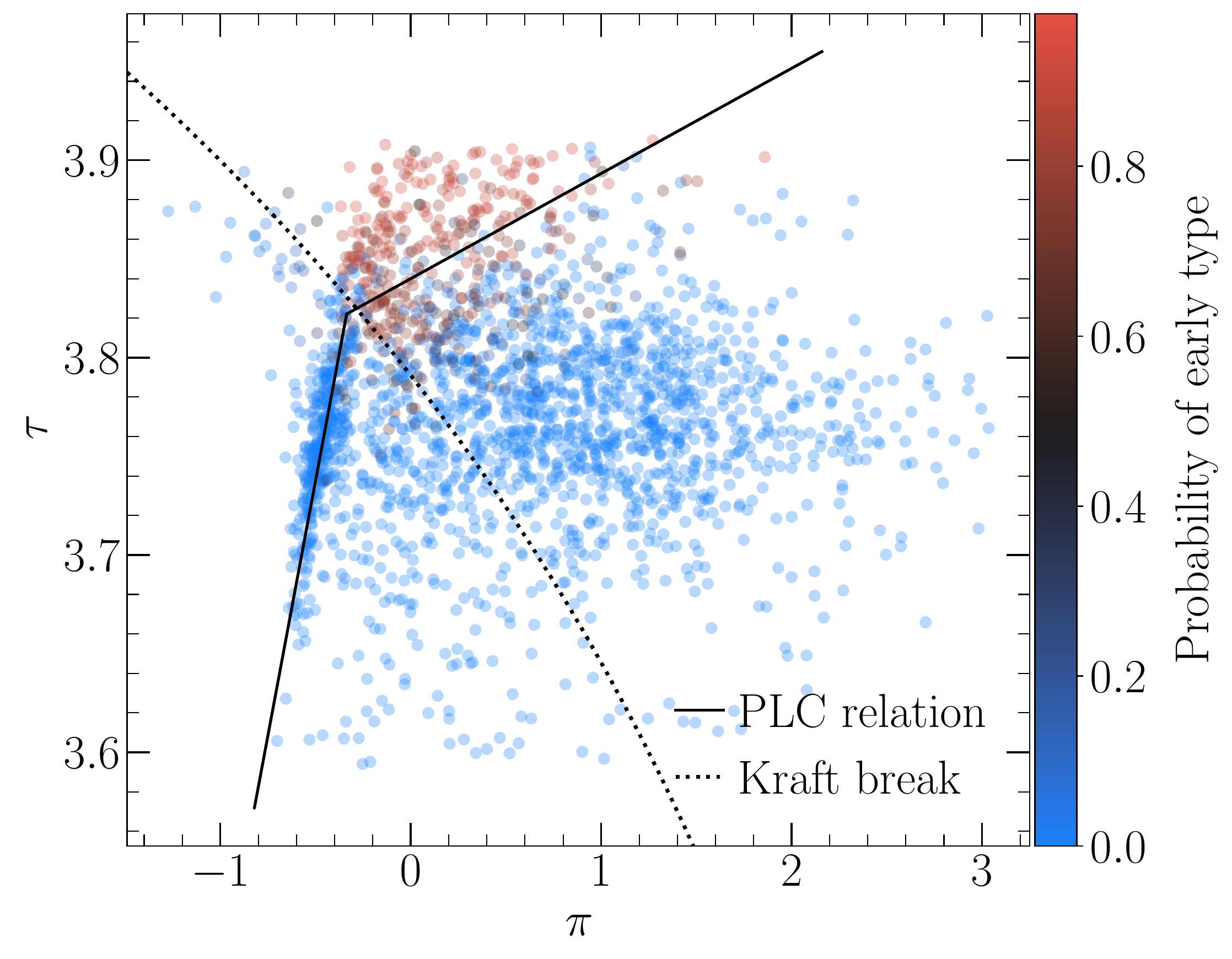}
\caption{Projection of our sample to the $\pi\tau$-plane. The color of each object corresponds to the probability of being a late-type (top panel) or early-type (bottom panel) contact binary star. The solid lines show our best-fit PLC relations, and the dotted line shows the position of the Kraft break from \citet{Jayasinghe_2020J}. \label{fig:2d_scatter_probability}}
\end{figure}

\begin{figure}
  \centering
  \resizebox{\hsize}{!}{\includegraphics{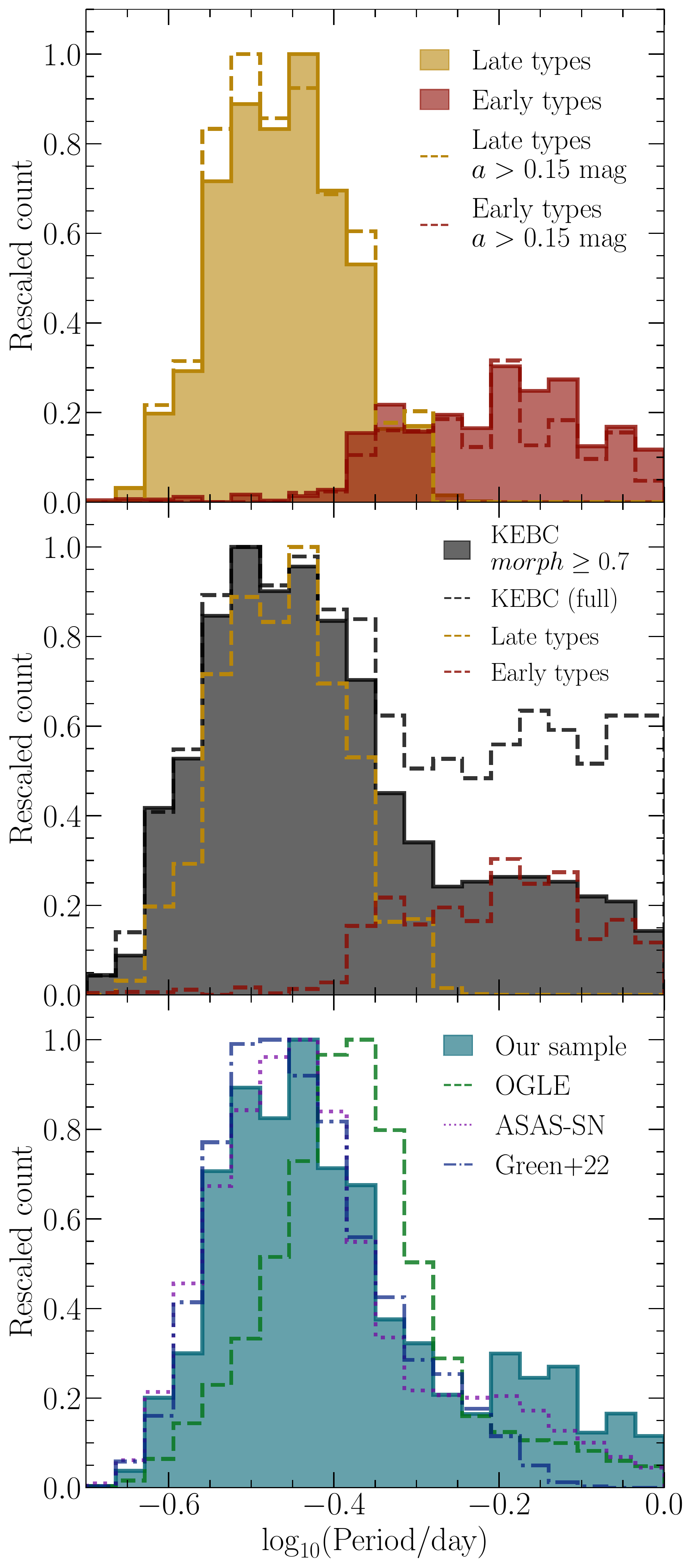}}
  \caption{Period distribution of contact binaries in our sample. Top panel: Our full late-type and early-type samples (filled histogram) and their high-amplitude subsamples (dashed lines). Middle panel: Comparison of our samples with the KEBC, where the dashed black line marks the full KEBC, and the filled histogram shows the period distribution of objects with $morph\ge0.7$. Bottom panel: Our sample in comparison to the samples from OGLE \citep{soszynski16}, ASAS-SN \citep{jayasinghe18,jayasinghe19,Jayasinghe_2020J}, and the recent catalog by \citet{green22}. The histograms in all panels are appropriately rescaled and, where available, weighted by the probability of being a contact binary of the respective type.}
  \label{fig:period_distribution}
\end{figure}

\begin{figure}
  \centering
  \includegraphics[width=0.48\textwidth]{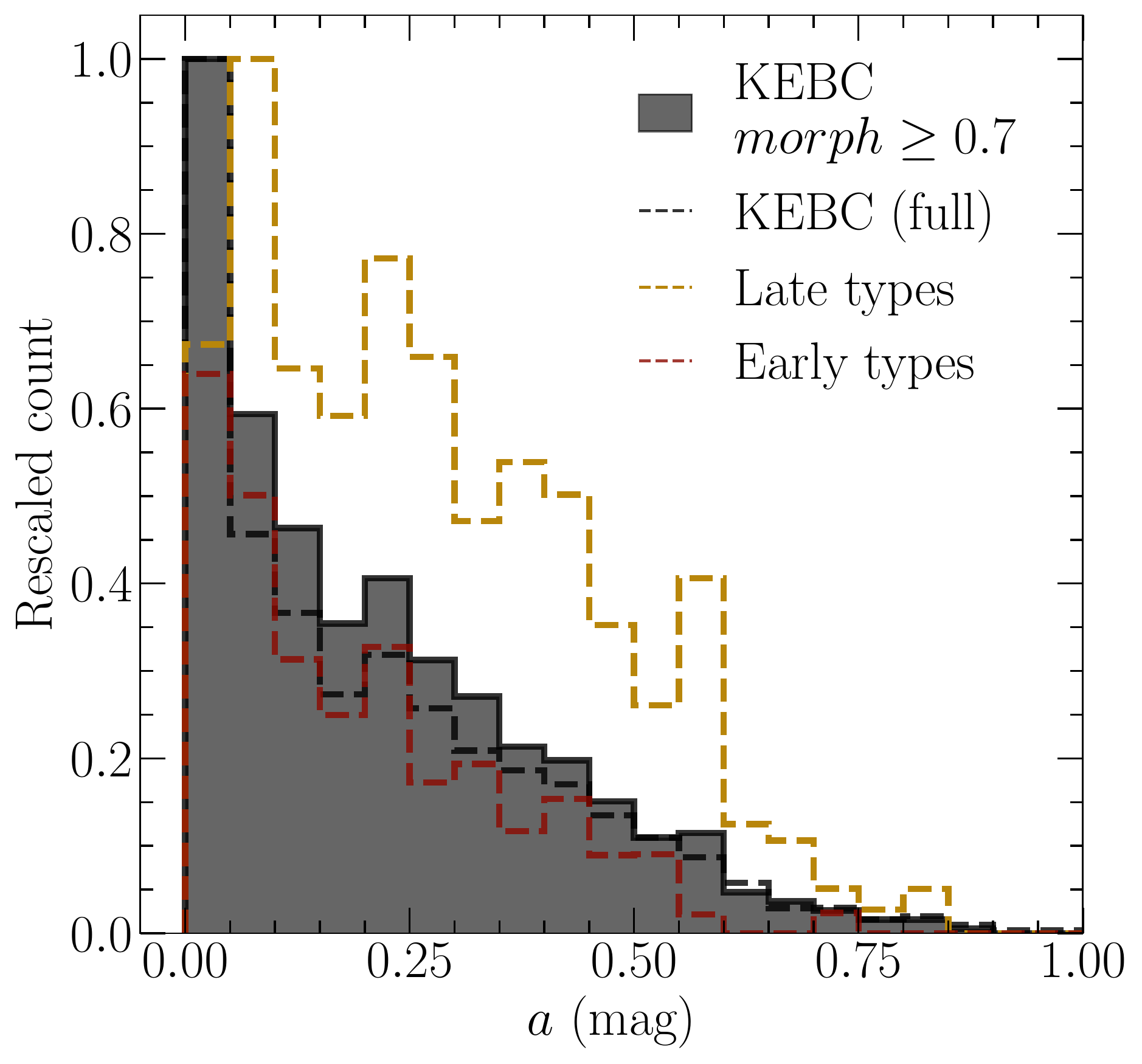}
  \caption{Amplitude distribution of the late-type (dashed orange line) and early-type (dashed red line) contact binaries in our cleaned sample. The dashed black line shows the amplitude distribution of the full KEBC, and the solid black line shows the distribution of the KEBC with $morph\ge0.7$. The late- and early-type histograms are weighted by the probability of being a contact binary of the respective type.}
  \label{fig:amplitude_distribution}
\end{figure}

In Table~\ref{tab:online_data} we show the photometric amplitudes and calculated probabilities for the objects in our sample. In Fig.~\ref{fig:2d_scatter_probability} we show a projection of our sample to the $\pi\tau$-plane, where the color of each point indicates the probability of being a contact binary of either late or early type. Our model assigns late-type contact binaries to a tight locus around the PLC relation, while for early-type binaries the scatter is significantly larger. This is consistent with previous results \citep[e.g.,][]{Jayasinghe_2020J}. Similarly, the position of the break in the PLC relation matches the results of \citet{Jayasinghe_2020J} remarkably well, even though no prior information other than the existence of a break along the PLC relation enters our model. This confirms our suspicion that the slope of the PLC relation changes at the Kraft break.

With these results, we can assess the quality of our sample and compare it to other existing samples of contact binaries. In Fig.~\ref{fig:period_distribution} we show the period distribution of our clean sample of contact binaries. In the top panel, we compare the period distributions of our late- and early-type samples with their high-amplitude subsamples ($a> 0.15$\,mag). The distributions of high-amplitude objects closely follow the distributions of the full samples. This implies that low-amplitude objects must also follow similar distributions, and therefore, they are consistent with being contact binaries. In the middle panel, we compare our samples with the unprocessed KEBC. The full KEBC substantially differs for periods longer than about $0.5$\,days, which is due to the presence of detached binaries. These can be efficiently removed by considering a cut on the \textit{morph} parameter. However, the modified KEBC still differs from our sample, especially at short periods. In the bottom panel of Fig.~\ref{fig:period_distribution}, we compare our combined late- and early-type sample with other contact binary samples from the literature. The Galactic bulge contact binary sample of \citet{soszynski16} peaks at noticeably longer periods than what we find. Since \citet{soszynski16} also reported some short-period objects, it is not clear whether the shift is entirely due to the greater distance of the Galactic bulge compared to the \textit{Kepler} field or if the population is truly different. The period shift is in contrast with the contact binary sample from ASAS-SN \citep{jayasinghe18,jayasinghe19,Jayasinghe_2020J,Pawlak_2019}, which peaks at around the same periods as our sample, and agrees relatively well with our sample even for $P\gtrsim0.6$ d, which are typical periods for early-type contact binaries. Recently, \citet{green22} published a sample of ellipsoidal and contact binaries, and their period distribution closely follows the distribution of our combined sample for $\log P\lesssim -0.2$. For longer periods, their sample appears to have fewer early-type objects than what we find. This is most likely due to a decreased efficiency of their selection algorithm for these systems.

In Fig.~\ref{fig:amplitude_distribution} we show the amplitude distributions of our two samples compared to the KEBC catalog. The figure shows that both the full KEBC and its $morph \ge 0.7$ subsample have a high fraction of objects with very small $a<0.05$\,mag, while our samples show a smaller fraction at these amplitudes. We suggest that this is due to the contaminants in the KEBC that were removed by our model. We emphasize that correctly obtaining the low-amplitude part of the distribution is important to model the mass-ratio distribution of contact binary stars properly.

\begin{figure}
  \centering
  \includegraphics[width=0.48\textwidth]{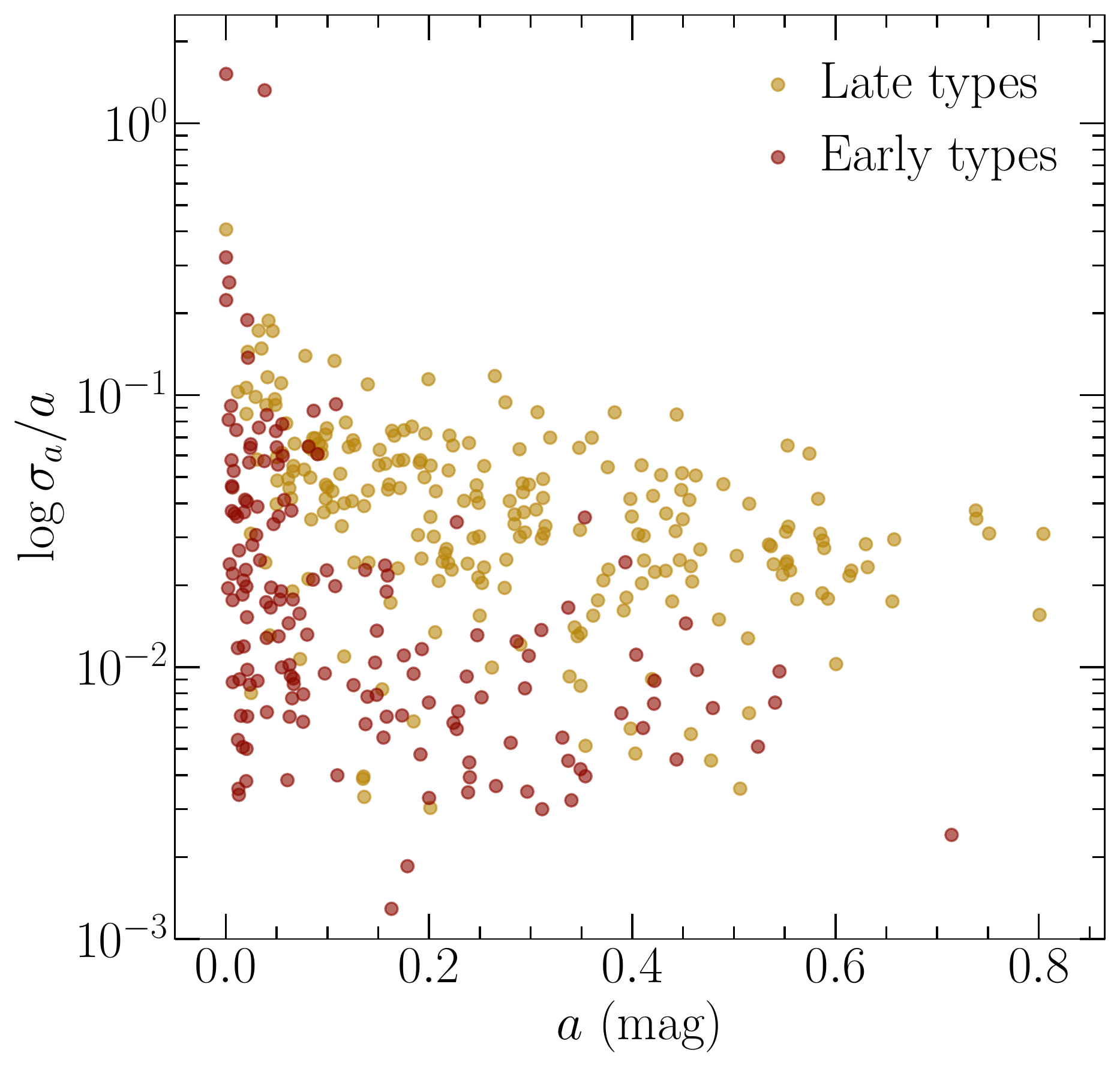}
  \caption{Relative amplitude uncertainty $\sigma_a /a$ as a function of amplitude $a$ for our late- and early-type samples. We show only objects with amplitude estimates in at least $\text{five}$ \textit{Kepler} quarters, $morph \ge 0.7$, and a probability of belonging to either type greater than $0.5$.}
  \label{fig:sigma_a_vs_a}
\end{figure}

In Fig.~\ref{fig:sigma_a_vs_a} we show the relative amplitude uncertainty $\sigma_a/a$ of our two samples. For the vast majority of objects, $\sigma_a/a < 0.1$. The relative uncertainty is higher only for several objects with very small amplitudes, which does not significantly affect our results. It is interesting to note that the late-type sample shows systematically higher amplitude uncertainties than the early-type sample. We can explain this observation by the appearance, disappearance, or migration of spots on the surfaces of late-type stars, which causes variations of the amplitude over time. We show an example of these amplitude variations in the middle panel of Fig.~\ref{fig:amplitude_calculation}. Unless spots on contact binaries exhibit strong variability on timescales longer than the duration of the \textit{Kepler} mission, our results in Fig.~\ref{fig:sigma_a_vs_a} imply that our method is not significantly affected by stellar spots.

\section{Results} \label{sec:results}
In this section, we present the results of our method for the inference of the mass-ratio distribution of contact binary stars. In Sect.~\ref{sec:population} we define various populations of contact binaries and provide an overview of all models that we investigated. Next, we compare Bayes factors of the individual models and select a fiducial model for each population (Sect.~\ref{sec:fiducial_models}). We present and discuss the mass-ratio distributions of contact binaries for the fiducial set of parameters in Sect.~\ref{sec:mass_ratio_distribution}. Finally, we discuss the dependence of our results on the probability cutoffs distinguishing contact binaries from contaminants (Sect.~\ref{sec:dependence_probability_cutoff}), on our choice of the default fill-out factor (Sect.~\ref{sec:dependence_fill_out}), on the splitting period for late-type binaries (Sect.~\ref{sec:dependence_period_cutoff}), and on the hyperparameters of our model (Sect.~\ref{sec:dependence_hyperparameters}).

\subsection{Populations of contact binary stars} \label{sec:population}

\begin{table}
\caption{List of samples constructed from our Bayesian model for the identification of contact binary stars.
\label{tab:samples}}
\begin{center}
\begin{tabular}{ccccc}
\hline\hline
Sample & Type & Prob. cutoff & $P$ (days) & Eff. size\\
\hline
CB1p50 & Late & 0.5 & --- & 258.99 \\
CB1p60 & Late & 0.6 & --- & 256.27 \\
CB1p70 & Late & 0.7 & --- & 249.00 \\
CB1p80 & Late & 0.8 & --- & 228.04 \\
CB2p10 & Late & 0.1 &$\le0.3$ & 62.40 \\
CB2p20 & Late & 0.2 &$\le0.3$ & 61.96 \\
CB2p30 & Late & 0.3 &$\le0.3$ & 61.38 \\
CB2p40 & Late & 0.4 &$\le0.3$ & 61.07 \\
CB2p50 & Late & 0.5 &$\le0.3$ & 60.59 \\
CB3p50 & Late & 0.5 &$>0.3$ & 198.41 \\
CB3p60 & Late & 0.6 &$>0.3$ & 197.30 \\
CB3p70 & Late & 0.7 &$>0.3$ & 192.69 \\
CB3p80 & Late & 0.8 &$>0.3$ & 177.76 \\
CB4p10 & Early & 0.1 & $<1$ & 106.42 \\
CB4p20 & Early & 0.2 & $<1$ & 105.56 \\
CB4p30 & Early & 0.3 & $<1$ & 105.56 \\
CB4p40 & Early & 0.4 & $<1$ & 104.91 \\
CB4p50 & Early & 0.5 & $<1$ & 104.42 \\
CB5p10 & Early & 0.1 & --- & 162.62 \\
CB5p20 & Early & 0.2 & --- & 161.55 \\
CB5p30 & Early & 0.3 & --- & 160.8 \\
CB5p40 & Early & 0.4 & --- & 159.49 \\
CB5p50 & Early & 0.5 & --- & 158.58 \\
\hline
\end{tabular}
\end{center}
\tablefoot{We require $morph \ge 0.7$ for all samples.}
\end{table}

\begin{figure}
\centering
\resizebox{\hsize}{!}{\includegraphics{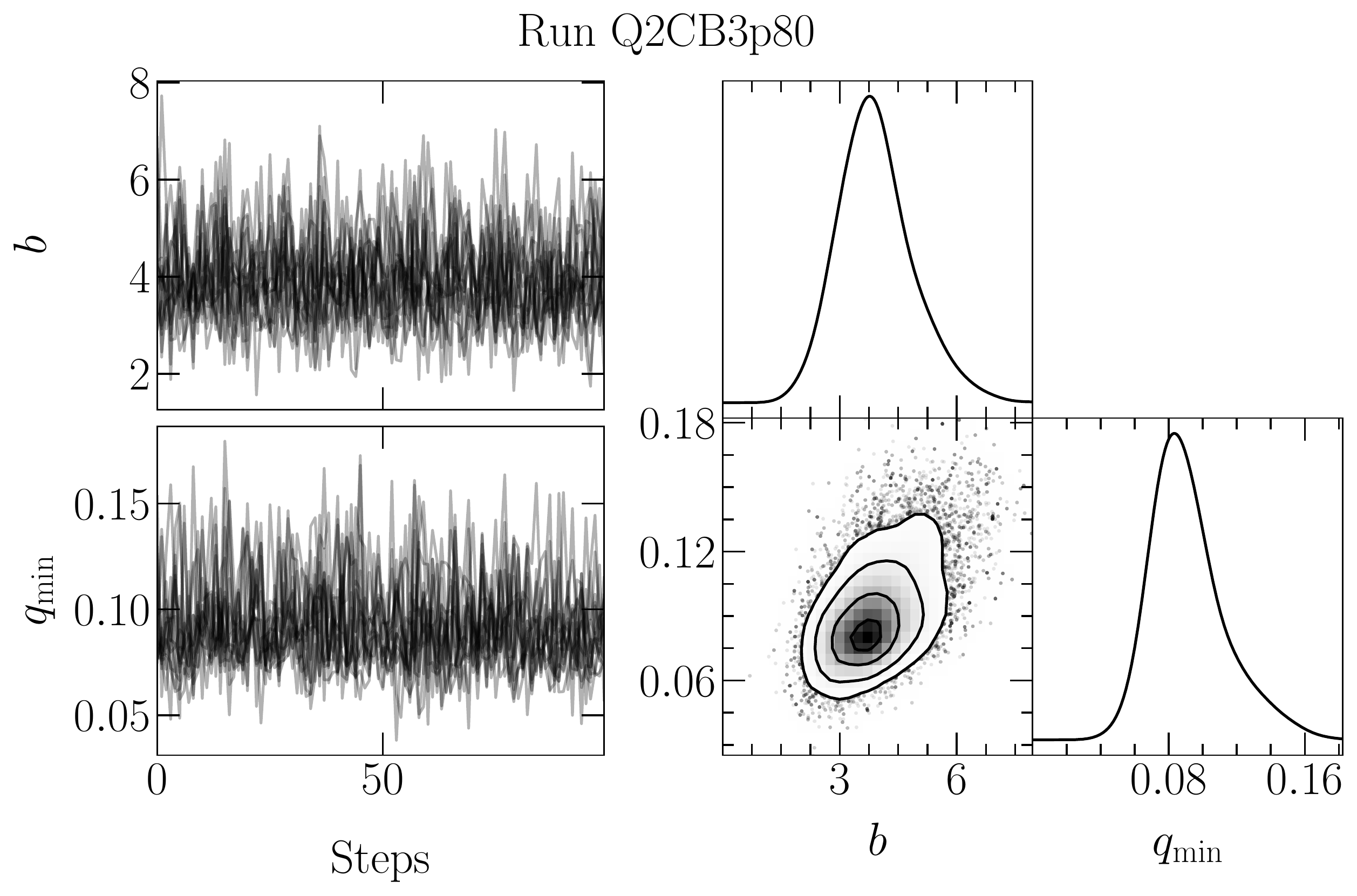}}
\caption{Chain plots and corner plots resulting from the run Q2CB3p80. We ran the sampler for a total of 2500 steps, but we discarded the first 500 steps as burn-in, and we thinned the chains by a factor of 20. \label{fig:run_Q2CB3p80}}
\end{figure}

We are interested in estimating the mass-ratio distribution for different populations of contact binary stars. For late-type binaries, we distinguish between the full population (CB1) and the populations of binaries with periods shorter (CB2) or longer (CB3) than $P_\text{split} = 0.3$\,days \citep[see][for a justification of this period division]{Stepien_2012}. For early-type binaries, we distinguish between a population of early-type contact binaries with $P<1$\,day (CB4) and an extended population containing all early-type contact binaries without a constraint on the period (CB5). These populations are separate, but overlap. By imposing various probability cutoffs on the individual populations, we obtain a number of samples with varying levels of contamination. In Table~\ref{tab:samples} we list the samples together with their definitions and their sizes. The effective sample size is calculated as $\sum_k p_{\text{CB},k}$ for $k$ belonging to the given sample and having $p_{\text{CB},k}$ higher than the probability cutoff. To maximize the size of the samples while keeping the contamination as low as possible, we limited the maximum probability cutoffs for populations CB2, CB4, and CB5 to $0.5$. In contrast, the high number of late-type contact binaries with periods around 0.37\,days \citep{Paczynski_2006} allowed us to consider cutoffs up to $0.8$ for populations CB1 and CB3. 

For each population CB1--CB5, we also investigated the dependence on the mass-ratio prescription, fill-out factor, and various model hyperparameters. In Table~\ref{tab:runs} we give a complete list of our model runs. The runs labeled Q1 and Q2 used the power-law prescriptions $Q_1$ and $Q_2$ , respectively, as defined in Eq.~\eqref{eq:Q}. All Q1 and Q2 runs were carried out with the default values of $f=0.25$, $h=0.02$, and $n=10000$. The runs labeled F investigated the dependence of the results on the fill-out factor (Sect.~\ref{sec:dependence_fill_out}), while the runs starting with S examined how the choice of $P_\text{split}$ affects the mass-ratio distribution of the two late-type subpopulations (Sect.~\ref{sec:dependence_period_cutoff}). Finally, the runs labeled H studied the dependence on the hyperparameters $h$ and $n$ (Sect.~\ref{sec:dependence_hyperparameters}).

\subsection{Fiducial models and Bayes factors} \label{sec:fiducial_models}
For each model, we determined the posterior distribution of $\qmin$ and $b$. In Fig.~\ref{fig:run_Q2CB3p80} we illustrate our results by showing the chain plots and posterior distributions for the run Q2CB3p80. The remaining Q1 and Q2 posteriors can be found in Figure~\ref{fig:posterior_Q1_Q2}. The number of steps is sufficient for the chains to converge and the parameters $\qmin$ and $b$ show no significant correlation.

To select a fiducial model for each population, we need to determine which power-law prescription for the mass-ratio distribution fits the observed data better. We achieved this by calculating the posterior Bayes factors \citep{Aitkin_1991} for each pair of Q1 and Q2 runs defined on the same sample. In other words, for each sample, we compared the goodness-of-fit of the two power laws by taking the ratio of the posterior average of the corresponding model likelihoods. We calibrated the Bayes factors according to the scale proposed by \citet{Aitkin_1991}, which suggests that posterior Bayes factors of 20, 100, or 1000 constitute a strong, very strong, or an overwhelming weight of sample evidence in favor of the model with the higher value. We present the results of the comparison in Table~\ref{tab:posterior_bayes_factors}. 

\begin{table}
\caption{Posterior Bayes factors for the two power-law prescriptions $Q_1$ and $Q_2$, different contact binary populations (CB1--CB5), and different probability cutoffs. \label{tab:posterior_bayes_factors}}
\begin{center}
\begin{tabular}{cc}
\hline\hline
Models & Posterior Bayes factor \\
\hline
Q2CB1p50 vs. Q1CB1p50 & 15.02 \\
Q2CB1p60 vs. Q1CB1p60 & 10.76 \\
Q2CB1p70 vs. Q1CB1p70 & 14.87 \\
Q2CB1p80 vs. Q1CB1p80 & 12.33 \\
\hline
Q2CB2p10 vs. Q1CB2p10 & 1.36 \\
Q2CB2p20 vs. Q1CB2p20 & 1.17 \\
Q2CB2p30 vs. Q1CB2p30 & 1.04 \\
Q2CB2p40 vs. Q1CB2p40 & 1.00 \\
Q2CB2p50 vs. Q1CB2p50 & 1.03 \\
\hline
Q2CB3p50 vs. Q1CB3p50 & 10.55 \\
Q2CB3p60 vs. Q1CB3p60 & 11.44 \\
Q2CB3p70 vs. Q1CB3p70 & 16.53 \\
Q2CB3p80 vs. Q1CB3p80 & 15.67 \\
\hline
Q2CB4p10 vs. Q1CB4p10 & 178.49 \\
Q2CB4p20 vs. Q1CB4p20 & 49.31 \\
Q2CB4p30 vs. Q1CB4p30 & 47.32 \\
Q2CB4p40 vs. Q1CB4p40 & 32.58 \\
Q2CB4p50 vs. Q1CB4p50 & 30.44 \\
\hline
Q2CB5p10 vs. Q1CB5p10 & 0.00 \\
Q2CB5p20 vs. Q1CB5p20 & 0.01 \\
Q2CB5p30 vs. Q1CB5p30 & 0.02 \\
Q2CB5p40 vs. Q1CB5p40 & 0.05 \\
Q2CB5p50 vs. Q1CB5p50 & 0.08 \\
\hline
\end{tabular}
\end{center}
\end{table}

In most cases, the second power-law prescription $Q_2$ is preferred over $Q_1$, or the comparison is inconclusive. The exception is population CB5, where $Q_1$ performs better than $Q_2$. However, the weight of sample evidence is reversed when the Bayes factor is evaluated on population CB4 (the evidence varies from very strong to strong, depending on the employed probability cutoff), suggesting that the preference of $Q_1$ is most likely due to an increased contamination of the CB5 samples in the long-period tail of the contact binary distribution. For populations CB1 and CB3, the Bayes factors are between $10$ and $16$, which gives substantial but not strong evidence in favor of $Q_2$. The analysis is inconclusive for population CB2, where the Bayes factors are very close to unity for all probability cutoffs. Overall, our results indicate a general preference for $Q_2$, which agrees with the results reported by \citet{Rucinski_2001}. Based on these results, the fiducial models for our three populations are Q2CB2p50, Q2CB3p80, and Q2CB4p50.

Now we address the issue of whether our separate treatment of short-period contact binaries is supported by the data. In other words, we wish to quantify whether fitting the two populations separately and doubling the number of free parameters gives better results than fitting the entire population with a single model. For late-type binaries, we calculated the posterior Bayes factor comparing the combined Q2CB2p50+Q2CB3p50 model with model Q2CB1p50. This is justified because the two samples CB2p50 and CB3p50 are disjoint. The calculation yields a Bayes factor of about 71, providing strong evidence in favor of treating short-period and long-period late-type contact binaries separately. For early-type binaries, the limited size of the CB5 samples unfortunately prevents us from performing a similar analysis for periods shorter and longer than one day. Taking also the increased contamination of our CB5 population into account, which likely occurs due to the strongly decreasing frequency of contact binaries with period, we completely discarded CB5 from our analysis. From this point on, we only consider three distinct contact binary populations: CB2, CB3, and CB4.

\begin{figure*}
\centering
\includegraphics[width=\textwidth]{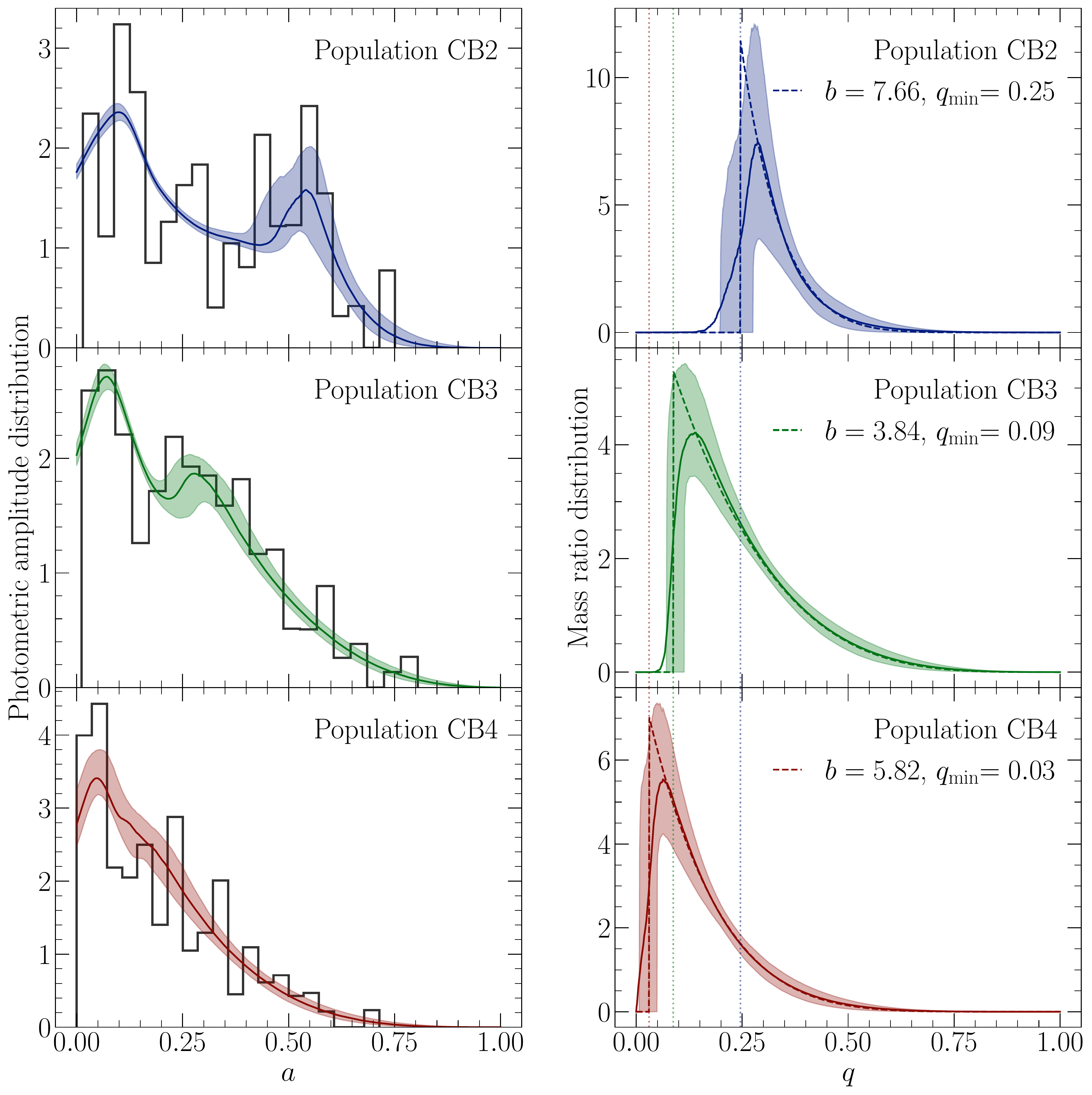}
\caption{Amplitude (left panel) and mass-ratio (right panel) distributions for our three populations CB2, CB3, and CB4. The solid black lines in the left panel show weighted histograms of the observed data. The solid blue, green, and red lines in both panels are obtained by marginalizing out the functional form of the $Q_2$ power law, and the dashed lines show $Q_2$ evaluated for the median values of $b$ and $\qmin$. The colored bands represent the $1\sigma$ credible intervals around the marginalized amplitude and mass-ratio distributions. The vertical dotted lines in the right panel compare the median values of $\qmin$ between the three populations.  \label{fig:fiducial_models_smoothed_Q}}
\end{figure*}

\begin{figure}
\centering
\resizebox{\hsize}{!}{\includegraphics{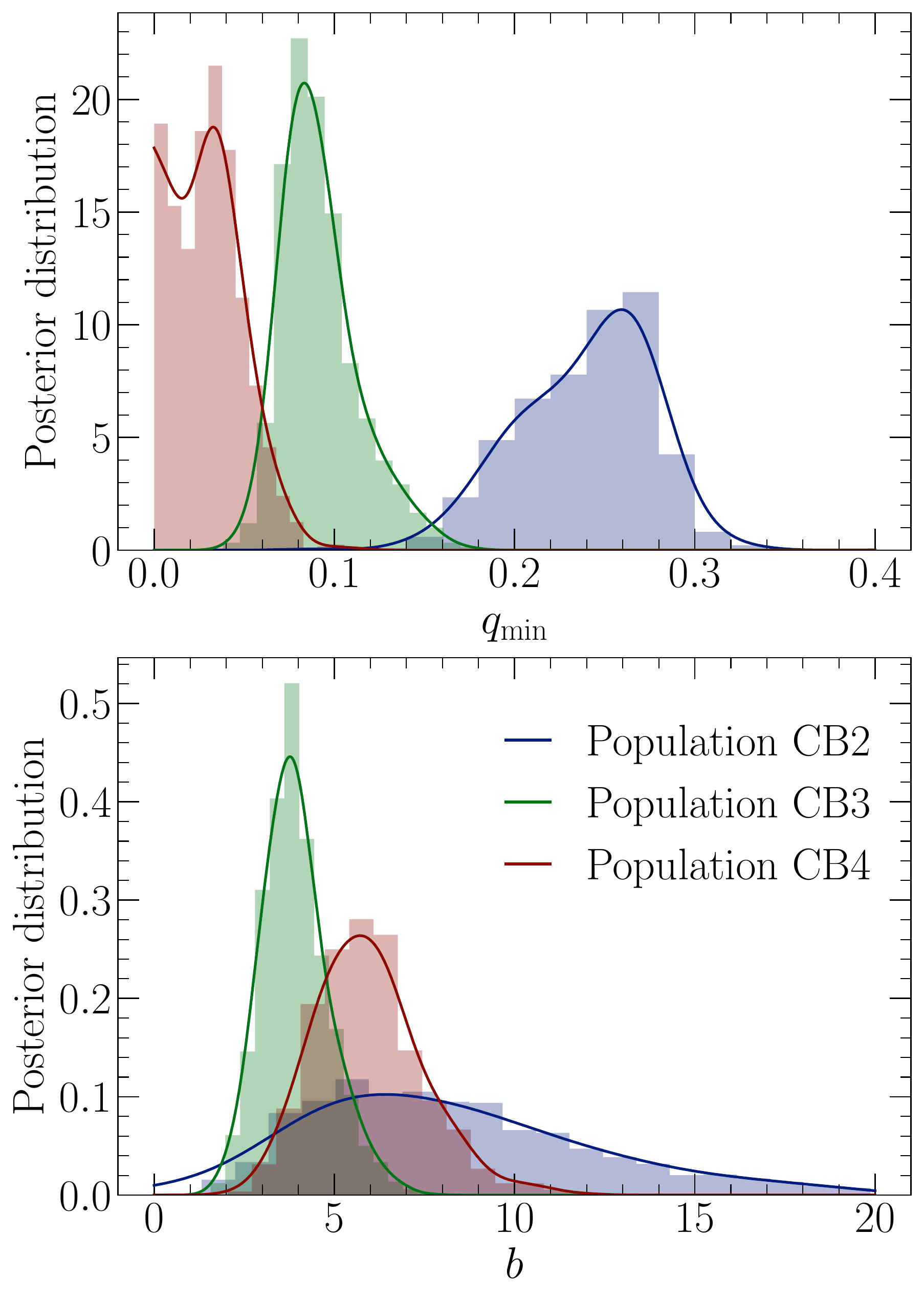}}
\caption{Comparison of the posteriors of $\qmin$ (top panel) and $b$ (bottom panel) resulting from the fiducial models for populations CB2, CB3, and CB4. \label{fig:fiducial_models_posterior_comparison}}
\end{figure}

\begin{figure*}
\centering
\includegraphics[width=\textwidth]{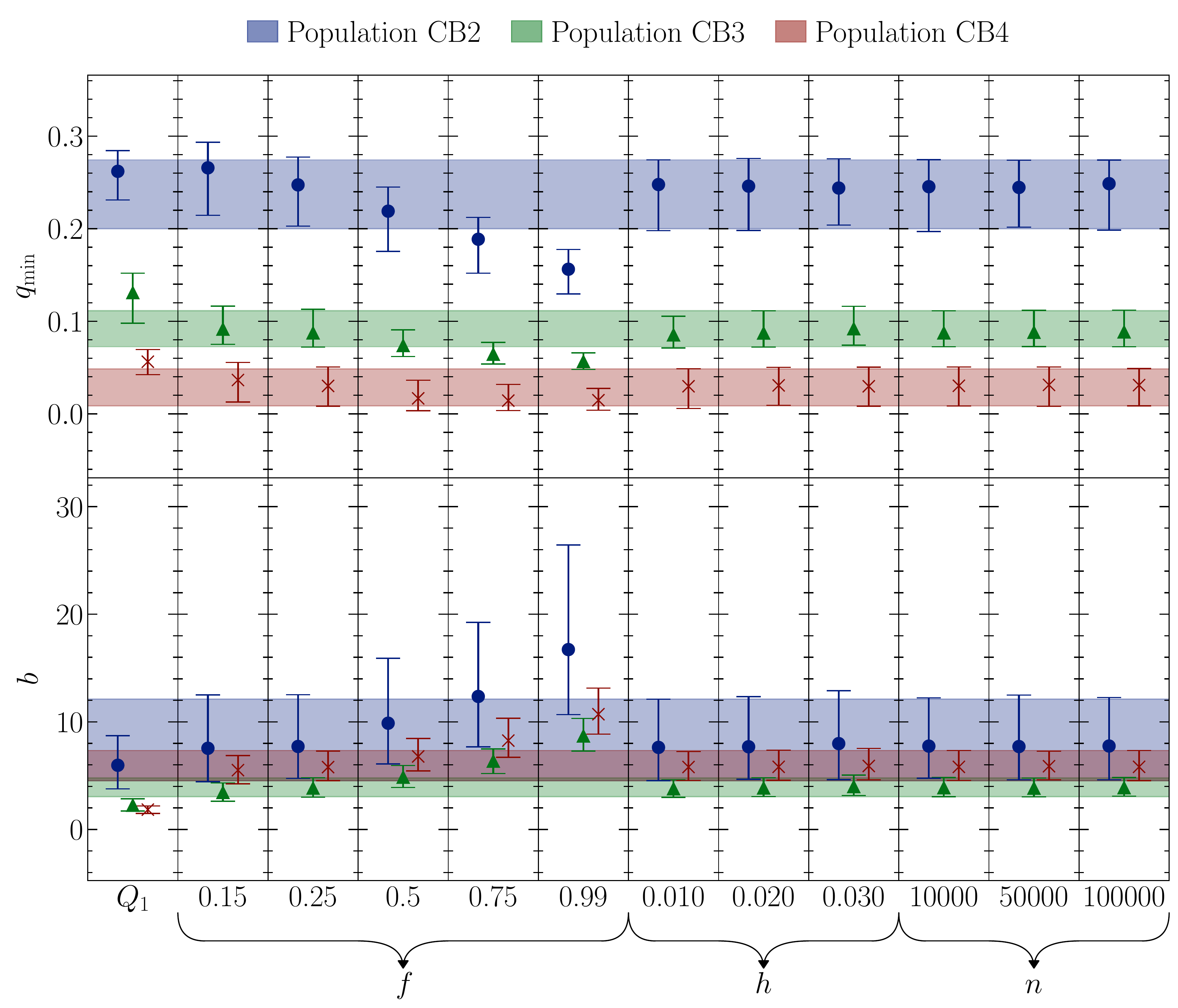}
\caption{Dependence of $\qmin$ and $b$ on the mass-ratio prescription $Q_1$, fill-out factor $f$, and hyperparameters of the model $h$ and $n$. The colored bands represent the $1\sigma$ credible intervals resulting from the fiducial models for the three populations. We show the full posterior distributions in Figs.~\ref{fig:posterior_Q1_Q2}, \ref{fig:posterior_distributions_F}, and \ref{fig:posterior_distributions_H}. \label{fig:summary_hyperparameters_q_min_b}}
\end{figure*}

\subsection{Mass-ratio distribution of contact binary stars} \label{sec:mass_ratio_distribution}
We now present our main results. In Fig.~\ref{fig:fiducial_models_smoothed_Q} we show the inferred amplitude and mass-ratio distributions for the three distinct contact binary populations CB2, CB3, and CB4. The distributions were obtained by marginalizing over the posteriors of the model parameters. We show the posterior distributions of $\qmin$ and $b$ for the three populations in Fig.~\ref{fig:fiducial_models_posterior_comparison}. In Fig.~\ref{fig:summary_hyperparameters_q_min_b} we compare the fiducial values of $\qmin$ and $b$ with the values obtained from models with different choices for some parameters, specifically, the mass-ratio distribution prescription ($Q_1$ vs.\ $Q_2$), fill-out factor, and hyperparameters $h$ and $n$. We show the full posterior distributions of all models in Appendix~\ref{app:posterior}. The fiducial models give for the minimum mass ratio
\begin{equation}
\qmin = \begin{cases}
0.246^{+0.029}_{-0.046} & \text{ CB2 (late-type binaries with } P \le 0.3\,\text{d)}, \\[5pt]
0.087^{+0.024}_{-0.015} & \text{ CB3 (late-type binaries with } P > 0.3\,\text{d)}, \\[5pt]
0.030^{+0.018}_{-0.022} & \text{ CB4 (early-type binaries with } P < 1\,\text{d)},
\end{cases}
\label{eq:summary_qmin}
\end{equation}
and for the slope of the mass-ratio distribution
\begin{equation}
b = \begin{cases}
7.66^{+4.45}_{-3.15} & \text{ CB2 (late-type binaries with } P \le 0.3\,\text{d)}, \\[5pt]
3.84^{+0.96}_{-0.80} & \text{ CB3 (late-type binaries with } P > 0.3\,\text{d)}, \\[5pt]
5.82^{+1.52}_{-1.30} & \text{ CB4 (early-type binaries with } P < 1\,\text{d)}.
\end{cases}
\label{eq:summary_b}
\end{equation}

Figs.~\ref{fig:fiducial_models_smoothed_Q} and \ref{fig:fiducial_models_posterior_comparison} show that $\qmin$ varies noticeably between our populations. There is a clear trend that $\qmin$ decreases with increasing $P$. The same holds for the mean values of $q$ calculated from the marginalized mass-ratio distributions, which go from $q_\text{mean}=0.33^{+0.21}_{-0.19}$ for population CB2 to $q_\text{mean}=0.25^{+0.06}_{-0.06}$ for population CB3 and $q_\text{mean}=0.16^{+0.04}_{-0.04}$ for population CB4. Using the PLC relation, we can translate the trend in $q_\text{min}$ into effective temperatures and luminosities: higher temperatures, luminosities, and larger radii imply lower values of $\qmin$. The shape of the CB4 fiducial posterior indicates that $\qmin$ for this populations is also consistent with being zero, but the limited size and the relatively high contamination of the CB4 fiducial sample prevent us from performing further tests of this hypothesis. We do not observe any clear trend in the values of the power-law exponent $b$. We discuss the astrophysical implications of our findings in Sect.~\ref{sec:conclusions}.

\subsection{Dependence on fill-out factor} \label{sec:dependence_fill_out}
Following \citet{Rucinski_2001}, we carried out all our Q1 and Q2 runs with $f=0.25$. To analyze the impact of $f$ on the fiducial models, we performed the runs FCB2--FCB4, which considered five different values of $f$: 0.15, 0.25, 0.5, 0.75, and 0.99. Fig.~\ref{fig:summary_hyperparameters_q_min_b} shows that the value of $\qmin$ strongly depends on $f$ in populations CB2 and CB3, with larger $f$ pushing $\qmin$ to lower values. Although the credible intervals overlap for $f \le 0.75$, the trend is clear. For population CB4, the values of $\qmin$ are consistent within the $1\sigma$ credible intervals across the whole range of $f$. Moreover, the power-law index $b$ grows with $f$ for all three populations.

In principle, we should be able to obtain the best-fitting value of $f$ by evaluating the posterior Bayes factors of the models. In this specific case, all Bayes factors are below 5, rendering the analysis inconclusive. Consequently, we are not able to infer the optimal value of $f$ from our data and we kept $f=0.25$ based on previous detailed models and theoretical considerations (Sect.~\ref{sec:method_overview}).

\subsection{Dependence on splitting period} \label{sec:dependence_period_cutoff}

\begin{figure}
\centering
\includegraphics[width=0.48\textwidth]{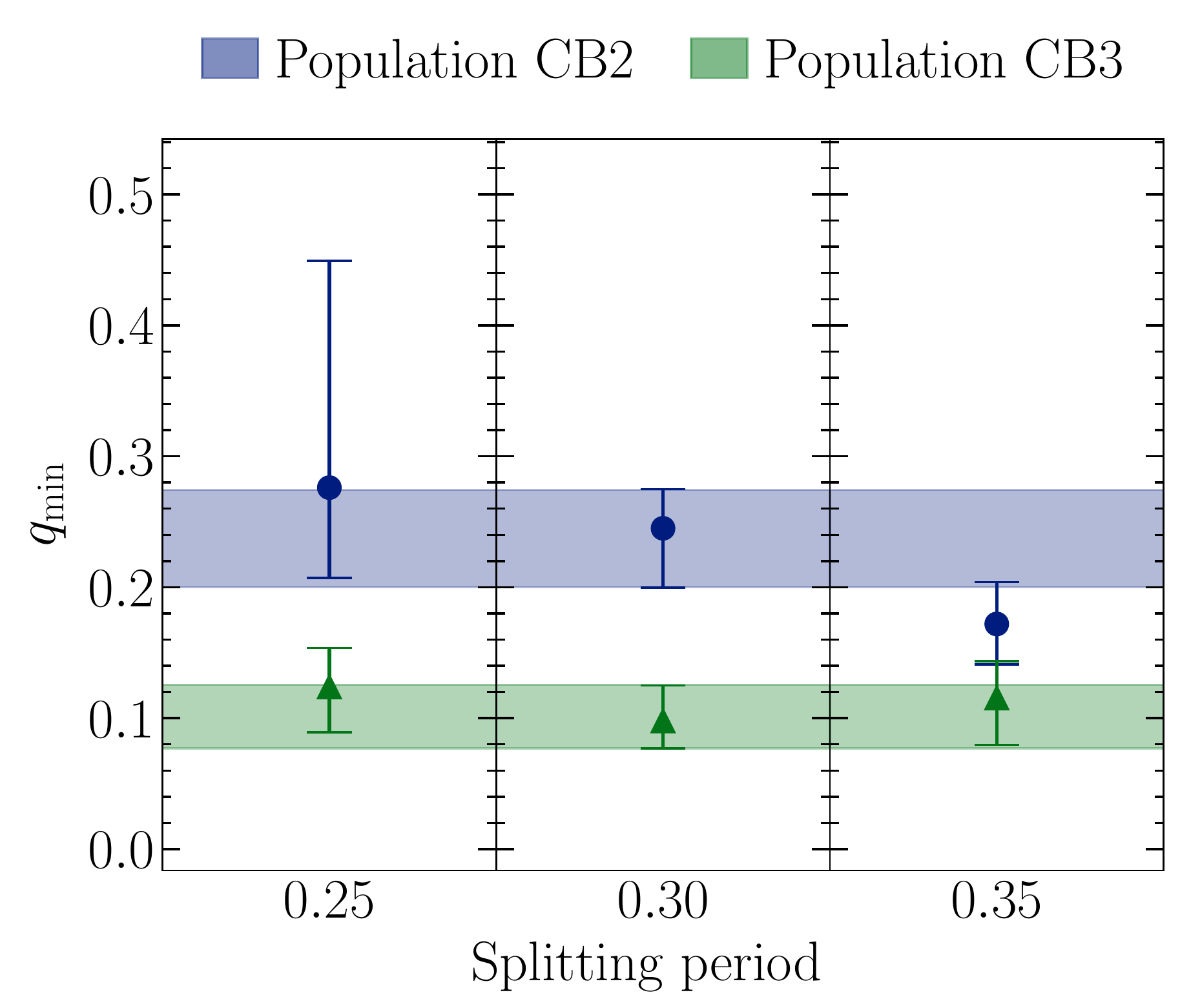}
\caption{Dependence of $\qmin$ on the splitting period for populations CB2 and CB3. The colored bands represent the $1\sigma$ credible intervals of the fiducial models. For the purpose of this plot, we lowered the probability cutoff of the fiducial model for population CB3 from 0.8 to 0.5. The full posterior is plotted in Fig.~\ref{fig:posterior_distributions_D}.
\label{fig:summary_q_min_period_cutoffs}}
\end{figure}

The choice to distinguish between the populations of late-type contact binaries with $P\le0.3$\,d and $P>0.3$\,d is motivated by \citet{Stepien_2012}, who argued that binaries with $P\lesssim 0.3$\,d do not live long enough to evolve to small $q$ , but instead merge at moderate $q$ due to the L2 overflow. Realistically, we expect a smooth transition between the two populations at around $P_\text{split}\approx 0.3$ d. If this is the case, $\qmin$ of population CB2 should gradually shift to lower values with increasing $P_\text{split}$ and increase or remain unchanged for $P_\text{split}<0.3$\,d. Conversely, $\qmin$ of population CB3 should not significantly change when $P_\text{split}$ is increased, but it should shift to higher values for $P_\text{split}<0.3$\,d. 

To investigate this hypothesis, we carried out runs SCB2 and SCB3, which examine how the fiducial results for populations CB2 and CB3 change when we shift $P_\text{split}$ from $0.30$\,d to $0.25$\,d or $0.35$\,d. To increase the size of the CB3 sample, we performed the analysis with a probability cutoff    of $0.5$ instead of the fiducial value $0.8$. We summarize the output from the runs in Fig.~\ref{fig:summary_q_min_period_cutoffs}, where we compare the resulting values of $\qmin$. The trends of $\qmin$ are consistent with our hypothesis overall, indicating that the observed difference between $\qmin$ of the two late-type populations is genuine and not just an artifact of the choice of $P_\text{split}$.

\subsection{Dependence on probability cutoffs} \label{sec:dependence_probability_cutoff}

\begin{figure}
\centering
\includegraphics[width=0.48\textwidth]{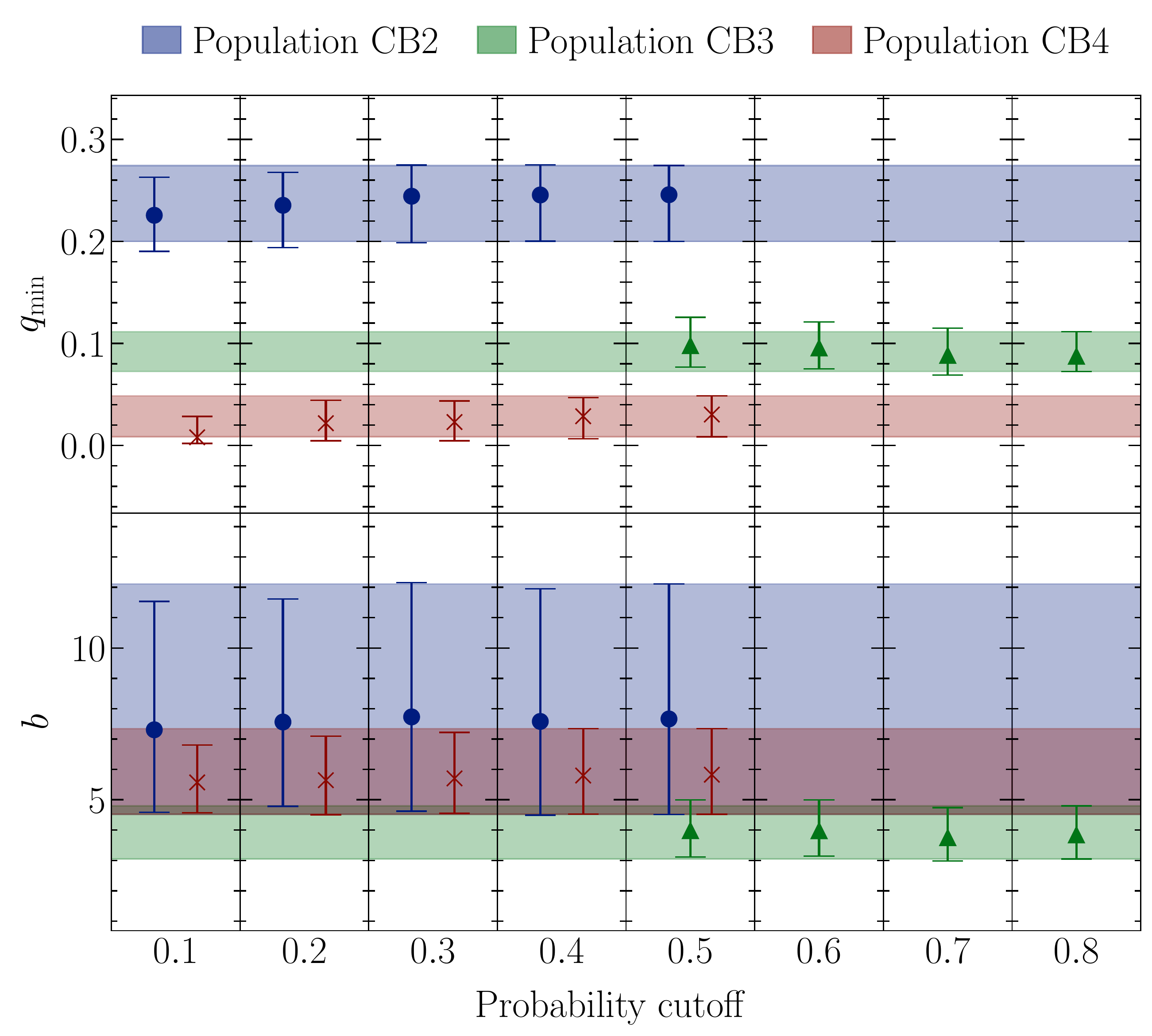}
\caption{Dependence of $\qmin$ and $b$ for populations CB2, CB3, and CB4 on different probability cutoffs separating contact binaries from contaminants. The colored bands represent the $1\sigma$ credible intervals of the fiducial models. We show the full posterior distributions in Fig.~\ref{fig:posterior_distributions_Q2}. \label{fig:summary_probability_cutoff_q_min_b}}
\end{figure}

In Fig.~\ref{fig:summary_probability_cutoff_q_min_b} we show how the fiducial values of $\qmin$ and $b$ change when we gradually decrease the probability cutoff that separates contact binaries from contaminants. The values are consistent with each other, and we do not observe any jumps or discontinuous changes in the full posterior distributions (Fig.~\ref{fig:posterior_distributions_Q2}). In most cases, as the probability cutoff increases, the posteriors simply become more concentrated toward the central point, which is the desired behavior. Population CB4 is an exception to this rule, with its $\qmin$ posteriors peaking close to zero or exactly at zero, depending on the employed probability cutoff. The CB4 posteriors of $\qmin$ appear to be cut off from the left, which is typically seen when the parameter is actually zero, but the prior enforces that it takes non-negative values.

\subsection{Dependence on hyperparameters} \label{sec:dependence_hyperparameters}
In defining the fiducial models, we had to make specific choices for the hyperparameters. To  ensure that the obtained results are robust with respect to these choices, we carried out a number of runs with different values of the hyperparameters. In Fig.~\ref{fig:summary_hyperparameters_q_min_b} we investigate the dependence of $\qmin$ and $b$ on $h$ and $n$. We tried three different values of $h$ (runs H1CB2--CB4 with $h=0.01$, $0.02$, and $0.03$) and three different values of $n$ (runs H2CB2--CB4 with $n=10000$, $50000$, and $n=100000$). The resulting $\qmin$ and $b$ are consistent with each other and almost perfectly overlap within their $1\sigma$ credible intervals. Detailed investigation of the posterior distributions in Fig.~\ref{fig:posterior_distributions_H} shows that the overall shapes and positions remain very similar. Our results indicate that $n=10000$ is already enough for the runs to converge to the correct solution, which justifies our fiducial choice of this value. 

\section{Discussions and conclusions} \label{sec:conclusions}
We have extended and reformulated the method for the estimation of the mass-ratio distribution of contact binary stars developed by \citet{Rucinski_2001}, which exploits the simplicity of contact binary light curves. Setting the fill-out factor to $f=0.25$ and assuming that binary orbits are randomly oriented in space, we obtained a semi-parametric mapping between the mass-ratio distribution and the photometric amplitude distribution (Sect.~\ref{sec:method}). We approximated the mass-ratio distribution as a power law with a slope $b$ and a sharp cutoff at $\qmin$, and using Bayesian inference, we obtained the posterior distributions of these parameters. This is possible because the position of the local maximum in the amplitude distribution is strongly correlated with the value of $\qmin$ (Fig.~\ref{fig:model_summary}). For the method to work, a sufficiently large sample of contact binaries is required that is complete for amplitudes $a\gtrsim0.1$\,mag or less. Such data sets have only recently become available from \textit{Kepler} and other space-based telescopes. The advantage of the method is that it infers $\qmin$ and $b$ purely from photometry, while other methods typically require much more costly spectroscopic observations or exhaustive modeling of stars one by one.

We constructed our sample from the Kepler Eclipsing Binary Catalog \citep{Prsa11,abdulmasih16,Kirk_2016}, which we combined with luminosities from \textit{Gaia} DR2 \citep{Gaia_DR2} and effective temperatures from \citet{Bai_2019} (Sect.~\ref{sec:data}). To filter out detached and semidetached binaries as well as other types of contaminating variable stars, we made use of the PLC relation (Sect.~\ref{sec:contaminants}). We distinguished between late- and early-type contact binaries, and we assumed that both types follow their own PLC relations, with a continuous transition between the two relations. We further assumed that the individual contact binaries are scattered around their respective PLC relations, and we modeled the contaminating noise as Gaussian (Sect.~\ref{sec:contaminants}). Employing Bayesian inference, we assigned a probability of being a contact binary of either late or early type to each object in the sample (Fig.~\ref{fig:2d_scatter_probability}). Late-type contact binaries have systematically larger amplitude scatter than early-type objects (Fig.~\ref{fig:sigma_a_vs_a}), which is most likely due to the presence of time-varying stellar spots in the atmospheres of late-type stars. Seeing that the relative amplitude uncertainty remains below $10\%$ for most objects, we conclude that this phenomenon does not significantly affect our method for the inference of the mass-ratio distribution.

Using different period cutoffs, we constructed five separate but overlapping populations of contact binary stars (Sect.~\ref{sec:population}): all late-type contact binaries (CB1), late-type contact binaries with $P\le0.3$\,d (CB2), late-type contact binaries with $P>0.3$\,d (CB3), early-type contact binaries with $P<1$\,d (CB4), and all early-type contact binaries with no constraint on the period (CB5). For each population, we defined multiple samples by imposing different cutoffs on the probability of being a contact binary of either type (Table~\ref{tab:samples}). We fit each sample with two different power-law prescriptions for the mass-ratio distribution, as defined in Eq.~\eqref{eq:Q}, and for each sample, we calculated the posterior Bayes factor comparing the goodness-of-fit of the two prescriptions (Sect.\ref{sec:fiducial_models}). In most cases, the second prescription $Q_2\propto(1-q)^\mathrm{b}$ yields a better fit than the first prescription $Q_1\propto q^{-b}$ (Tab.~\ref{tab:posterior_bayes_factors}), but the evidence is not strong, with most Bayes factors at or below 20. A notable exception is the CB5 population, where $Q_1$ outperformed $Q_2$. This result is most likely affected by the increased contamination of the CB5 samples. The population of late-type contact binaries with $P\le0.3$\,d (CB2) does not favor either prescription. Only the CB4 population gives conclusive evidence in support of $Q_2$ against $Q_1$. In conclusion, we observe a marginally strong evidence in support of $Q_2$, which agrees with the previous results of \citet{Rucinski_2001}. 

Adopting $Q_2$ as the mass-ratio distribution of contact binary stars, we justified the separate treatment of populations CB2 and CB3 by calculating the posterior Bayes factor of the combined CB2+CB3 model and the model for CB1. We found very strong evidence in support of the combined model (Sect.~\ref{sec:fiducial_models} and \ref{sec:dependence_period_cutoff}). Consequently, we investigated the two populations CB2 and CB3 separately, and we discarded the combined population CB1. We also discarded CB5 due to the increased contamination of its samples. In summary, we were left with three distinct contact binary populations CB2, CB3, and CB4. 

Our results for $\qmin$ and $b$ are summarized in Sect.~\ref{sec:mass_ratio_distribution}. We find that $\qmin$ decreases with increasing orbital period. For late-type binaries with $P\le0.3$\,d, we find a relatively high $\qmin=0.246^{+0.029}_{-0.046}$. For normal late-type binaries, we find $\qmin=0.087^{+0.024}_{-0.015}$. For early-type binaries with $P<1$\,d, we find $\qmin=0.030^{+0.018}_{-0.022}$. Our results are compatible with theoretical predictions of $\qmin$. Specifically, our $\qmin$ for late-type binaries with $P > 0.3$\,d agrees with theoretical values for solar-type stars, where \citet{Rasio_1995} predicted $\qmin = 0.08$ for an $n=3$ polytrope. It is also known that $\qmin$ scales with the stellar gyration radius, which is relatively small for early-type stars \citep{Rasio_1995,wadhwa21,blagorodnova21}. This agrees with our very small $\qmin$ for this population. However, given the credible interval of our result, we cannot definitely claim detection of the signature of the Darwin instability in the early-type population.

The trend of decreasing $\qmin$ with increasing orbital period agrees with the conclusions of \citet{Stepien_2012}, who argued that this is due to the different timescales of mass transfer and angular momentum loss in low-mass contact binaries and more massive systems. The relatively moderate mass transfer in low-mass (short-period) contact binaries is insufficient to make the binary unstable to the Darwin instability, but instead, it leads to the overflow of the outer Roche lobe, resulting in the loss of mass angular momentum through the L2 point and culminating with merger at comparatively larger $q$ than in the case of the Darwin instability. In contrast, \citet{Kobulnicky_2022} argued for an opposite trend, where $\qmin$ increases with period for $P \gtrsim 0.8$\,d. Their model assumed that new contact binary systems form with $q\approx 1$ and conservatively evolve toward longer periods and smaller $q$ until the onset of the Darwin instability.

We find different values of the power-law index $b$ for different populations, but unlike for $\qmin$, we do not observe a clear trend with the orbital period. For $Q_2$, \citet{Rucinski_2001} reported $b=6 \pm 2$, which is consistent with our results for all three populations (CB2: $b=7.66^{+4.45}_{-3.15}$, CB3: $b=3.84^{+0.96}_{-0.80}$, and CB4: $b=5.82^{+1.52}_{-1.30}$). We note that \citet{Rucinski_2001} did not distinguish between late- and early-type contact binaries and that their sample is complete only for $a\gtrsim 0.3$\,mag. Our relative uncertainties in $b$ for populations CB3 and CB4 are only mildly smaller than those reported by \citet{Rucinski_2001}, which is understandable given the similar sample sizes. Larger samples of contact binaries are required to better constrain $b$. This could be quite rewarding because $b$ encodes physical processes such as nuclear evolution, magnetic braking, and thermal relaxation oscillations \citep{vilhu81}.  \citet{Rucinski_2001} indeed suggested that $b$ is related to the thermal timescale of the secondary star and thus to the exponent of its mass--luminosity relation.

Our results show that $\qmin$ noticeably depends on the value of the fill-out factor $f$ (Sect.~\ref{sec:dependence_fill_out} and Fig.~\ref{fig:summary_hyperparameters_q_min_b}), but our analysis of the posterior Bayes factors was inconclusive due to the insufficient evidence in favor of any specific $f$ (all factors were below $5$). Consequently, we were not able to constrain $f$ from our data. Nonetheless, thermal relaxation oscillations theory suggests that $f$ should be small and similar to our default value $f=0.25$ \citep{Lucy_1973,Rucinski_1973,Rucinski_1997,Paczynski_2006}. Still, there is some evidence that $f$ is different for late- and early-type binaries \citep{mochnacki81}. In addition to the fill-out factor, we also verified that our estimates of $b$ and $\qmin$ are fairly robust with respect to the splitting period $P_\text{split}$ between populations CB2 and CB3 (Sect.~\ref{sec:dependence_period_cutoff} and Fig.~\ref{fig:summary_q_min_period_cutoffs}), the probability cutoff (Sect.~\ref{sec:dependence_probability_cutoff} and Fig.~\ref{fig:summary_probability_cutoff_q_min_b}), and the KDE bandwidth $h$ and number of Gaussians $n$ involved in the construction of the amplitude distribution (Sect.~\ref{sec:dependence_hyperparameters} and Fig.~\ref{fig:summary_hyperparameters_q_min_b}).

The method presented here can easily be extended to the large samples of contact binaries expected from TESS and other space-based telescopes. In addition to giving better estimates for the parameters of the current model, these samples will enable characterization of more complex models that better capture the underlying mass-ratio distribution of contact binaries. One way to improve the current model is to include the splitting period as a parameter in the generative distribution constructed in Sect.~\ref{sec:generative_model}. With this modification, we could fit the mass-ratio distributions of populations CB2 and CB3 simultaneously, and by marginalizing out the exact location of the split, we would obtain $P_\text{split}$-free estimates of $\qmin$ for the two populations. 

A straightforward improvement of our approach would come from using more precise values for effective temperatures and luminosities. The recently released \textit{Gaia} DR3 \citep{gaia_dr3} provides a significant improvement over DR2, but unfortunately, the physical parameters continue to be based on single-star models \citep{creevey22}. We showed here that this assumption does not significantly affect our results, but improvements in this area could provide better distinction of contact binaries from various contaminants.

Another exciting possibility comes from combining space-borne all-sky photometry from TESS or \textit{Gaia} with data from massive spectroscopic surveys such as SDSS-V \citep{kollmeier19}, WEAVE \citep{dalton12}, 4MOST \citep{dejong19}, LAMOST \citep{zhao12}, or \textit{Gaia} RVS \citep{katz22}. These spectroscopic surveys often secure several spectra of each object. Although obtaining complete orbital and physical solution is still hard with these data alone \citep[e.g.,][]{pricewhelan18}, even a constraint with a low signal-to-noise ratio of the radial velocity amplitude or the flux ratio of the two components might greatly increase the statistical power of our model by excluding ranges of possible inclinations for each binary. Operationally, we would simultaneously fit the model to the observed amplitude and mass-ratio distributions, effectively yielding a nonuniform prior on the parameters of the power law. Ultimately, the scalability and flexibility of our method make it a powerful tool for the inference of the mass-ratio distribution and the minimum mass ratio of contact binary stars.

\begin{acknowledgements}
We thank Matthew Green for sharing their sample of contact binaries and our referee, Panagiota-Eleftheria Christopoulou, for her helpful comments. This work has been supported by INTER-EXCELLENCE grant LTAUSA18093 from the Ministry of Education, Youth, and Sports. The research of OP has been supported also by Horizon 2020 ERC Starting Grant ‘Cat-In-hAT’ (grant agreement no. 803158). This research made use of the cross-match service provided by CDS, Strasbourg.
\end{acknowledgements}

\bibliographystyle{aa}
\bibliography{bibliography}

\onecolumn
\begin{appendix}

\section{Evaluation of likelihood} \label{app:likelihood}
When  the amplitude distribution in Sect.~\ref{sec:method} was evaluated, we were only able to construct $\hat{A}(a;\Theta)$, which is an analytical approximation to $A(a;\Theta)$. The approximation involves performing KDE on a finite number of randomly drawn amplitudes, which introduces a stochastic element into the process, transforming $\hat{A}(a;\Theta)$ into a random variable. This means that the likelihood in Eq.~\eqref{eq:Theta_likelihood} does not yield a unique value for a given $\Theta$ and $\{a_k\}_{k=1}^N$, but is actually a random variable itself.

\begin{figure*}[h]
\centering
\includegraphics[width=0.95\textwidth]{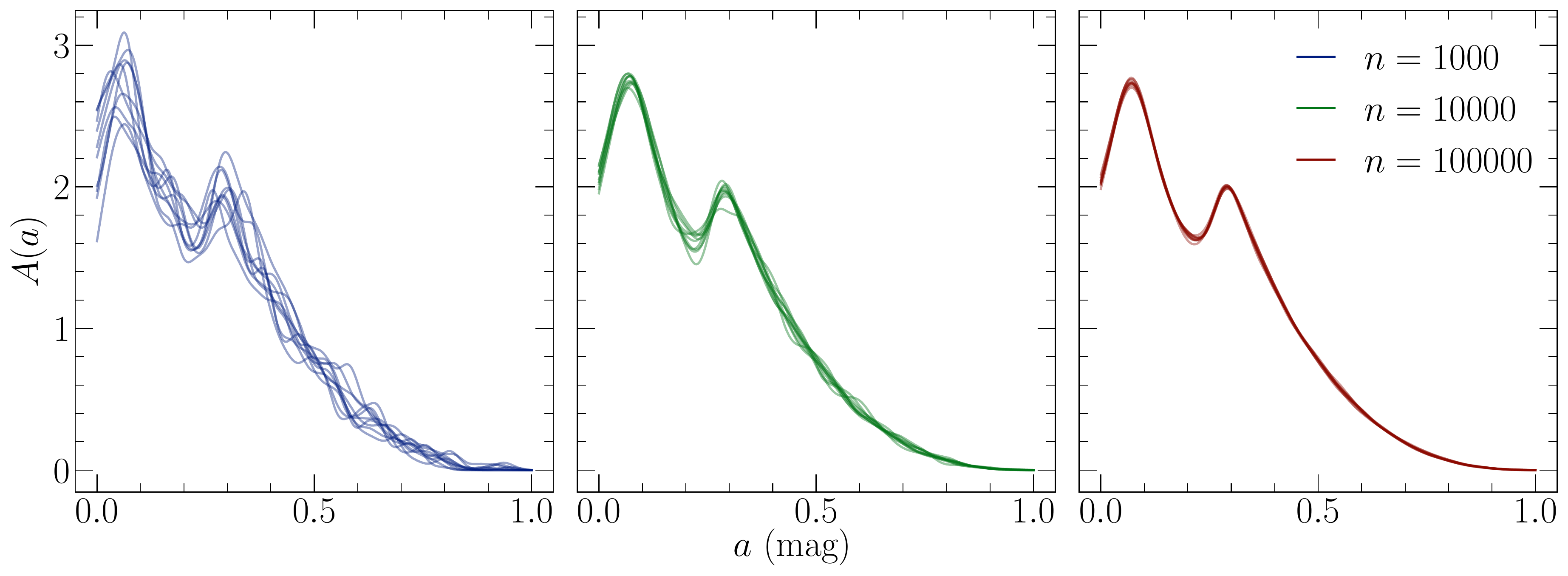}
\caption{Comparison of synthetic contact binary amplitude distributions resulting from the kernel density estimation performed on samples of different sizes. The distribution is rather noisy for $n=1000$, but the noise is already significantly reduced for $n=10000$. When we increase the number of samples to $100000$, the distribution effectively converges to the correct shape. \label{fig:amplitude_distribution_kde_comparison}}
\end{figure*}

As illustrated in Fig.~\ref{fig:amplitude_distribution_kde_comparison}, the noise in $\hat{A}(a;\Theta)$ can be significantly reduced by employing a sufficiently large number of samples, but for the stochastic behavior to completely disappear, we would have to use the same input for KDE in each evaluation of $\hat{A}(a,\Theta)$. Unfortunately, none of these options are feasible; increasing the number of drawn amplitudes comes at huge computational cost due to the repeated log-likelihood evaluation during an MCMC run, and fixing the KDE input requires an analysis of which amplitude sample leads to the most accurate representation of $A(a)$, which cannot be achieved in any practical way.

Instead of trying to minimize the stochastic effect, we fully embraced the nondeterministic nature of $\hat{A}(a;\Theta)$ and modeled it as a sampling noise in $A(a;\Theta)$. We note that the scatter in $A(a;\Theta)$ is different from the scatter in the PLC relation that we investigated in Sect.~\ref{sec:intrinsic_scatter}. The scatter in $A(a;\Theta)$ smears the distribution itself, while the scatter in the PLC relation affects an originally exact relation and transforms it into a distribution. In principle, the extent to which $A(a;\Theta)$ is smeared depends on the value of $\Theta$, which further adds to the complexity of the problem. The smearing can be equivalently viewed as an implicit dependence of $\hat{A}(a;\Theta)$ on an additional $2n$ parameters corresponding to the $(i,q)$ positions of the $n$ samples entering the KDE algorithm. Denoting the individual parameters by $Y_l$, with $l$ going from $1$ to $2n$, we can write the likelihood as
\begin{ceqn}
\begin{equation}
  \hat{\mathscr{L}}(\Theta,\{Y_l\}_{l=1}^{2n}|\{a_k\}_{k=1}^N) = \prod_{k=1}^N p_{\mathrm{CB},k} \int \hat{A}(a_k;\Theta,\{Y_l\}_{l=1}^{2n}) \mathscr{N}(a;a_k,\sigma_{a_k})\text{d}a.
\end{equation} 
\end{ceqn}

By including these parameters, we remove the stochasticity and the likelihood becomes deterministic again. The additional parameters are distributed according to the joint distribution of $i$ and $q$, which is given by $I(i)\times Q(q;\Theta)$ and serves as the conditional prior for these parameters. Since we are only interested in the posterior of $\Theta$, we did not actively sample the additional parameters and their prior did not directly enter the Bayes theorem. Instead, in each step of the MCMC run, we updated $\Theta$ according to the chosen step-proposal strategy (e.g., stretch move or differential evolution) and the additional $2n$ parameters are simply drawn from the prior. This is equivalent to sampling the full posterior,
\begin{ceqn}
\begin{equation} \label{eq:full_posterior}
  p(\Theta,\{Y_l\}_{l=1}^{2n}|\{a_k\}_{k=1}^N)=\frac{\mathscr{L}(\Theta,\{Y_l\}_{l=1}^{2n}|\{a_k\}_{k=1}^N)p(\{Y_l\}_{l=1}^{2n}|\Theta)p(\Theta)}{p(\{a_k\}_{k=1}^N)},
\end{equation}
\end{ceqn}
and marginalizing out the additional parameters, yielding the marginalized posterior probability distribution of $\Theta$, or $p(\Theta|\{a_k\}_{k=1}^N)$.

This approach does not yield the posterior for the additional parameters, which is needed for the marginalization of $\hat{A}(a;\Theta,\{Y_l\}_{l=1}^{2n})$ or the calculation of the Bayes factors. To reconstruct the full posterior, we substituted the posterior of the additional parameters with the prior. This is justifiable because in the limit of $N\xrightarrow{}\infty$, the stochastic amplitude distribution $\hat{A}(a;\Theta,\{Y_l\}_{l=1}^{2n})$ converges to $A(a;\Theta)$, causing the posterior of the additional parameters to converge to the prescribed prior.

\clearpage

\section{Additional tables and figures for the identification of sample contamination} \label{app:additional}
\begin{figure*}[h]
\centering
\includegraphics[width=0.95\textwidth]{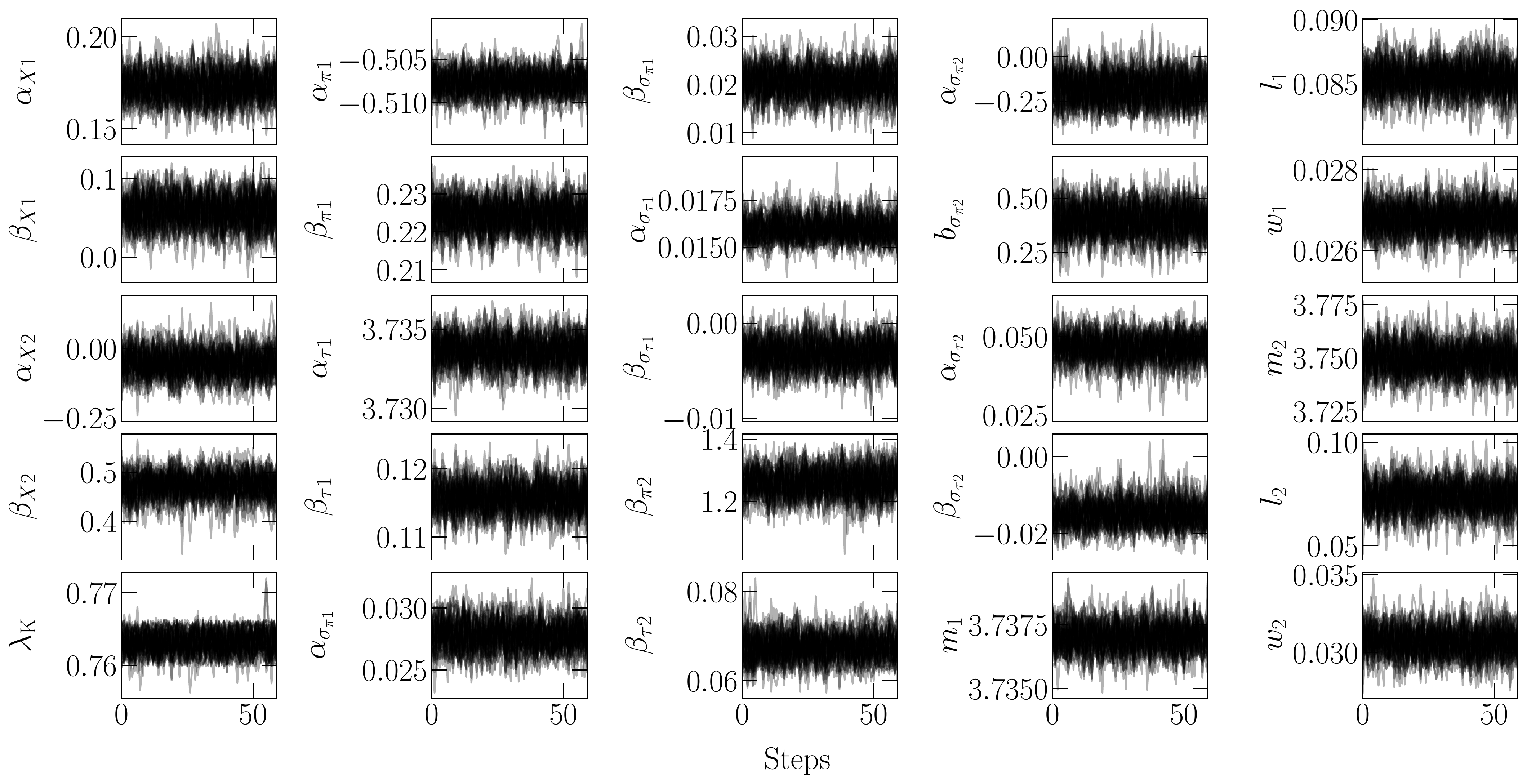}
\caption{Chain plots resulting from the \emph{emcee} run of our Bayesian model for the identification of contact binary stars. We ran the sampler for a total of 160 000 steps, but we discarded the first 10000 as burn-in, and we thinned the chains by a factor of 300. Visual inspection of the plot confirms that the number of steps was sufficient for the chains to converge. \label{fig:data_model_chains}}
\end{figure*}

\begin{figure*}[h]
\centering
\includegraphics[width=0.95\textwidth]{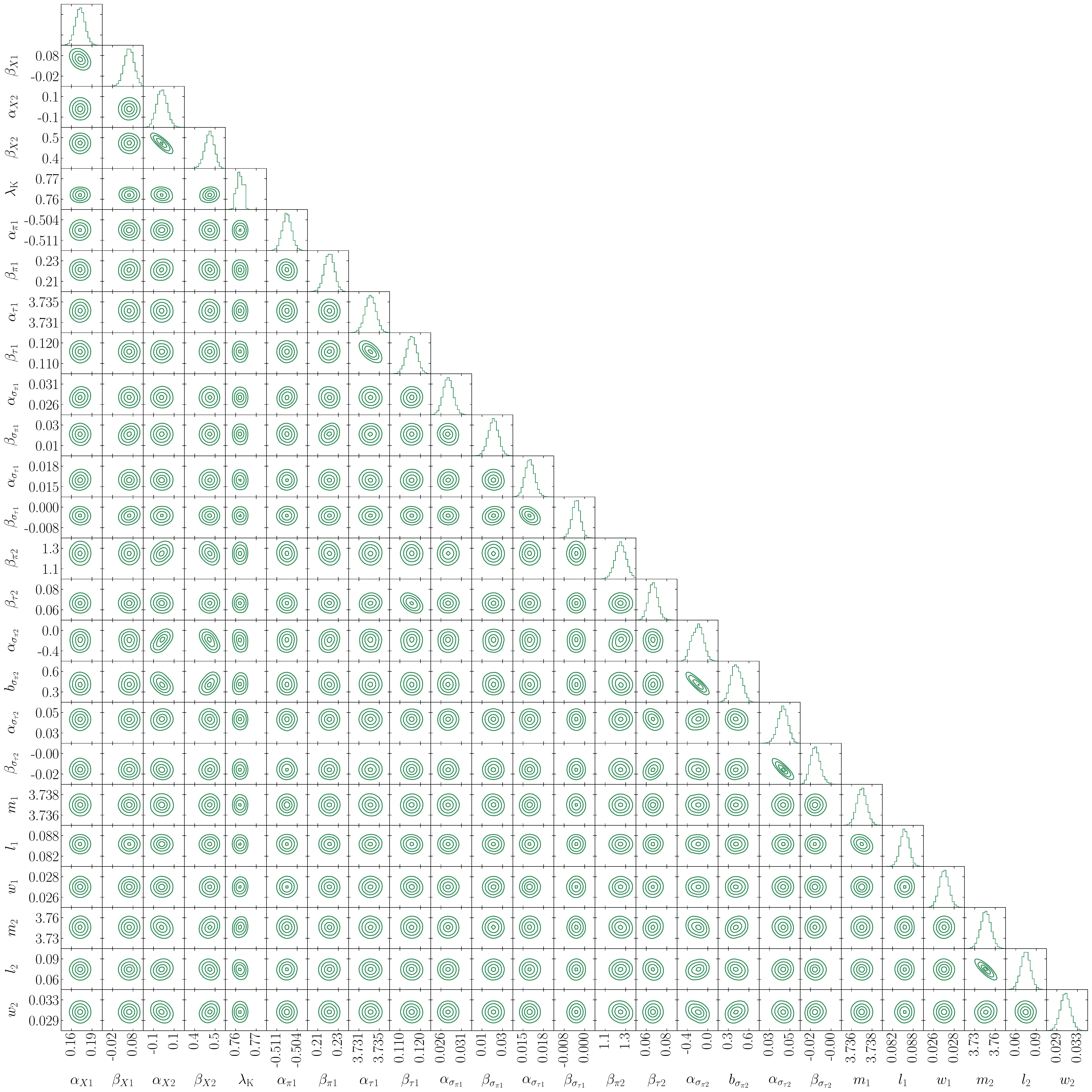}
\caption{Corner plot resulting from the \emph{emcee} run of our Bayesian model for the identification of contact binary stars. We ran the sampler for a total of 160 000 steps, but we discarded the first 10000 as burn-in, and we thinned the chains by a factor of 300. Visual inspection of the plot confirms that the number of steps was sufficient for the chains to converge. \label{fig:data_model_corner}}
\end{figure*}

Here, we present a list of the parameters of our Bayesian model for removing contaminants in the sample of contact binaries (Tab.~\ref{tab:model_parameters}), plots of the MCMC chains resulting from the model (Fig.~\ref{fig:data_model_chains}), and visualization of the posterior distribution of the model parameters (Fig.~\ref{fig:data_model_corner}).

\begin{table}[h]
\caption{List of the parameters of our Bayesian model for the identification of contact binary stars. \label{tab:model_parameters}}
\begin{center}
\begin{tabular}{ccr}
\hline\hline
Parameter & Definition & Value \\
\hline
\multicolumn{3}{c}{Global parameters}\\
\hline
$\lambdak$ & $\lambda$-location of the Kraft break along the PLC relation & $0.7631^{+0.0021}_{-0.0017}$\\
$\alpha_{X1}$ & \multirow[c]{2}{*}{$X=\alpha_{X1}+\beta_{X1} \lambda,\quad \lambda\le \lambdak$} & $0.1727^{+0.0086}_{-0.0086}$\\
$\beta_{X1}$ & & $0.0591^{+0.0219}_{-0.0225}$\\
$\alpha_{X2}$ & \multirow[c]{2}{*}{$X=\alpha_{X2}+\beta_{X2} \lambda,\quad \lambda> \lambdak$} & $-0.0538^{+0.0537}_{-0.0520}$\\
$\beta_{X2}$ & & $0.4750^{+0.0262}_{-0.0283}$\\
\hline
\multicolumn{3}{c}{PLC parameters}\\
\hline
$\alpha_{\pi1}$ & \multirow[c]{2}{*}{$\mu_{\mathrm{S}\pi}=\alpha_{\pi1}+\beta_{\pi1} \lambda,\quad \lambda\le \lambdak$} & $-0.5077^{+0.0015}_{-0.0015}$\\
$\beta_{\pi1}$ & & $0.2243^{+0.0043}_{-0.0044}$\\
$\beta_{\pi2}$ & $\mu_{\mathrm{S}\pi}=\alpha_{\pi1}+(\beta_{\pi1}-\beta_{\pi2}) \lambdak + \beta_{\pi2}\lambda,\quad \lambda> \lambdak$ & $1.2614^{+0.0490}_{-0.0484}$\\
$\alpha_{\tau1}$ & \multirow[c]{2}{*}{$\mu_{\mathrm{S}\tau}=\alpha_{\tau1}+\beta_{\tau1} \lambda,\quad \lambda\le \lambdak$} & $3.7337^{+0.0010}_{-0.0010}$\\
$\beta_{\tau1}$ & & $0.1159^{+0.0023}_{-0.0023}$\\
$\beta_{\tau2}$ & $\mu_{\mathrm{S}\pi}=\alpha_{\tau1}+(\beta_{\tau1}-\beta_{\tau2}) \lambdak + \beta_{\tau2}\lambda,\quad \lambda> \lambdak$ & $0.0672^{+0.0033}_{-0.0031}$\\
$\alpha_{\sigma_{\pi1}}$ & \multirow[c]{2}{*}{$\sigma_{\mathrm{S}\pi}=\alpha_{\sigma_{\pi1}}+\beta_{\sigma_{\pi1}} \lambda,\quad \lambda\le \lambdak$} & $0.0277^{+0.0013}_{-0.0013}$\\
$\beta_{\sigma_{\pi1}}$ & & $0.0206^{+0.0033}_{-0.0033}$\\
$\alpha_{\sigma_{\pi2}}$ & \multirow[c]{2}{*}{$\sigma_{\mathrm{S}\pi}=\alpha_{\sigma_{\pi2}}+\beta_{\sigma_{\pi2}} \lambda,\quad \lambda> \lambdak$} & $-0.1754^{+0.0986}_{-0.1081}$\\
$\beta_{\sigma_{\pi2}}$ & & $0.3961^{+0.0891}_{-0.0820}$\\
$\alpha_{\sigma_{\tau1}}$ & \multirow[c]{2}{*}{$\sigma_{\mathrm{S}\tau}=\alpha_{\sigma_{\tau1}}+\beta_{\sigma_{\tau1}} \lambda,\quad \lambda\le \lambdak$} & $0.0159^{+0.0007}_{-0.0007}$\\
$\beta_{\sigma_{\tau1}}$ & & $-0.0032^{+0.0016}_{-0.0017}$\\
$\alpha_{\sigma_{\tau2}}$ & \multirow[c]{2}{*}{$\sigma_{\mathrm{S}\tau}=\alpha_{\sigma_{\tau2}}+\beta_{\sigma_{\tau2}} \lambda,\quad \lambda> \lambdak$} & $0.0470^{+0.0045}_{-0.0048}$\\
$\beta_{\sigma_{\tau2}}$ & & $-0.0153^{+0.0039}_{-0.0035}$\\
\hline
\multicolumn{3}{c}{Background noise parameters}\\
\hline
$m_1$ & \multirow[c]{2}{*}{$\mu_{\mathrm{B}\tau}=m_1+l_1 \lambda,\quad \lambda\le \lambdak$} & $3.7370^{+0.0006}_{-0.0006}$\\
$l_1$ & & $0.0855^{+0.0013}_{-0.0013}$\\
$m_2$ & \multirow[c]{2}{*}{$\mu_{\mathrm{B}\tau}=m_2+l_2 \lambda,\quad \lambda> \lambdak$} & $3.7496^{+0.0077}_{-0.0076}$\\
$l_2$ & & $0.0738^{+0.0077}_{-0.0078}$\\
$w_1$ & $\sigma_{\mathrm{B}\tau}=w_1,\quad \lambda\le \lambdak$ & $0.0268^{+0.0004}_{-0.0004}$\\
$w_2$ & $\sigma_{\mathrm{B}\tau}=w_2,\quad \lambda> \lambdak$ & $0.0307^{+0.0010}_{-0.0010}$\\
\hline
\end{tabular}
\end{center}
\tablefoot{We assumed that contact binaries are scattered around the PLC relation, parametrically expressed as $(\lambda,\mu_{\mathrm{S}\pi},\mu_{\mathrm{S}\tau})$ in the log-luminosity $\lambda$ vs. log-period $\pi$ vs. log-effective temperature $\tau$ space. The $\sigma_{\mathrm{S}}$ parameters control the level of scatter around the relation. We modeled the background noise as though it were generated from a thick plane $(\lambda,\pi,\mu_\mathrm{B})$ with its thickness controlled by the $\sigma_\mathrm{B}$ parameters. We present the values of the parameters with their $1\sigma$ credible intervals.}
\end{table}

\newpage
\null

\vspace*{\fill}
\vspace*{\fill}

\newpage
\null

\vspace*{\fill}
\vspace*{\fill}

\clearpage

\section{Additional tables and figures for the mass-ratio distribution} \label{app:posterior}
\begin{figure}[h]
\centering
\includegraphics[width=0.9\textwidth]{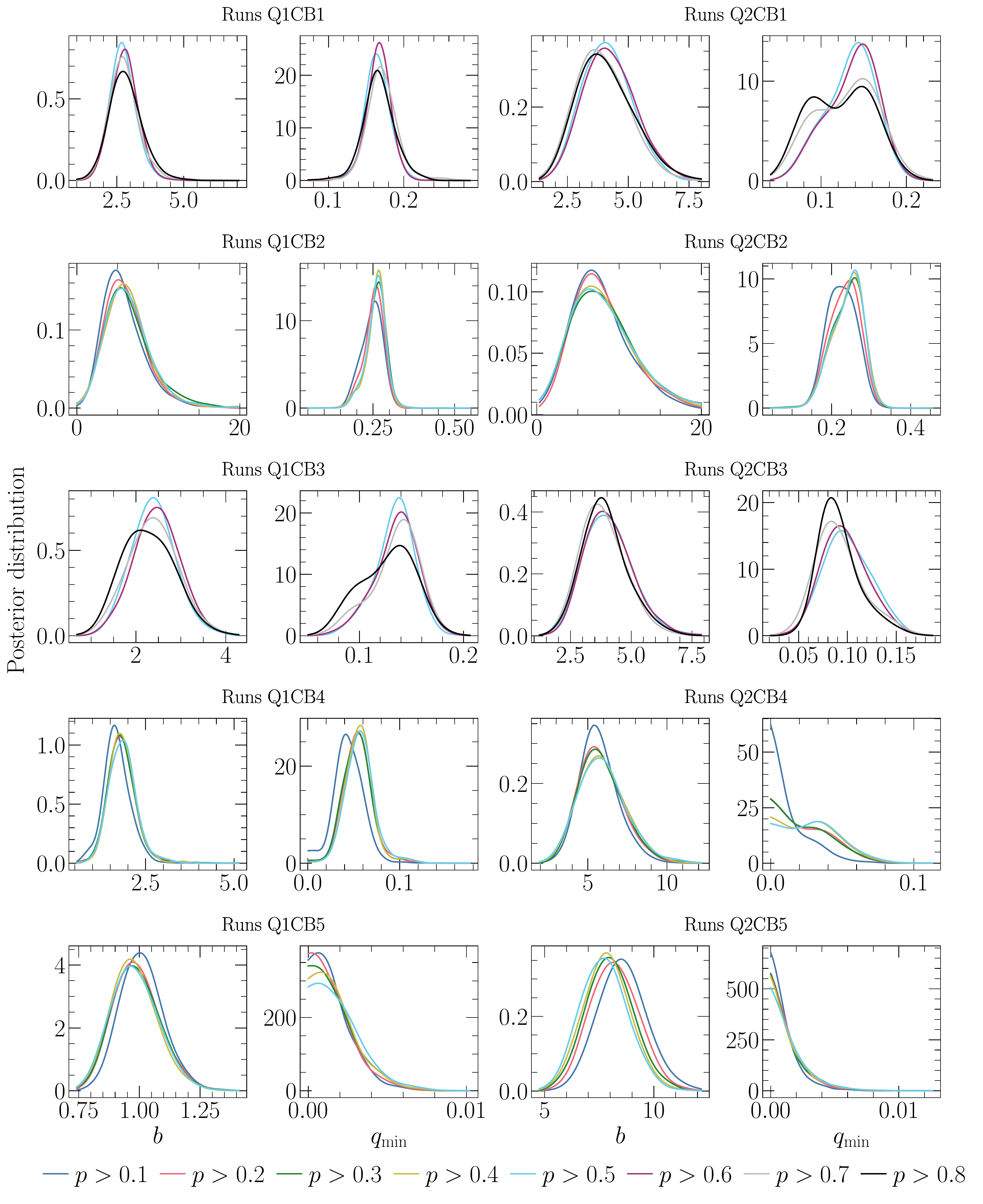}
\caption{Posterior distributions of $b$ and $q_\mathrm{min}$ conditional on $Q_1$ (left) and $Q_2$ (right) for populations CB1--CB5 and different probability cutoffs.
\label{fig:posterior_Q1_Q2}}
\end{figure}

\begin{figure}
\centering
\includegraphics[width=0.48\textwidth]{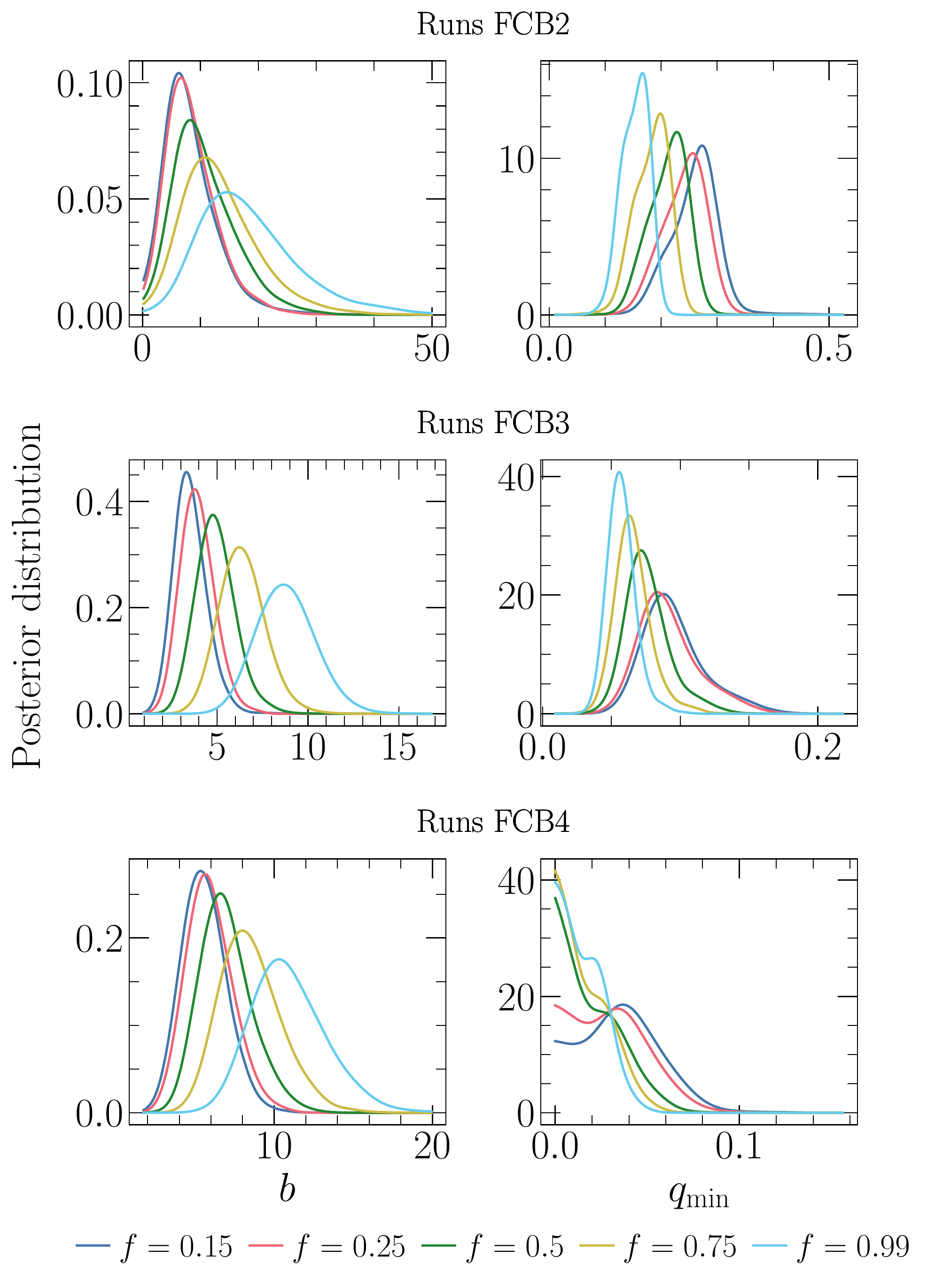}
\caption{Dependence of the fiducial posterior distributions of $b$ (left) and $q_\mathrm{min}$ (right) for populations CB2--CB4 on different fill-out factors. \label{fig:posterior_distributions_F}}
\end{figure}

\begin{figure}
\centering
\includegraphics[width=0.45\textwidth]{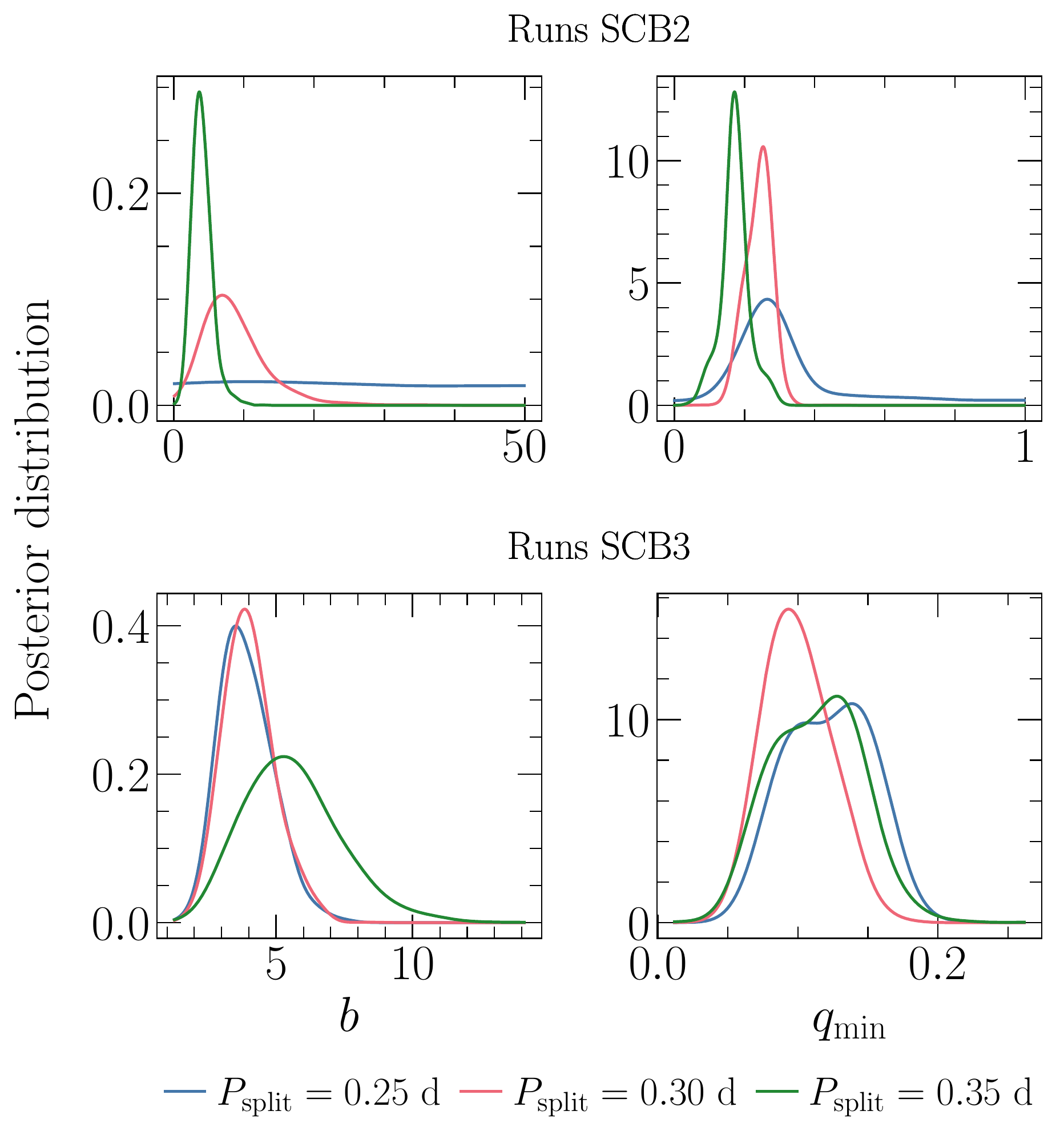}
\caption{Dependence of the fiducial posterior distributions of $b$ (left) and $q_\mathrm{min}$ (right) for populations CB2 and CB3 on different splitting periods between the two populations. \label{fig:posterior_distributions_D}}
\end{figure}

\begin{figure}
\centering
\includegraphics[width=0.45\textwidth]{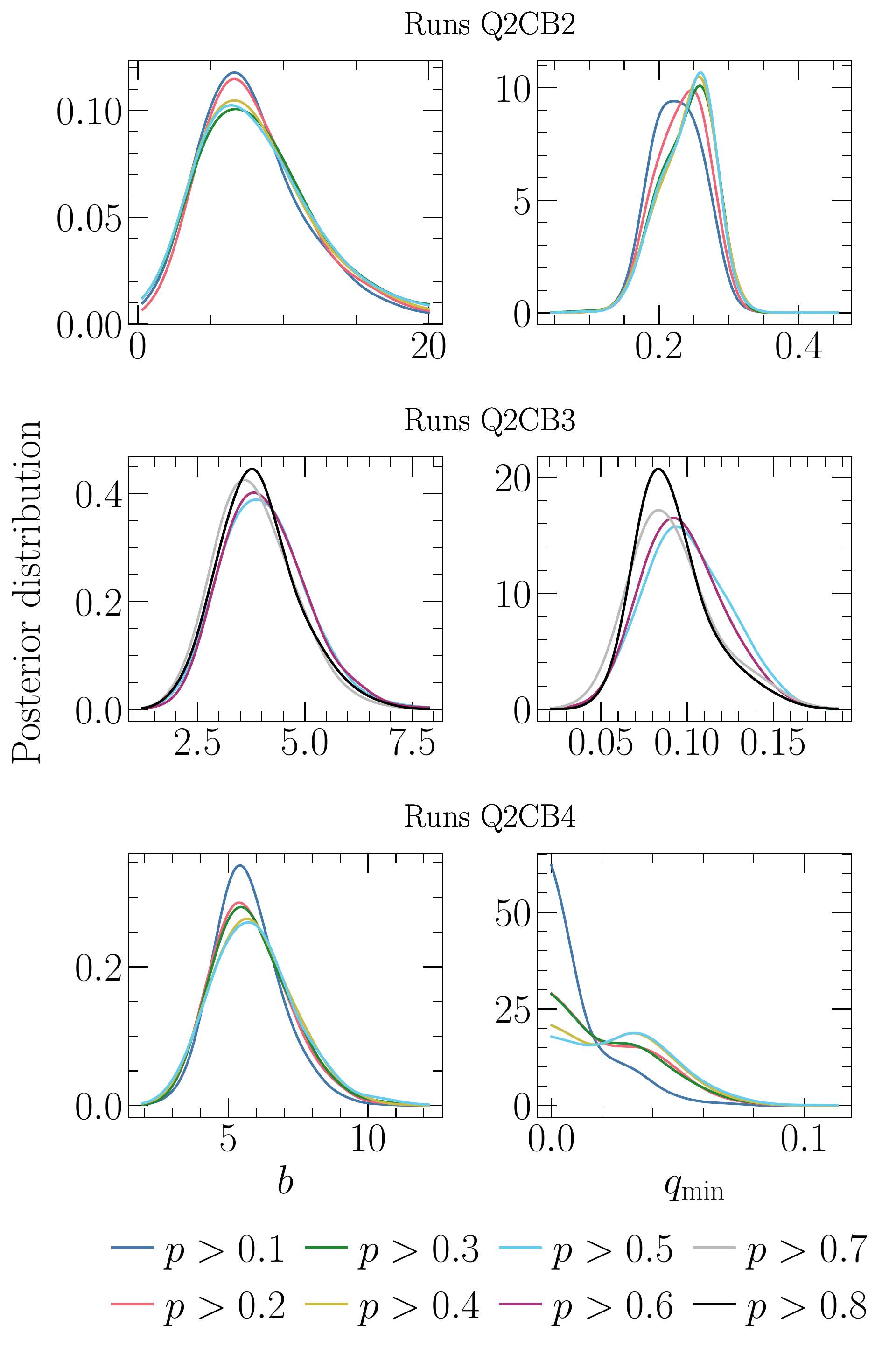}
\caption{Dependence of the fiducial posterior distributions of $b$ (left) and $q_\mathrm{min}$ (right) for populations CB2--CB4 on different probability cutoffs. \label{fig:posterior_distributions_Q2}}
\end{figure}

\begin{figure*}
\centering
\includegraphics[width=0.85\textwidth]{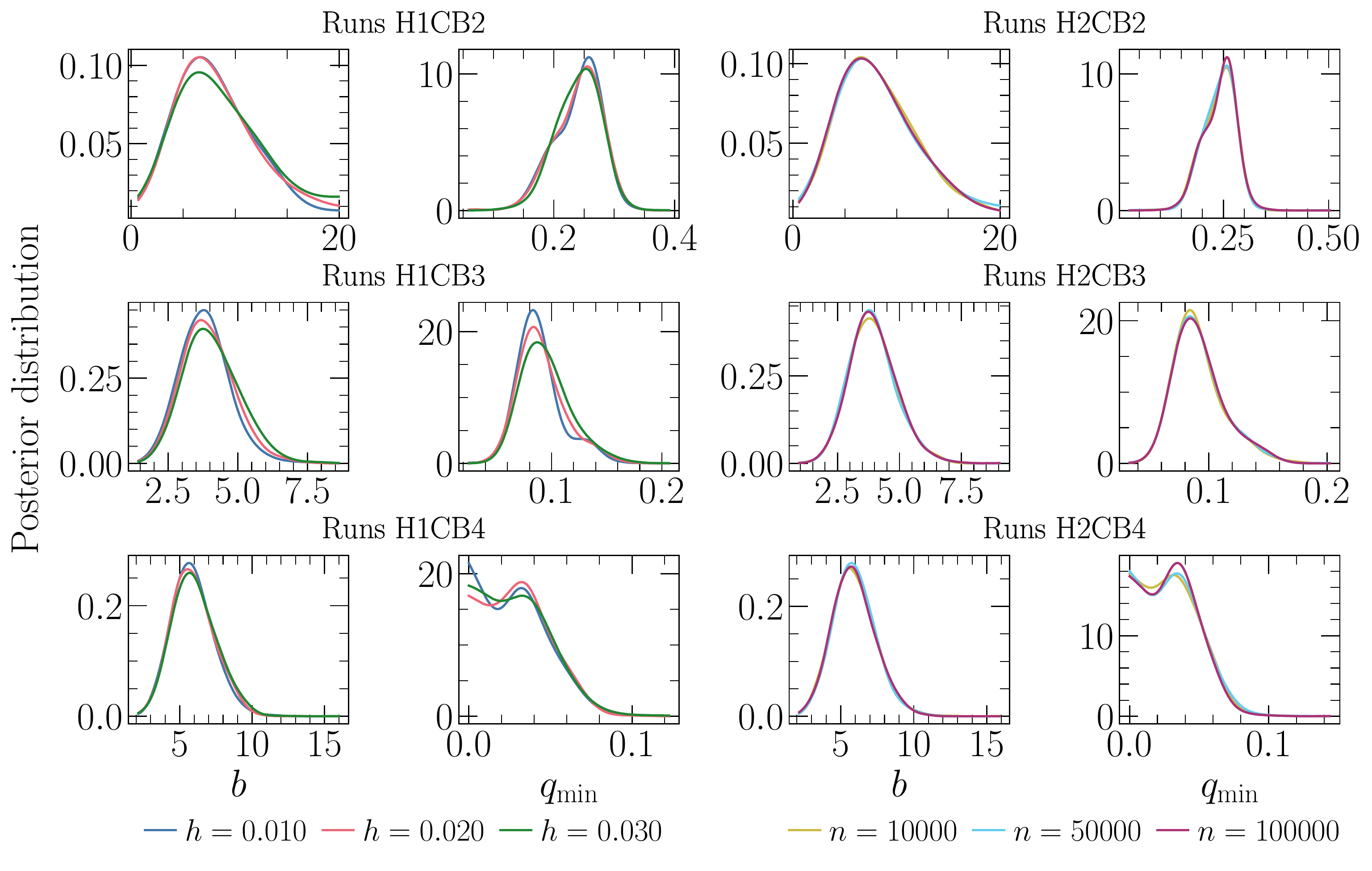}
\caption{Dependence of the fiducial posterior distributions of $b$ and $q_\mathrm{min}$ for populations CB2--CB4 on different values of $h$ (left) and $n$ (right). \label{fig:posterior_distributions_H}}
\end{figure*}

We show how the posterior distributions of $b$ and $\qmin$ depend on the two mass-ratio distribution parameterizations (Fig.~\ref{fig:posterior_Q1_Q2}), fill-out factors (Fig.~\ref{fig:posterior_distributions_F}), splitting periods (Fig.~\ref{fig:posterior_distributions_D}), probability cutoffs (Fig.~\ref{fig:posterior_distributions_Q2}), and model hyperparameters (Fig.~\ref{fig:posterior_distributions_H}). We also present a complete list of all our \emph{emcee} runs together with the resulting values of $b$ and $\qmin$ (Tab.~\ref{tab:runs}).

\vspace*{\fill}
\vspace*{\fill}

\onecolumn
\begin{landscape}
\begin{longtable}{rcccccccccccc}
\caption{Full list of all \emph{emcee} runs sampling the posterior distributions of the parameters of the mass-ratio distribution for different power-law prescriptions, probability cutoffs, fill-out factors, and hyperparameters. \label{tab:runs}}\\
\hline\hline
Run & $Q$ & $f$ & $h$ & $n$ & Sample & Type & Prob. cutoff & $P$ (days) & Eff. size & Autocorr. time & $b$ & $\qmin$\\
\hline
\endfirsthead
\caption{continued.}\\
\hline\hline
Run & $Q$ & $f$ & $h$ & $n$ & Sample & Type & Prob. cutoff & $P$ (days) & Eff. size & Autocorr. time & $b$ & $\qmin$\\
\hline
\endhead
\hline
\endfoot
Q1CB1p50 & $Q_1$ & $0.25$ & 0.02 & 10000 & CB1p50 & Late & $0.5$ & --- & 258.99 & 11.25 & $2.69^{+0.48}_{-0.43}$ & $0.164^{+0.016}_{-0.015}$\\
Q1CB1p60 & $Q_1$ & $0.25$ & 0.02 & 10000 & CB1p60 & Late & $0.6$ & --- & 256.27 & 11.47 & $2.79^{+0.46}_{-0.48}$ & $0.167^{+0.014}_{-0.015}$\\
Q1CB1p70 & $Q_1$ & $0.25$ & 0.02 & 10000 & CB1p70 & Late & $0.7$ & --- & 249.00 & 10.94 & $2.74^{+0.53}_{-0.48}$ & $0.170^{+0.019}_{-0.017}$\\
Q1CB1p80 & $Q_1$ & $0.25$ & 0.02 & 10000 & CB1p80 & Late & $0.8$ & --- & 228.04 & 1.11 & $2.79^{+0.62}_{-0.52}$ & $0.166^{+0.020}_{-0.017}$\\
Q1CB2p10 & $Q_1$ & $0.25$ & 0.02 & 10000 & CB2p10 & Late & $0.1$ & $\le0.3$ & 62.40 & 13.24 & $5.29^{+2.70}_{-1.83}$ & $0.250^{+0.027}_{-0.037}$\\
Q1CB2p20 & $Q_1$ & $0.25$ & 0.02 & 10000 & CB2p20 & Late & $0.2$ & $\le0.3$ & 61.96 & 15.8 & $5.69^{+2.55}_{-1.99}$ & $0.256^{+0.024}_{-0.032}$\\
Q1CB2p30 & $Q_1$ & $0.25$ & 0.02 & 10000 & CB2p30 & Late & $0.3$ & $\le0.3$ & 61.38 & 16.97 & $5.90^{+3.04}_{-2.12}$ & $0.263^{+0.023}_{-0.033}$\\
Q1CB2p40 & $Q_1$ & $0.25$ & 0.02 & 10000 & CB2p40 & Late & $0.4$ & $\le0.3$ & 61.07 & 14.61 & $6.01^{+2.59}_{-2.22}$ & $0.263^{+0.020}_{-0.032}$\\
Q1CB2p50 & $Q_1$ & $0.25$ & 0.02 & 10000 & CB2p50 & Late & $0.5$ & $\le0.3$ & 60.59 & 15.69 & $5.96^{+2.75}_{-2.18}$ & $0.262^{+0.022}_{-0.031}$\\
Q1CB3p50 & $Q_1$ & $0.25$ & 0.02 & 10000 & CB3p50 & Late & $0.5$ & $>0.3$ & 198.41 & 1.9 & $2.38^{+0.46}_{-0.47}$ & $0.136^{+0.015}_{-0.020}$\\
Q1CB3p60 & $Q_1$ & $0.25$ & 0.02 & 10000 & CB3p60 & Late & $0.6$ & $>0.3$ & 197.30 & 9.88 & $2.46^{+0.50}_{-0.48}$ & $0.138^{+0.017}_{-0.023}$\\
Q1CB3p70 & $Q_1$ & $0.25$ & 0.02 & 10000 & CB3p70 & Late & $0.7$ & $>0.3$ & 192.69 & 10.79 & $2.38^{+0.53}_{-0.55}$ & $0.139^{+0.017}_{-0.028}$\\
Q1CB3p80 & $Q_1$ & $0.25$ & 0.02 & 10000 & CB3p80 & Late & $0.8$ & $>0.3$ & 177.76 & 10.39 & $2.25^{+0.61}_{-0.53}$ & $0.131^{+0.021}_{-0.033}$\\
Q1CB4p10 & $Q_1$ & $0.25$ & 0.02 & 10000 & CB4p10 & Early & $0.1$ & $<1$ & 106.42 & 10.55 & $1.66^{+0.38}_{-0.29}$ & $0.044^{+0.015}_{-0.013}$\\
Q1CB4p20 & $Q_1$ & $0.25$ & 0.02 & 10000 & CB4p20 & Early & $0.2$ & $<1$ & 105.56 & 13.19 & $1.81^{+0.36}_{-0.32}$ & $0.054^{+0.013}_{-0.015}$\\
Q1CB4p30 & $Q_1$ & $0.25$ & 0.02 & 10000 & CB4p30 & Early & $0.3$ & $<1$ & 105.56 & 11.9 & $1.80^{+0.36}_{-0.33}$ & $0.054^{+0.013}_{-0.015}$\\
Q1CB4p40 & $Q_1$ & $0.25$ & 0.02 & 10000 & CB4p40 & Early & $0.4$ & $<1$ & 104.91 & 9.56 & $1.82^{+0.35}_{-0.31}$ & $0.055^{+0.012}_{-0.014}$\\
Q1CB4p50 & $Q_1$ & $0.25$ & 0.02 & 10000 & CB4p50 & Early & $0.5$ & $<1$ & 104.42 & 11.77 & $1.84^{+0.35}_{-0.35}$ & $0.056^{+0.013}_{-0.014}$\\
Q1CB5p10 & $Q_1$ & $0.25$ & 0.02 & 10000 & CB5p10 & Early & $0.1$ & --- & 162.62 & 25.9 & $1.01^{+0.09}_{-0.08}$ & $0.001^{+0.001}_{-0.001}$\\
Q1CB5p20 & $Q_1$ & $0.25$ & 0.02 & 10000 & CB5p20 & Early & $0.2$ & --- & 161.55 & 24.71 & $0.98^{+0.10}_{-0.08}$ & $0.001^{+0.001}_{-0.001}$\\
Q1CB5p30 & $Q_1$ & $0.25$ & 0.02 & 10000 & CB5p30 & Early & $0.3$ & --- & 160.80 & 29.96 & $0.98^{+0.10}_{-0.09}$ & $0.001^{+0.002}_{-0.001}$\\
Q1CB5p40 & $Q_1$ & $0.25$ & 0.02 & 10000 & CB5p40 & Early & $0.4$ & --- & 159.49 & 30.56 & $0.97^{+0.09}_{-0.08}$ & $0.002^{+0.002}_{-0.001}$\\
Q1CB5p50 & $Q_1$ & $0.25$ & 0.02 & 10000 & CB5p50 & Early & $0.5$ & --- & 158.58 & 28.57 & $0.98^{+0.10}_{-0.09}$ & $0.002^{+0.002}_{-0.001}$\\
Q2CB1p50 & $Q_2$ & $0.25$ & 0.02 & 10000 & CB1p50 & Late & $0.5$ & --- & 258.99 & 10.61 & $4.09^{+1.01}_{-0.95}$ & $0.138^{+0.023}_{-0.035}$\\
Q2CB1p60 & $Q_2$ & $0.25$ & 0.02 & 10000 & CB1p60 & Late & $0.6$ & --- & 256.27 & 11.86 & $4.17^{+1.09}_{-0.95}$ & $0.141^{+0.022}_{-0.038}$\\
Q2CB1p70 & $Q_2$ & $0.25$ & 0.02 & 10000 & CB1p70 & Late & $0.7$ & --- & 249.00 & 12.11 & $3.84^{+1.16}_{-0.92}$ & $0.135^{+0.029}_{-0.047}$\\
Q2CB1p80 & $Q_2$ & $0.25$ & 0.02 & 10000 & CB1p80 & Late & $0.8$ & --- & 228.04 & 12.55 & $3.96^{+1.23}_{-0.98}$ & $0.127^{+0.033}_{-0.044}$\\
Q2CB2p10 & $Q_2$ & $0.25$ & 0.02 & 10000 & CB2p10 & Late & $0.1$ & $\le0.3$ & 62.40 & 14.76 & $7.30^{+4.23}_{-2.72}$ & $0.226^{+0.037}_{-0.035}$\\
Q2CB2p20 & $Q_2$ & $0.25$ & 0.02 & 10000 & CB2p20 & Late & $0.2$ & $\le0.3$ & 61.96 & 13.47 & $7.56^{+4.05}_{-2.78}$ & $0.235^{+0.032}_{-0.041}$\\
Q2CB2p30 & $Q_2$ & $0.25$ & 0.02 & 10000 & CB2p30 & Late & $0.3$ & $\le0.3$ & 61.38 & 15.25 & $7.73^{+4.43}_{-3.11}$ & $0.244^{+0.031}_{-0.046}$\\
Q2CB2p40 & $Q_2$ & $0.25$ & 0.02 & 10000 & CB2p40 & Late & $0.4$ & $\le0.3$ & 61.07 & 13.95 & $7.58^{+4.37}_{-3.09}$ & $0.246^{+0.029}_{-0.045}$\\
Q2CB2p50 & $Q_2$ & $0.25$ & 0.02 & 10000 & CB2p50 & Late & $0.5$ & $\le0.3$ & 60.59 & 12.93 & $7.66^{+4.45}_{-3.15}$ & $0.246^{+0.029}_{-0.046}$\\
Q2CB3p50 & $Q_2$ & $0.25$ & 0.02 & 10000 & CB3p50 & Late & $0.5$ & $>0.3$ & 198.41 & 10.25 & $3.98^{+1.01}_{-0.87}$ & $0.098^{+0.028}_{-0.021}$\\
Q2CB3p60 & $Q_2$ & $0.25$ & 0.02 & 10000 & CB3p60 & Late & $0.6$ & $>0.3$ & 197.30 & 9.51 & $3.98^{+1.02}_{-0.84}$ & $0.096^{+0.026}_{-0.020}$\\
Q2CB3p70 & $Q_2$ & $0.25$ & 0.02 & 10000 & CB3p70 & Late & $0.7$ & $>0.3$ & 192.69 & 9.75 & $3.75^{+0.99}_{-0.77}$ & $0.088^{+0.027}_{-0.019}$\\
Q2CB3p80 & $Q_2$ & $0.25$ & 0.02 & 10000 & CB3p80 & Late & $0.8$ & $>0.3$ & 177.76 & 9.66 & $3.84^{+0.96}_{-0.80}$ & $0.087^{+0.024}_{-0.015}$\\
Q2CB4p10 & $Q_2$ & $0.25$ & 0.02 & 10000 & CB4p10 & Early & $0.1$ & $<1$ & 106.42 & 20.35 & $5.57^{+1.23}_{-1.00}$ & $0.008^{+0.020}_{-0.006}$\\
Q2CB4p20 & $Q_2$ & $0.25$ & 0.02 & 10000 & CB4p20 & Early & $0.2$ & $<1$ & 105.56 & 16.96 & $5.64^{+1.45}_{-1.14}$ & $0.022^{+0.023}_{-0.017}$\\
Q2CB4p30 & $Q_2$ & $0.25$ & 0.02 & 10000 & CB4p30 & Early & $0.3$ & $<1$ & 105.56 & 15.21 & $5.70^{+1.52}_{-1.15}$ & $0.023^{+0.021}_{-0.018}$\\
Q2CB4p40 & $Q_2$ & $0.25$ & 0.02 & 10000 & CB4p40 & Early & $0.4$ & $<1$ & 104.91 & 13.68 & $5.80^{+1.54}_{-1.27}$ & $0.029^{+0.018}_{-0.022}$\\
Q2CB4p50 & $Q_2$ & $0.25$ & 0.02 & 10000 & CB4p50 & Early & $0.5$ & $<1$ & 104.42 & 13.28 & $5.82^{+1.52}_{-1.30}$ & $0.030^{+0.018}_{-0.022}$\\
Q2CB5p10 & $Q_2$ & $0.25$ & 0.02 & 10000 & CB5p10 & Early & $0.1$ & --- & 162.62 & 25.37 & $8.52^{+1.08}_{-1.02}$ & $0.001^{+0.001}_{-0.001}$\\
Q2CB5p20 & $Q_2$ & $0.25$ & 0.02 & 10000 & CB5p20 & Early & $0.2$ & --- & 161.55 & 23.00 & $8.17^{+1.10}_{-1.06}$ & $0.001^{+0.001}_{-0.001}$\\
Q2CB5p30 & $Q_2$ & $0.25$ & 0.02 & 10000 & CB5p30 & Early & $0.3$ & --- & 160.80 & 30.26 & $7.98^{+1.03}_{-1.06}$ & $0.001^{+0.002}_{-0.001}$\\
Q2CB5p40 & $Q_2$ & $0.25$ & 0.02 & 10000 & CB5p40 & Early & $0.4$ & --- & 159.49 & 25.45 & $7.85^{+1.01}_{-1.01}$ & $0.001^{+0.002}_{-0.001}$\\
Q2CB5p50 & $Q_2$ & $0.25$ & 0.02 & 10000 & CB5p50 & Early & $0.5$ & --- & 158.58 & 24.71 & $7.71^{+1.05}_{-1.01}$ & $0.001^{+0.002}_{-0.001}$\\
FCB2f015 & $Q_2$ & $0.15$ & 0.02 & 10000 & CB2p50 & Late & $0.5$ & $\le0.3$ & 60.59 & 20.83 & $7.54^{+4.96}_{-3.11}$ & $0.266^{+0.028}_{-0.051}$\\
FCB2f025 & $Q_2$ & $0.25$ & 0.02 & 10000 & CB2p50 & Late & $0.5$ & $\le0.3$ & 60.59 & 16.90 & $7.71^{+4.81}_{-2.97}$ & $0.247^{+0.030}_{-0.044}$\\
FCB2f050 & $Q_2$ & $0.50$ & 0.02 & 10000 & CB2p50 & Late & $0.5$ & $\le0.3$ & 60.59 & 18.91 & $9.87^{+6.04}_{-3.78}$ & $0.219^{+0.026}_{-0.043}$\\
FCB2f075 & $Q_2$ & $0.75$ & 0.02 & 10000 & CB2p50 & Late & $0.5$ & $\le0.3$ & 60.59 & 19.59 & $12.36^{+6.88}_{-4.69}$ & $0.189^{+0.024}_{-0.037}$\\
FCB2f099 & $Q_2$ & $0.99$ & 0.02 & 10000 & CB2p50 & Late & $0.5$ & $\le0.3$ & 60.59 & 18.10 & $16.72^{+9.72}_{-6.04}$ & $0.156^{+0.021}_{-0.027}$\\
FCB3f015 & $Q_2$ & $0.15$ & 0.02 & 10000 & CB3p80 & Late & $0.8$ & $>0.3$ & 177.76 & 11.34 & $3.44^{+0.89}_{-0.81}$ & $0.092^{+0.025}_{-0.016}$\\
FCB3f025 & $Q_2$ & $0.25$ & 0.02 & 10000 & CB3p80 & Late & $0.8$ & $>0.3$ & 177.76 & 10.79 & $3.84^{+0.97}_{-0.84}$ & $0.087^{+0.026}_{-0.015}$\\
FCB3f050 & $Q_2$ & $0.50$ & 0.02 & 10000 & CB3p80 & Late & $0.8$ & $>0.3$ & 177.76 & 10.39 & $4.85^{+1.09}_{-0.95}$ & $0.074^{+0.017}_{-0.012}$\\
FCB3f075 & $Q_2$ & $0.75$ & 0.02 & 10000 & CB3p80 & Late & $0.8$ & $>0.3$ & 177.76 & 1.12 & $6.34^{+1.14}_{-1.15}$ & $0.064^{+0.013}_{-0.010}$\\
FCB3f099 & $Q_2$ & $0.99$ & 0.02 & 10000 & CB3p80 & Late & $0.8$ & $>0.3$ & 177.76 & 11.95 & $8.69^{+1.62}_{-1.40}$ & $0.057^{+0.009}_{-0.009}$\\
FCB4f015 & $Q_2$ & $0.15$ & 0.02 & 10000 & CB4p50 & Early & $0.5$ & $<1$ & 104.42 & 12.25 & $5.51^{+1.35}_{-1.27}$ & $0.037^{+0.019}_{-0.024}$\\
FCB4f025 & $Q_2$ & $0.15$ & 0.02 & 10000 & CB4p50 & Early & $0.5$ & $<1$ & 104.42 & 12.64 & $5.79^{+1.49}_{-1.26}$ & $0.030^{+0.020}_{-0.022}$\\
FCB4f050 & $Q_2$ & $0.15$ & 0.02 & 10000 & CB4p50 & Early & $0.5$ & $<1$ & 104.42 & 20.48 & $6.78^{+1.66}_{-1.34}$ & $0.017^{+0.020}_{-0.013}$\\
FCB4f075 & $Q_2$ & $0.15$ & 0.02 & 10000 & CB4p50 & Early & $0.5$ & $<1$ & 104.42 & 19.77 & $8.26^{+2.07}_{-1.55}$ & $0.014^{+0.018}_{-0.011}$\\
FCB4f099 & $Q_2$ & $0.15$ & 0.02 & 10000 & CB4p50 & Early & $0.5$ & $<1$ & 104.42 & 14.56 & $10.71^{+2.43}_{-1.85}$ & $0.015^{+0.013}_{-0.011}$\\
H1CB2h001 & $Q_2$ & $0.25$ & 0.01 & 10000 & CB2p50 & Late & $0.5$ & $\le0.3$ & 60.59 & 16.76 & $7.62^{+4.48}_{-3.09}$ & $0.248^{+0.027}_{-0.050}$\\
H1CB2h002 & $Q_2$ & $0.25$ & 0.02 & 10000 & CB2p50 & Late & $0.5$ & $\le0.3$ & 60.59 & 14.10 & $7.68^{+4.66}_{-3.03}$ & $0.246^{+0.030}_{-0.048}$\\
H1CB2h003 & $Q_2$ & $0.25$ & 0.03 & 10000 & CB2p50 & Late & $0.5$ & $\le0.3$ & 60.59 & 14.74 & $7.98^{+4.91}_{-3.34}$ & $0.244^{+0.032}_{-0.040}$\\
H1CB3h001 & $Q_2$ & $0.25$ & 0.01 & 10000 & CB3p80 & Late & $0.8$ & $>0.3$ & 177.76 & 1.12 & $3.78^{+0.84}_{-0.81}$ & $0.085^{+0.020}_{-0.014}$\\
H1CB3h002 & $Q_2$ & $0.25$ & 0.02 & 10000 & CB3p80 & Late & $0.8$ & $>0.3$ & 177.76 & 1.10 & $3.85^{+0.96}_{-0.79}$ & $0.087^{+0.024}_{-0.015}$\\
H1CB3h003 & $Q_2$ & $0.25$ & 0.03 & 10000 & CB3p80 & Late & $0.8$ & $>0.3$ & 177.76 & 10.39 & $3.98^{+1.07}_{-0.83}$ & $0.092^{+0.024}_{-0.018}$\\
H1CB4h001 & $Q_2$ & $0.25$ & 0.01 & 10000 & CB4p50 & Early & $0.5$ & $<1$ & 104.42 & 15.33 & $5.80^{+1.44}_{-1.25}$ & $0.030^{+0.019}_{-0.024}$\\
H1CB4h002 & $Q_2$ & $0.25$ & 0.02 & 10000 & CB4p50 & Early & $0.5$ & $<1$ & 104.42 & 13.40 & $5.82^{+1.56}_{-1.26}$ & $0.031^{+0.019}_{-0.022}$\\
H1CB4h003 & $Q_2$ & $0.25$ & 0.03 & 10000 & CB4p50 & Early & $0.5$ & $<1$ & 104.42 & 11.51 & $5.90^{+1.63}_{-1.28}$ & $0.030^{+0.021}_{-0.021}$\\
H2CB2n10k & $Q_2$ & $0.25$ & 0.02 & 10000 & CB2p50 & Late & $0.5$ & $\le0.3$ & 60.59 & 17.3 & $7.74^{+4.49}_{-2.98}$ & $0.245^{+0.029}_{-0.048}$\\
H2CB2n50k & $Q_2$ & $0.25$ & 0.02 & 50000 & CB2p50 & Late & $0.5$ & $\le0.3$ & 60.59 & 14.76 & $7.70^{+4.78}_{-3.09}$ & $0.245^{+0.029}_{-0.043}$\\
H2CB2n100k & $Q_2$ & $0.25$ & 0.02 & 100000 & CB2p50 & Late & $0.5$ & $\le0.3$ & 60.59 & 14.73 & $7.75^{+4.51}_{-3.14}$ & $0.249^{+0.025}_{-0.050}$\\
H2CB3n10k & $Q_2$ & $0.25$ & 0.02 & 10000 & CB3p80 & Late & $0.8$ & $>0.3$ & 177.76 & 1.11 & $3.87^{+0.95}_{-0.84}$ & $0.087^{+0.024}_{-0.015}$\\
H2CB3n50k & $Q_2$ & $0.25$ & 0.02 & 50000 & CB3p80 & Late & $0.8$ & $>0.3$ & 177.76 & 9.22 & $3.84^{+0.94}_{-0.82}$ & $0.088^{+0.024}_{-0.015}$\\
H2CB3n100k & $Q_2$ & $0.25$ & 0.02 & 100000 & CB3p80 & Late & $0.8$ & $>0.3$ & 177.76 & 9.75 & $3.88^{+0.95}_{-0.78}$ & $0.088^{+0.024}_{-0.016}$\\
H2CB4n10k & $Q_2$ & $0.25$ & 0.02 & 10000 & CB4p50 & Early & $0.5$ & $<1$ & 104.42 & 12.42 & $5.81^{+1.52}_{-1.26}$ & $0.030^{+0.020}_{-0.022}$\\
H2CB4n50k & $Q_2$ & $0.25$ & 0.02 & 50000 & CB4p50 & Early & $0.5$ & $<1$ & 104.42 & 11.28 & $5.87^{+1.40}_{-1.27}$ & $0.031^{+0.019}_{-0.023}$\\
H2CB4h100k & $Q_2$ & $0.25$ & 0.02 & 100000 & CB4p50 & Early & $0.5$ & $<1$ & 104.42 & 11.26 & $5.82^{+1.52}_{-1.28}$ & $0.031^{+0.018}_{-0.022}$\\
SCB2P025 & $Q_2$ & $0.25$ & 0.02 & 10000 & CB2p50 & Late & $0.5$ & $\le0.25$ & 9.96 & 39.71 & $23.37^{+18.37}_{-15.40}$ & $0.276^{+0.173}_{-0.069}$\\
SCB2P030 & $Q_2$ & $0.25$ & 0.02 & 10000 & CB2p50 & Late & $0.5$ & $\le0.30$ & 60.59 & 20.52 & $7.87^{+4.47}_{-3.15}$ & $0.245^{+0.030}_{-0.045}$\\
SCB2P035 & $Q_2$ & $0.25$ & 0.02 & 10000 & CB2p50 & Late & $0.5$ & $\le0.35$ & 135.36 & 11.7 & $3.86^{+1.38}_{-1.14}$ & $0.172^{+0.032}_{-0.031}$\\
SCB3P025 & $Q_2$ & $0.25$ & 0.02 & 10000 & CB3p50 & Late & $0.5$ & $>0.25$ & 249.04 & 1.11 & $3.79^{+1.09}_{-0.80}$ & $0.124^{+0.030}_{-0.035}$\\
SCB3P030 & $Q_2$ & $0.25$ & 0.02 & 10000 & CB3p50 & Late & $0.5$ & $>0.30$ & 198.41 & 10.51 & $3.94^{+0.95}_{-0.83}$ & $0.098^{+0.027}_{-0.021}$\\
SCB3P035 & $Q_2$ & $0.25$ & 0.02 & 10000 & CB3p50 & Late & $0.5$ & $>0.35$ & 123.63 & 12.50 & $5.45^{+1.83}_{-1.57}$ & $0.115^{+0.028}_{-0.036}$\\
\end{longtable}
\end{landscape}

\end{appendix}
\end{document}